\documentclass[aps,pra,amsmath,amssymb,twocolumn,showpacs,letterpaper,superscriptaddress]{revtex4-1}

\usepackage{hyperref}
\usepackage{graphicx}
\usepackage{dcolumn}

\newcommand{\ftm}[1]{\footnotemark[#1]}

\begin{document}

\title{A cluster-based mean-field and perturbative description of
  strongly correlated fermion systems. Application to the 1D and 2D
  Hubbard model}

\author{Carlos A. Jim\'enez-Hoyos}
\email{jimenez.hoyos@gmail.com}
\affiliation{Department of Chemistry, Rice University, Houston, TX
  77005}

\author{Gustavo E. Scuseria}
\affiliation{Department of Chemistry, Rice University, Houston, TX
  77005}
\affiliation{Department of Physics and Astronomy, Rice University,
  Houston, TX 77005}

\date{\today}

\pacs{71.27.+a, 74.20.Pq, 71.10.Fd}

\begin{abstract}
We introduce a mean-field and perturbative approach, based on
clusters, to describe the ground state of fermionic
strongly-correlated systems. In cluster mean-field, the ground state
wavefunction is written as a simple tensor product over optimized
cluster states. The optimization of the single-particle basis where
the cluster mean-field is expressed is crucial in order to obtain
high-quality results. The mean-field nature of the ansatz allows us to
formulate a perturbative approach to account for inter-cluster
correlations; other traditional many-body strategies can be easily
devised in terms of the cluster states. We present benchmark
calculations on the half-filled 1D and (square) 2D Hubbard model, as
well as the lightly-doped regime in 2D, using cluster mean-field and
second-order perturbation theory. Our results indicate that, with
sufficiently large clusters or to second-order in perturbation theory,
a cluster-based approach can provide an accurate description of the
Hubbard model in the considered regimes. Several avenues to improve
upon the results presented in this work are discussed.
\end{abstract}

\maketitle

\section{Introduction}
\label{sec:introduction}

Despite some substantial recent progress, an accurate and efficient
description of the ground and excited states of low-dimensional
strongly-correlated fermionic systems represents an open problem in
condensed matter physics and quantum chemistry. A common feature in
strongly-correlated systems is the collective behavior displayed by
fermions in low-lying states. Accordingly, approaches based on
composite particles have been proposed for treating these systems. One
notorious example is the resonating valence bond as a ground state
candidate for high-$T_c$ superconductors, as suggested by Anderson
\cite{anderson-1987}.

In this work, we use composite many-fermion cluster states to describe
the ground state of strongly-correlated systems. Here, a cluster is
simply a subset of all available single-fermion states that we group
(generally using a criterion of proximity in real space). We presume
that an accurate zero-th order description of the ground state of the
full system can be prepared as a product of cluster states, each being
many fermion in character. Two key aspects in obtaining an accurate
description are: 1) the many-fermion state in each cluster is
determined {\em in the presence} of other clusters, and 2) the
single-particle basis used to determine the grouping into clusters is
fully optimized. This optimization could in principle break the real
space localization criterion but in practice it generally does
not. The resulting cluster mean-field (cMF) state is, by construction,
guaranteed to provide a variational estimate of the ground state
energy that is lower than Hartree--Fock (HF), {\em i.e.}, the standard
mean-field of single-particles. The optimization provides not only the
optimal cMF state, but also a renormalized Hamiltonian expressed in
term of cluster states. Traditional many-body approaches can then be
used, on this renormalized Hamiltonian, to account for the missing
inter-cluster correlations.

Our work is inspired by McWeeny \cite{mcweeny-1959,mcweeny-1960}, who
first considered the properties of wavefunctions written as a tensor
product of the state of several subsystems (or groups), which are
mutually orthogonal. McWeeny realized that the density matrix of
cluster product states can be easily expressed in terms of the density
matrices of the individual clusters. In related work, Isaev, Ortiz,
and Dukelsky \cite{isaev-2009} considered, in their hierarchical
mean-field (HMF) approach, a similar ansatz to ours for the 2D
Heisenberg Hamiltonian. We note, nevertheless, that the generalization
to fermionic systems of the HMF approach used in
Ref.~\onlinecite{isaev-2009} is not straightforward if the full Fock
space within each cluster is treated.  An attempt was performed in
Ref.~\onlinecite{zhao-2014} to split the Fock space in each cluster
according to its number parity: states with {\em even} ({\em odd})
parity where treated as bosonic (fermionic) degrees of freedom. An
ansatz for the ground state was constructed in
Ref.~\onlinecite{zhao-2014} as a tensor product of the bosonic and
fermionic degrees of freedom; this decoupling, however, may not be
justified in all cases and can potentially result in unphysical
states.

Our approach differs from that used in Ref.~\onlinecite{isaev-2009},
aside from the application to fermionic systems, in not requiring the
individual clusters to share the same ground state. That is, the
ground state of each cluster is optimized independently allowing for
(translational and spin) symmetry-broken solutions. In addition, we
here consider Rayleigh-Schr\"odinger perturbation theory (RS-PT)
\cite{schrodinger-1926} to second-order as a means to obtain a
correlated approach defined in terms of clusters.

A closely related cluster product approach was also proposed by Li
\cite{li-2004}. In that work, the ground state of each cluster was,
nonetheless, not optimized in the presence of other clusters. The
author did, on the other hand, go beyond perturbation theory into a
coupled-cluster ansatz (so-called block-correlated coupled cluster
(BCCC)) as a way to correlate the cluster product state. The BCCC
approach has been used with high success in quantum chemistry to
describe strongly-correlated molecular systems using either a complete
active-space \cite{fang-2008,shen-2009} or generalized valence-bond
\cite{xu-2013} reference states.

A cluster product state is naturally connected with the tensor network
(TN) techniques that have been gaining popularity for treating
strongly-correlated systems \cite{cirac-2009,orus-2014}. In essence,
the cluster product state is the simplest possible TN, a simple scalar
(the bond or ancillary index dimension is set to 1), although the
elements defining the network are chosen as cluster states rather than
single-particle degrees of freedom as often done. The consequence of
using a scalar product is that the clusters become disentangled; more
general TNs such as the matrix product states (MPS) used within the
density matrix renormalization group algorithm (DMRG)
\cite{white-1992,white-1993} introduce entanglement in the ansatz and
can provide highly accurate solutions for strongly-correlated
systems. The optimization of TN states beyond the simple MPS is,
however, non-trivial and remains an area of active research
\cite{orus-2014}.

Yet other wavefunction ans\"atze that are related to the cluster
product states are the correlator product states (CPS)
\cite{changlani-2009} or entangled plaquette states (EPS)
\cite{mezzacapo-2009,mezzacapo-2010}. Here, a variational ansatz is
expressed in terms of entangled cluster products, as opposed to a
simple uncorrelated product. The price to pay is that the evaluation
of matrix elements becomes more cumbersome, and it often has to be
carried out by stochastic means (within a variational Monte Carlo
framework).  We note that Ref. \onlinecite{neuscamman-2011} proposed a
non-stochastic algorithm for optimizing CPS.

At this point, we want to clarify why we have decided to use simple
cluster product states even when more powerful ans\"atze are already
available (such as more general TNs or CPS). In our perspective, the
power of a cluster mean-field approach has not been fully realized. In
particular, symmetry breaking and orbital optimization can partially
account for inter-cluster correlations (when expressed in terms of the
original set of fermion states). Moreover, the fact that the cluster
mean-field state constitutes the ground state of a mean-field
Hamiltonian of which the full set of eigenstates can be easily
constructed has often been overlooked. This allows us to formulate a
perturbative strategy to introduce the missing inter-cluster
correlations. Yet more powerful many-body approaches (such as
coupled-cluster theory) can be easily introduced as done by Li in the
BCCC method.

Our objective with this work is thus two-fold. First, we present the
cMF formalism as well as provide details of the RS-PT formulation we
use. (We refer to the second-order perturbative result as cPT2 in the
remainder of this paper.) We describe in some detail the strategy used
to optimizate the one-electron basis in which the cMF state is
expressed. Our second objective is to apply these techniques to the
simplest paradigm of strongly-correlated fermionic systems: the
Hubbard model \cite{hubbard-1963} in one- (1D) and two-dimensions (2D)
in a square lattice. The 1D case is exactly solvable \cite{lieb-1968},
while the 2D model has been extensively studied. We refer the reader
to Refs.~\onlinecite{scalapino,leblanc-2015} for a survey of numerical
methods applied to the 2D Hubbard model. We compare our results with
second-order unrestricted M\o ller-Plesset (UMP2) \cite{moller-1934}
and unrestricted coupled-cluster with singles and doubles (UCCSD)
\cite{bartlett}, as well as with perturbative triples (UCCSD(T)),
calculations. MP2 constitutes second-order RS-PT based on a HF
wavefunction (using canonical orbital and orbital energies), while
coupled-cluster constitutes a non-perturbative approach that involves
an infinite-order resummation of diagrams. Our results show that cMF
(cPT2) significantly improves upon HF (MP2) and can provide an
accurate description of the ground state of the Hubbard model.

The remainder of this work is organized as follows. In
Sec.~\ref{sec:formalism} we present the formalism behind cMF and
cPT2. Section~\ref{sec:comp_details} provides some practical
computational details regarding the calculations presented in this
work. In Sec.~\ref{sec:results} we present the results of cMF and cPT2
calculations for the half-filled 1D and 2D Hubbard model, as well as
for the lightly-doped 2D regime. A brief discussion following the
results is presented in Sec.~\ref{sec:discussion}, along with some
ideas as to how to improve the calculations here presented. Lastly,
Sec.~\ref{sec:conclusions} is devoted to some general conclusions. In
Appendix \ref{sec:appendix_pert} we show higher order perturbation
results in a small lattice, while in Appendix \ref{sec:appendix_dimer}
we discuss the applicability of cMF to weakly correlated systems.

\section{Formalism}
\label{sec:formalism}

\subsection{Hubbard model}
\label{sec:hubbard}

In this work, we focus our attention on the Hubbard model in one- and
two-dimensions (in a square lattice). The Hubbard model
\cite{hubbard-1963} describes a collection of electrons in a lattice
(of finite size $L$) interacting through the Hamiltonian
\begin{equation}
  \hat{H} = -t \sum_{\langle ij \rangle,\sigma} (c^\dagger_{i,\sigma}
  \, c_{j,\sigma} + \textrm{h.c.}) + U \sum_i n_{i,\uparrow} \,
  n_{i,\downarrow},
\end{equation}
where the notation $\langle ij \rangle$ implies interaction only among
nearest-neighbors. Here, $c^\dagger_{i,\sigma}$ ($c_{i,\sigma}$)
creates (annihilates) an electron with spin $\sigma$ on site $i$ of
the lattice and $n_{i,\sigma} \equiv c^\dagger_{i,\sigma} \,
c_{i,\sigma}$. The Hamiltonian contains one-electron hopping terms and
an on-site repulsion ($U>0$) term. The hopping amplitude $t$ is used
to set the energy scale.

\subsection{Cluster Mean-Field}
\label{sec:cmf}

Consider a set of single-fermion states $\{|k \rangle, \, k=1,\ldots,M
\}$, where $M$ is the size of the basis, that satisfies the
appropriate set of boundary conditions for the system. These
single-fermion states may be different than the ones used to define
the problem; in the case of the Hubbard model, they may be obtained by
a rotation of the lattice (on-site) basis states:
\begin{align}
  |k \rangle \equiv &\, a^\dagger_k |- \rangle, \\
  a^\dagger_k = &\, \hat{R} \, c^\dagger_k \, \hat{R}^{-1}.
\end{align}
We assume that $a^\dagger_k$ and $a_k$ satisfy standard
anti-commutation rules (implying orthonormality of $\{ |k \rangle
\}$). A basis for many-fermion states can be constructed from properly
antisymmetrized products of such single-fermion states.

Let the single-fermion states be grouped, according to some criterion
(such as proximity in real space), into clusters of size $l_1, l_2,
\ldots l_n$, where $n$ is the number of such clusters. The Fock space
in each cluster can be constructed using the single-fermion states
that define it. Due to the orthogonality of single-fermion states in
different clusters, the Fock space of the full system is simply given
by the tensor product of the Fock spaces of all clusters.

A second-quantized formulation in terms of cluster product states can
also be established. Let $A^\dagger_{I,c}$ ($A_{I,c}$) create
(annihilate) the $I$-th many-fermion state in cluster $c$. This $I$-th
state is a linear combination of many-fermion basis states $\{ |\mu
\rangle_c \}$ (possibly mixing states with different number of
fermions) constructed as antisymmetrized products of the
single-fermion states in the cluster. We formally write
\begin{equation}
  |I \rangle_c = A^\dagger_{I,c} |- \rangle_c,
\end{equation}
where $|- \rangle_c$ is the vacuum state in cluster $c$. (We emphasize
here that $|- \rangle_c$ does {\em not} correspond to the state with
no fermions in the cluster, but is simply a useful abstract
construct.)

An arbitrary state $|\Psi \rangle$ in the Fock space of the entire
system can be formed as
\begin{equation}
  |\Psi \rangle = \sum_I \sum_J \cdots \sum_Z c_{I1;J2;\ldots;Zn} \,
  |I \rangle_1 |J \rangle_2 \cdots |Z \rangle_n,
\end{equation}
where $c_{I1;J2;\ldots;Zn}$ are linear coefficients. Here, the sum
over $I$ spans the full Fock space in cluster 1, and so on. Each state
in the expansion above constitutes a cluster product state. Formally,
each cluster product state is built as
\begin{equation}
  |I \rangle_1 |J \rangle_2 \cdots |Z \rangle_n \equiv A^\dagger_{I,1}
  \, A^\dagger_{J,2} \cdots A^\dagger_{Z,n} |- \rangle,
\end{equation}
where $|- \rangle$ is an abstract vacuum state for the full system.

In this work, we consider a cluster product (mean-field) state as a
variational ansatz for the ground state wavefunction. That is, the
ansatz $|\Phi_0 \rangle$ for the ground state is given by
\begin{equation}
  |\Phi_0 \rangle = |0 \rangle_1 |0 \rangle_2 \cdots |0 \rangle_n,
\end{equation}
where we have indicated that the ground state (hence the $0$ label) in
each cluster is used to build the product state. The optimal cMF state
is obtained by a variational minimization scheme, as outlined in
Sec. \ref{sec:optimization}.

Having defined a ground state cluster product configuration, excited
configurations can also be considered. We write them as
\begin{align}
  |\Phi_{Ii} \rangle = &\, |0 \rangle_1 \cdots |I \rangle_i \cdots |0
  \rangle_n, \\
  |\Phi_{Ii;Jj} \rangle = &\, |0 \rangle_1 \cdots |I \rangle_i \cdots
  |J \rangle_j \cdots |0 \rangle_n, \\
  |\Phi_{Ii;Jj;Kk} \rangle = &\, |0 \rangle_1 \cdots |I \rangle_i \cdots
  |J \rangle_j \cdots |K \rangle_k \cdots |0 \rangle_n,
\end{align}
for singly-, doubly-, and triply-excited clusters. A full
configuration expansion can be written in terms of excited
configurations as
\begin{align}
  |\Psi \rangle = &\, c_0 |\Phi_0 \rangle + \sum_i \sum_{I \neq 0}
  c_{Ii} |\Phi_{Ii} \rangle \nonumber \\
  &\, + \sum_{i<j} \sum_{I \neq 0} \sum_{J \neq 0} c_{Ii;Jj}
  |\Phi_{Ii;Jj} \rangle + \ldots
\end{align}
This provides exact eigenstates for the full system.

Before proceeding further, let us comment on the nature of the cluster
product states considered in this work. We indicated above that the
ground state of each cluster is expressed as a linear combination of
the many-fermion basis states in it. That is,
\begin{equation}
  |0 \rangle_c = \sum_\mu d_{0,c}^\mu |\mu \rangle_c,
  \label{eq:cluster_exp}
\end{equation}
with $d_{0,c}^\mu = {}_c \langle \mu | 0 \rangle_c$, where $\mu$ is a
compound index of occupation numbers in the subset of single-fermion
states of the cluster. The expansion over states $\{ |\mu \rangle_c
\}$ can be restricted, {\em i.e.}, some of the $\{ d_{0,c}^\mu \}$
coefficients may be set to 0. We note that if a given cluster state is
expanded in terms of even- and odd-number parity (here referring to an
even or odd number of fermions) states, commutation rules between the
operators $A^\dagger_{I,c}$ and $A_{J,c'}$ are not simple,
complicating the evaluation of matrix elements as hinted below. All
the calculations included in this work restrict the expansion of
cluster states to a given $n_\uparrow$ and $n_\downarrow$ (or,
equivalently, $n$ and $m_s$) sector within the cluster, but include
the full Hilbert subspace with those quantum numbers. This was done in
order for the cluster product state $|\Phi_0 \rangle$ to be an
eigenfunction of $\hat{N}_{\uparrow}$ and $\hat{N}_{\downarrow}$.

Consider the determinantal expansion of the full system. The exact
ground state wavefunction is expressed as
\begin{equation}
  |\Psi \rangle = \sum_{\mu \nu \lambda \cdots} c_{\mu \nu \lambda
    \cdots} |\mu \nu \lambda \cdots \rangle,
\end{equation}
including all possible many-fermion states $\{ |\mu \rangle_c \}$
within the cluster. In cMF, the coefficients in the expansion above
are not independent, but are parametrized according to
\begin{equation}
  c_{\mu \nu \lambda \cdots} = d_{0,1}^{\mu} \, d_{0,2}^{\nu} \,
  d_{0,3}^{\lambda} \cdots .
  \label{eq:parametrize}
\end{equation}
This parametrization permits us to put cMF in the context of TN states
\cite{cirac-2009,orus-2014} and CPS
\cite{changlani-2009,mezzacapo-2009,mezzacapo-2010}. In cMF the
coefficient of each determinant is parametrized as a scalar product of
cluster states (with compound indices). That is, the cluster product
state has intra-cluster correlations, but lacks inter-cluster
ones. This is in contrast to a TN state, where ancillary or bond
indices include explicit entanglement in the ansatz. While in cMF the
compound indices $\mu$, $\nu$, etc. refer to different orbital
subspaces, CPS use overlapping indices as a means of introducing
entanglement in the ansatz. The optimization of TN states and CPS is,
nonetheless, more involved than that of cMF states.

A cMF state constitutes a generalization of a single Slater
determinant, and thus HF can be written as a cMF where the orbitals
are grouped into clusters, as we next describe. It should be stressed
that this is only possible in the optimized single-particle basis (or
a unitary rotation of it that does not mix particles with holes). The
HF state is recovered in different ways: as a product of two clusters
(one fully occupied and one fully empty), or as a product of $M$
clusters (the holes being occupied and the particles being empty), or
arbitrary related constructions.

The model also contains other wavefunction ans\"atze commonly used in
quantum chemistry such as the antisymmetrized product of strongly
orthogonal geminals (APSG)
\cite{parks-1958,surjan,rassolov-2002}. Here, each cluster would
contain two-electrons in a subspace of orbitals that define each
geminal. It also encompasses the multi-configuration
self-consistent-field (MC-SCF) \cite{hartree-1939,shepard} model as
well as the complete-active-space (CAS) \cite{roos-1980,roos} variant
of it. The latter can be considered as a three-cluster state: the core
is fully occupied, the virtual set of orbitals is fully empty, and the
state in the active space is expressed as an optimized linear
combination of all possible many-electron basis states in the
appropriate Hilbert space.

\subsection{Matrix Elements}
\label{sec:cmf_matel}

We now turn to the evaluation of matrix elements over cMF states. The
(two-body) fermionic Hamiltonian is expressed in second-quantized form
(in the basis that defines the clusters) as
\begin{equation}
  \hat{H} = \sum_{pr} \langle p | \hat{t} | r \rangle a^\dagger_p \,
  a_r + \frac{1}{2} \sum_{pqrs} \langle pq | \hat{v} | rs \rangle
  a^\dagger_p \, a^\dagger_q \, a_s \, a_r,
\end{equation}
where $\langle p | \hat{t} | r \rangle$ are one-body and $\langle pq |
\hat{v} | rs \rangle$ are two-body integrals. The Hamiltonian can be
expressed as a sum of single, two, three, and four-cluster
interactions:
\begin{equation}
  \hat{H} = \sum_i \hat{H}_i + \sum_{i<j} \hat{H}_{ij} + \sum_{i<j<k}
  \hat{H}_{ijk} + \sum_{i<j<k<l} \hat{H}_{ijkl}.
\end{equation}
Here, for instance,
\begin{align}
  \hat{H}_{ij} = &\, \sum_{p \in i, r \in j} \Big( \langle p | \hat{t}
  | r \rangle a^\dagger_p \, a_r + \langle r | \hat{t} | p \rangle
  a^\dagger_r \, a_p \Big) \nonumber \\
  &\, + \frac{1}{2} \sum_{pr \in i, qs \in j} \langle pq | \hat{v} |
  rs \rangle \, a^\dagger_p \, a^\dagger_q \, a_s \, a_r + \ldots, \\
  \hat{H}_{ijkl} = &\, \frac{1}{2} \sum_{\substack{p \in i, q \in j,\\r
      \in k, s \in l}} \langle pq | \hat{v} | rs \rangle \,
  a^\dagger_p \, a^\dagger_q \, a_s \, a_r + \ldots.
\end{align}

Given that fermion operators $\{ a^\dagger_p, a_p \}$ act on specific
clusters, the matrix elements can be evaluated straightforwardly if
all the cluster states have a well defined number parity (though care
has to be taken to respect fermionic anti-commutation rules). For
instance,
\begin{equation}
  \mathrm{if} \, p \in 2 \qquad a^\dagger_p |\Phi_0 \rangle = \pm |0
  \rangle_1 \, a^\dagger_p |0 \rangle_2 |0 \rangle_3 \cdots,
\end{equation}
where the sign depends on the number parity of $|0 \rangle_1$. (The
action of $a^\dagger_p$ ($a_p$) on a specific cluster can be easily
expressed in the occupation number basis $\{ |\mu \rangle_c \}$ within
each cluster.) If $|0 \rangle_c$ is of mixed number parity, the
evaluation becomes more cumbersome. For instance,
\begin{align}
  \mathrm{if} \, p \in 2 \qquad a^\dagger_p |\Phi_0 \rangle =
  &+ \, |0 \rangle_1^+ \, a^\dagger_p |0 \rangle_2 |0 \rangle_3 \cdots
  \nonumber \\
  &- \, |0 \rangle_1^- \, a^\dagger_p |0 \rangle_2 |0 \rangle_3 \cdots,
\end{align}
where $| 0 \rangle_1^+$ denotes the even-number parity projection out
of $|0 \rangle_1$, {\em i.e.}, $| 0 \rangle_1^+ \equiv |+ \rangle
\langle + | 0 \rangle_1$.

We close this section by noting that, if the ground state in each
cluster preserves number parity, then expectation values of
single-fermion operators (such as ${}_c \langle 0 | a_p^\dagger |0
\rangle_c$, for $p \in c$) vanish. This further implies that all
three- and four-cluster interactions vanish in $\langle \Phi_0 |
\hat{H} | \Phi_0 \rangle$.  If the number of fermion states within
each cluster is fixed, then the expectation value $\langle \Phi_0 |
\hat{H} | \Phi_0 \rangle$ can be fully expressed in terms of the one-
and two-particle reduced density matrices within each cluster, as
first noted by McWeeny \cite{mcweeny-1959}. (Note that this implies
that the cost of evaluating the energy of a cMF state scales as
$\mathcal{O}(n^2)$, where $n$ is the number of clusters used.)

\subsection{cMF Optimization}
\label{sec:optimization}

In this section we discuss how the cMF state is optimized, that is,
how the ground state $|0 \rangle_c$ in each cluster $c$ is
determined. We use a diagonalization strategy akin to that used in HF
or multi-configuration self-consistent-field (MC-SCF) methods.

The optimal set of coefficients $\{ d_{0,c}^\mu \}$ ({\em cf.}
Eq.~\ref{eq:cluster_exp}) can be found by minimization of the energy
subject to the constraint that the state $|0 \rangle_c$ remains
normalized:
\begin{equation}
  \frac{\partial}{\partial d_{0,c}^{\mu \ast}} \langle \Phi_0 |
  \hat{H} | \Phi_0 \rangle - \epsilon_{0,c} d_{0,c}^{\mu} = 0.
\end{equation}
where $\epsilon_{0,c}$ is introduced as a Lagrange multiplier. The
above equation can be cast as an eigenvalue equation that, at the same
time, defines a zero-th order Hamiltonian $\hat{H}^0_c$ within the
cluster, {\em i.e.},
\begin{equation}
  \hat{H}^0_c \, d_{0,c}^{\mu} \equiv \frac{\partial}{\partial
    d_{0,c}^{\mu \ast}} \langle \Phi_0 | \hat{H} | \Phi_0 \rangle.
\end{equation}
The ground state of the cluster Hamiltonian is obtained as its lowest
energy eigenvector. (Note that this also gives $\epsilon_{0,c}$ the
physical meaning of the energy in cluster $c$.) The cluster
Hamiltonian can be found trivially. As an example, if all cluster
ground states $\{ |0 \rangle_c \}$ have a fixed number of fermions,
$\hat{H}^0_c$ is given by
\begin{align}
  \hat{H}^0_c = &\, \sum_{pr \in c} \langle p | \hat{t} | r \rangle
  a^\dagger_p \, a_r + \frac{1}{2} \sum_{pqrs \in c} \langle pq |
  \hat{v} | rs \rangle a^\dagger_p \, a^\dagger_q \, a_s \, a_r
  \nonumber \\
  + &\, \sum_{pr \in c} a^\dagger_p \, a_r \sum_{c'} \sum_{qs}
  \rho^{c'}_{sq} (\langle pq | \hat{v} | rs \rangle \, - \langle pq |
  \hat{v} | sr \rangle).
\end{align}
Here, $\rho^{c'}_{sq} = {}_{c'} \langle 0 | a^\dagger_q \, a_s | 0
\rangle_{c'}$ is the one-particle density matrix in cluster $c'$.
Because the cluster Hamiltonian $\hat{H}^0_c$ depends on the ground
state density matrices of other clusters, the equations must be solved
self-consistently. This represents a generalization of the MC-SCF
method, where there are several active subspaces (the clusters) each
with its own multi-configurational expansion.

\subsection{Orbital Optimization}
\label{sec:orb_optim}

As discussed in the introduction, in order to realize the full
capability of cMF states it is necessary to include the optimization
of the single-fermion basis in which the grouping into clusters is
defined, which we refer to as an orbital optimization. Otherwise, a
cMF state may yield an energy that is even above HF, despite having
significantly more flexibility in the ansatz. We describe in this
section how this is accomplished in our work. We note that the orbital
optimization in cMF states is akin to the same process performed in
traditional MC-SCF (and CAS, by extension) calculations in quantum
chemistry.

Given the single-particle basis $\{ |k \rangle \}$ in which the cMF
state is constructed, we aim to rotate this to a new basis $\{
|\bar{k} \rangle \}$ in order to lower the energy. We relate the two
basis by a unitary transformation (parametrized as the exponential of
an anti-Hermitian operator),
\begin{align}
  \bar{a}^\dagger_k = &\, \exp(\hat{\kappa}) \, a_k \,
  \exp(-\hat{\kappa}), \\
  \hat{\kappa} = &\, \sum_{p<q} (\kappa_{pq} \, a^\dagger_p \, a_q -
  \mathrm{h.c.}).
\end{align}
In particular, we define an energy functional
\begin{equation}
  E[\boldsymbol{\kappa}] = \langle \Phi_0 | \exp(-\hat{\kappa}) \,
  \hat{H} \, \exp(\hat{\kappa}) | \Phi_0 \rangle,
\end{equation}
where the (complex) elements $\{ \kappa_{pq} \}$ serve as variational
parameters.

With the optimized cMF state $|\Phi_0 \rangle$ at hand, we can compute
the gradient with respect to orbital rotations at $\boldsymbol{\kappa}
= \mathbf{0}$ ({\em i.e.}, the gradient evaluated at zero-rotation) as
\begin{equation}
  G_{pq} \equiv \left. \frac{\partial E}{\partial \kappa_{pq}^\ast}
  \right\vert _{\boldsymbol{\kappa} = \boldsymbol{0}} = - \Big\langle
  \Phi_0 \Big| \big[ \hat{H}, a_q^\dagger \, a_p \big] \Big| \Phi_0
  \Big\rangle.
\end{equation}
Similarly, the Hessian can be constructed as
\begin{equation}
  \mathbf{H} = \begin{pmatrix} \mathbf{A} & \mathbf{B}
    \\ \mathbf{B}^\ast & \mathbf{A}^\ast \end{pmatrix},
\end{equation}
with
\begin{align}
   A_{pq,rs} \equiv &\, \left. \frac{\partial^2 E}{\partial
     \kappa_{ij}^\ast \, \partial \kappa_{kl}} \right\vert
   _{\boldsymbol{\kappa} = \boldsymbol{0}} \nonumber \\
   =&\, - \frac{1}{2} \Big\langle \Phi_0 \Big| \big[ \big[ \hat{H},
       a_q^\dagger \, a_p \big], a_r^\dagger \, a_s \big] \Big| \Phi_0
   \Big\rangle \nonumber \\
   &\, + q,p \leftrightarrow r,s, \\[4pt]
   B_{pq,rs} \equiv &\, \left. \frac{\partial^2 E}{\partial
     \kappa_{pq}^\ast \, \partial \kappa_{rs}^\ast} \right\vert
   _{\boldsymbol{\kappa} = \boldsymbol{0}} \nonumber \\
   = &\, + \frac{1}{2} \Big\langle
   \Phi_0 \Big| \big[ \big[ \hat{H}, a_q^\dagger \, a_p \big],
     a_s^\dagger \, a_r \big] \Big| \Phi_0 \Big\rangle \nonumber \\
   &\, + q,p \leftrightarrow s,r.
\end{align}
The gradient (and Hessian) can be used to find a direction of energy
lowering with respect to orbital rotations. Several comments are in
order:
\begin{itemize}
  \item The energy and Hessian can be evaluated following the strategy
    described in Sec. \ref{sec:cmf_matel} for the evaluation of matrix
    elements. It can be shown that the cost of building the full
    gradient and Hessian scales as $\mathcal{O}(n^3)$ and
    $\mathcal{O}(n^4)$, respectively, where $n$ is the number of
    clusters in the system.

  \item Once a direction of energy lowering is found (defining a
    non-zero $\boldsymbol{\kappa}$), we perform a
    finite rotation ($\alpha \boldsymbol{\kappa}$) of the Hamiltonian
    integrals. The energy functional is re-parametrized with respect
    to orbital rotations in terms of the new single-particle basis.

  \item The above strategy (re-parametrizing the energy functional at
    each step) is necessary as the evaluation of the orbital gradient
    (and Hessian) is not as simple when $\boldsymbol{\kappa} \neq
    \mathbf{0}$. This also prevents us from using a quasi-Newton
    strategy to perform the orbital optimization.

  \item Care has to be taken of handling linear dependencies between
    the orbital rotations and the coefficients in each cluster
    expansion. For instance, if a cluster state $|0 \rangle_c$ is
    expanded in terms of a full Hilbert (or Fock) subspace, then
    orbital rotations within the subset of orbitals that define the
    cluster $c$ do not lower the energy.
\end{itemize}

In this work, as we use a full Hilbert subspace to describe each
cluster, the orbital gradient and Hessian for intra-cluster rotations
are not considered as degrees of freedom.

\subsection{Perturbation Theory}
\label{sec:rspt}

In standard RS-PT \cite{schrodinger-1926}, we aim to solve for the
eigenstates of the Hamiltonian $\hat{H}$ given the simpler Hamiltonian
$\hat{H}^0$, for which all eigenstates are known. In RS-PT, the
second-order correction to the ground state energy is evaluated as
\begin{equation}
  E^{(2)} = \sum_{\mu \neq 0} \frac{|V_{0 \mu}|^2}{\varepsilon_0 -
    \varepsilon_\mu},
\end{equation}
where $\hat{V} = \hat{H} - \hat{H}^0$ and $V_{0 \mu} = \langle \Phi_0
| \hat{V} | \mu \rangle$. Here, $\mu$ labels the eigenstates of
$\hat{H}_0$ and $\varepsilon_\mu$ are the corresponding eigenvalues.

The cMF state, as outlined in Sec. \ref{sec:optimization}, provides a
natural zero-th order Hamiltonian of which all eigenstates can be
easily constructed. This is expressed as a direct sum of the zero-th
order Hamiltonians of each cluster
\begin{equation}
  \hat{H}^0 = \hat{H}^0_1 + \hat{H}^0_2 + \ldots
\end{equation}
The eigenstates of such Hamiltonian are given by
\begin{align}
  \hat{H}^0 |I \rangle_1 |J \rangle_2 \cdots |Z \rangle_n, = &\,
  \varepsilon_{I1;J2;\cdots;Zn} |I \rangle_1 |J \rangle_2 \cdots |Z
  \rangle_n \\
  \varepsilon_{I1;J2;\cdots;Zn} = &\, \epsilon_{1,I} + \epsilon_{2,J}
  + \cdots + \epsilon_{n,Z}.
\end{align}

As described in Sec. \ref{sec:cmf_matel}, $\hat{H}$ has up to
four-cluster interactions, while $\hat{H}^0$ is, by construction,
single-cluster in character. If a full Hilbert subspace in each
cluster is used, matrix elements between the ground state $|\Phi_0
\rangle$ and singly-excited (cluster) configurations vanish due to a
generalized-Brillouin condition. Therefore, only two, three, and
four-cluster interactions contribute to the second-order energy. The
evaluation of the corresponding matrix elements can be carried out in
a similar fashion as the evaluation of $\langle \Phi_0 | \hat{H}
|\Phi_0 \rangle$. Naturally, computing the four-cluster interactions
is the most expensive step in evaluating the second-order energy, with
a computational scaling of $\mathcal{O}(n^4)$ in the number of
clusters.

As described in Sec. \ref{sec:cmf}, in this work we have chosen to use
cluster ground states which preserve the number of $\uparrow$- and
$\downarrow$-electrons. In that case, several two, three, and
four-cluster interaction channels can be identified in the
Hamiltonian, as summarized in Tab. \ref{tab:pert_channel}. The cMF
ground state $|\Phi_0 \rangle$ interacts with excited cluster
configurations following these channels.
\begin{table*}[!htb]
  \caption{Two-, three-, and four-cluster interaction
    channels when the ground state in each cluster has well
    defined $n_\uparrow$ and $n_\downarrow$. \label{tab:pert_channel}}
  \begin{ruledtabular}
  \begin{tabular}{r l l l}
    \# clusters & type\footnote{CT denotes charge transfer.} & sample
    interaction\footnote{Only two-body interactions are shown. Here, a
      generic two-fermion interaction takes the form
      $a^\dagger_{p,\sigma} \, a^\dagger_{q,\sigma'} \, a_{s,\sigma'}
      \, a_{r,\sigma}$, with $\sigma' = \sigma$ or $-\sigma$.} &
    restrictions \\ \hline
    2 & one-electron CT
    & $a^\dagger_{p,\sigma} \, a^\dagger_{q,\sigma'} \, a_{s,\sigma'} \,
    a_{r,\sigma}$
    & $pqs \in i$, $r \in j$ \\
    2 & two-electron, opp spin, CT
    & $a^\dagger_{p,\sigma} \, a^\dagger_{q,-\sigma} \, a_{s,-\sigma} \,
    a_{r,\sigma}$
    & $pq \in i$, $sr \in j$ \\
    2 & two-electron, same spin, CT
    & $a^\dagger_{p,\sigma} \, a^\dagger_{q,\sigma} \, a_{s,\sigma} \,
    a_{r,\sigma}$
    & $pq \in i$, $sr \in j$ \\
    2 & two-cluster spin flip
    & $a^\dagger_{p,\sigma} \, a^\dagger_{q,-\sigma} \, a_{s,-\sigma} \,
    a_{r,\sigma}$
    & $ps \in i$, $qr \in j$ \\
    2 & two-cluster dispersion
    & $a^\dagger_{p,\sigma} \, a^\dagger_{q,\sigma'} \, a_{s,\sigma'} \,
    a_{r,\sigma}$
    & $pr \in i$, $qs \in j$ \\
    3 & two-electron, opp spin, CT
    & $a^\dagger_{p,\sigma} \, a^\dagger_{q,-\sigma} \, a_{s,-\sigma} \,
    a_{r,\sigma}$
    & $pq \in i$, $s \in j$, $r \in k$ \\
    3 & two-electron, same spin, CT
    & $a^\dagger_{p,\sigma} \, a^\dagger_{q,\sigma} \, a_{s,\sigma} \,
    a_{r,\sigma}$
    & $pq \in i$, $s \in j$, $r \in k$ \\
    3 & single-cluster spin flip
    & $a^\dagger_{p,\sigma} \, a^\dagger_{q,-\sigma} \, a_{s,-\sigma} \,
    a_{r,\sigma}$
    & $ps \in i$, $q \in j$, $r \in k$ \\
    3 & one-electron CT + dispersion
    & $a^\dagger_{p,\sigma} \, a^\dagger_{q,\sigma'} \, a_{s,\sigma'} \,
    a_{r,\sigma}$
    & $pr \in i$, $q \in j$, $s \in k$ \\
    4 & two-electron scattering
    & $a^\dagger_{p,\sigma} \, a^\dagger_{q,\sigma'} \, a_{s,\sigma'} \,
    a_{r,\sigma}$ 
    & $p \in i$, $q \in j$, $s \in k$, $r \in l$ \\
  \end{tabular}
  \end{ruledtabular}
\end{table*}

At this point, we clarify that in this work the zero-th order cluster
Hamiltonian is used, without any modification, to generate the full
Fock space within the cluster. It is possible to tweak the definition
of the non-interacting Hamiltonian ({\em e.g.}, by adding a
level-shift) in specific Hilbert space subsectors in order to improve
the convergence properties of the perturbation series.

\section{Computational Details}
\label{sec:comp_details}

The cMF and cPT2 calculations presented in this work were carried out
with a locally prepared code. Most of the results use an unrestricted
cMF (U-cMF) formalism, where $\uparrow$-orbitals are allowed to have a
different spatial distribution than $\downarrow$-ones. Some of the
results in 1D lattices use a restricted (R-cMF) formalism, where the
spatial distribution is required to be the same. Real orbitals are
used in both cases. In all the calculations we use the same number of
$\uparrow$- and $\downarrow$-orbitals in each cluster, which we denote
as $l$ and refer to as the {\em size of the cluster}. The number of
$\uparrow$- and $\downarrow$-electrons in each cluster was held fixed
(thus preserving $n$ and $m_s$ within each cluster).\footnote{Some
  exploratory calculations were carried out using the full Fock space
  within each cluster. For half-filled systems in the on-site basis,
  the additional flexibility in the ansatz does not result in a lower
  variational estimate of the ground state energy. This, however, may
  not be true for doped systems, or if a full orbital optimization is
  carried out.} Although not enforced from the outset, R-cMF
calculations resulted in spin singlet eigenstates within each cluster.

The full relevant $m_s$ sector of Hilbert space within each cluster
was used in constructing the cluster ground state $|0 \rangle_c$. For
small cluster sizes, the ground state in each cluster was found by a
standard diagonalization of the local cluster Hamiltonian. For larger
cluster sizes, a Lanczos \cite{lehoucq} or a Jacobi-Davidson
\cite{davidson-1975,sleijpen-1996} algorithm was used to solve for the
ground state.

The orbital optimization was carried out using a pseudo Newton-Raphson
approach. After optimizing the cluster mean-field state, a Newton step
was taken in the direction of energy lowering (using the orbital
gradient and Hessian). A finite rotation provided a new
single-particle basis in which the cluster mean-field was
reoptimized. These two steps were alternated until convergence was
achieved in both the cMF state and the orbitals. This is akin to the
most common methods of optimizing MC-SCF wavefunctions in quantum
chemistry \cite{yeager-1979,douady-1980,siegbahn-1981}. We note that a
full Newton-Raphson approach (with the mean-field and the orbital
optimization carried out concomitantly) should be preferred
\cite{werner-1980}, but we have not used it in this work.\footnote{The
  alternating optimization strategy adopted may have poor convergence
  if the coefficients in the cluster mean-field state couple strongly
  to the orbital optimization degrees of freedom. This problem was
  indeed encountered for certain systems at low $U/t$.} A globally
convergent algorithm was used to guarantee that the variational cMF
energy is reduced in each orbital optimization step. As described in
detail below, for 2D lattices several local minima can be found in the
orbital optimization process. We have not attempted to use an
algorithm to locate the global minimum.

In U-cPT2 calculations, all relevant cluster states were used in
computing the second-order energy for small cluster sizes ($l=2$ and
$l=3$). For $l=4$ and $l=5$, the four-cluster contributions were
computed using only 16 states in each Hilbert subspace of a cluster,
while two- and three-tile contributions used all available states. In
$l=6$ calculations we truncated the number of states in each Hilbert
subspace in three- (four)-tile interactions to 64 (16), while no
truncation was done in computing two-tile interactions. An
energy-based criterion for the cluster states was used to carry out
the truncation. We should point out that the second-order energy
appears to be converged in all cases with respect to the number of
states included.

UCCSD and UCCSD(T) calculations were carried out with a locally
modified version of the MRCC code \cite{kallay-2001,mrcc}. Exact
solutions to the 1D Hubbard lattice were obtained by solving the
Lieb-Wu \cite{lieb-1968} equations.

\section{Results}
\label{sec:results}

In this section we present results of cMF and cPT2 calculations on the
1D and 2D Hubbard models. We start by providing an illustrative
example in Sec.~\ref{sec:results_illust}, where we get into some
practical details regarding the optimization of cMF states and the way
in which other results are presented. In Sec.~\ref{sec:results_1d} we
consider the 1D half-filled case, for which exact solutions are
available. We then proceed to study the 2D half-filled case in
Sec.~\ref{sec:results_2d}, and finally consider the lightly-doped 2D
case in Sec.~\ref{sec:results_2d_dop}. Our 2D results are compared to
highly accurate numerical estimates from
Refs.~\onlinecite{leblanc-2015,zheng-2015}.

\subsection{Illustrative example}
\label{sec:results_illust}

\begin{figure*}[!htb]
  \includegraphics[height=3.025cm]{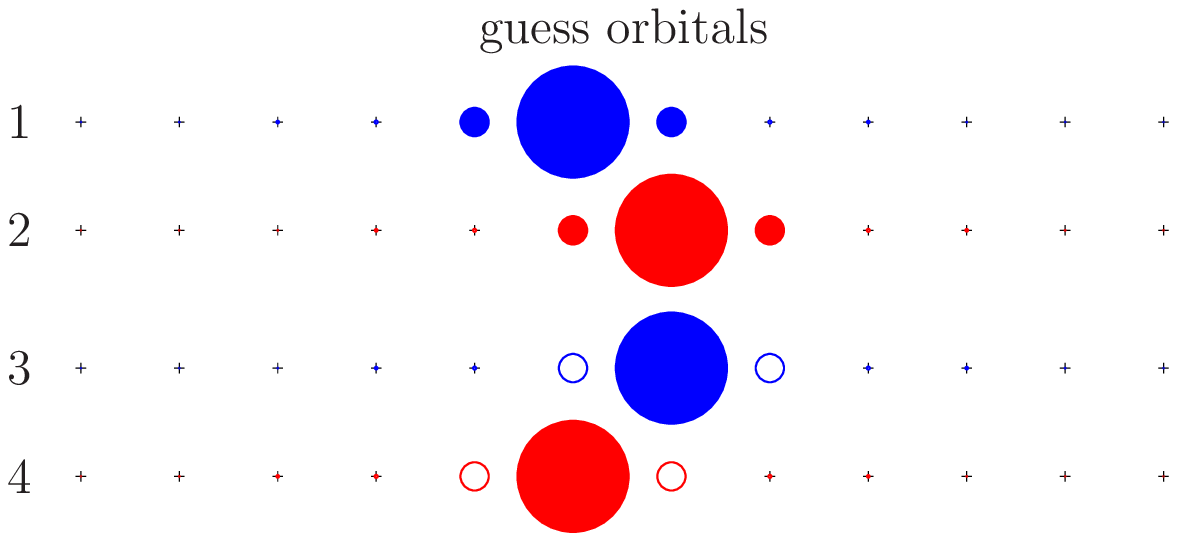}
  \hspace{0.5cm}
  \includegraphics[height=3.025cm]{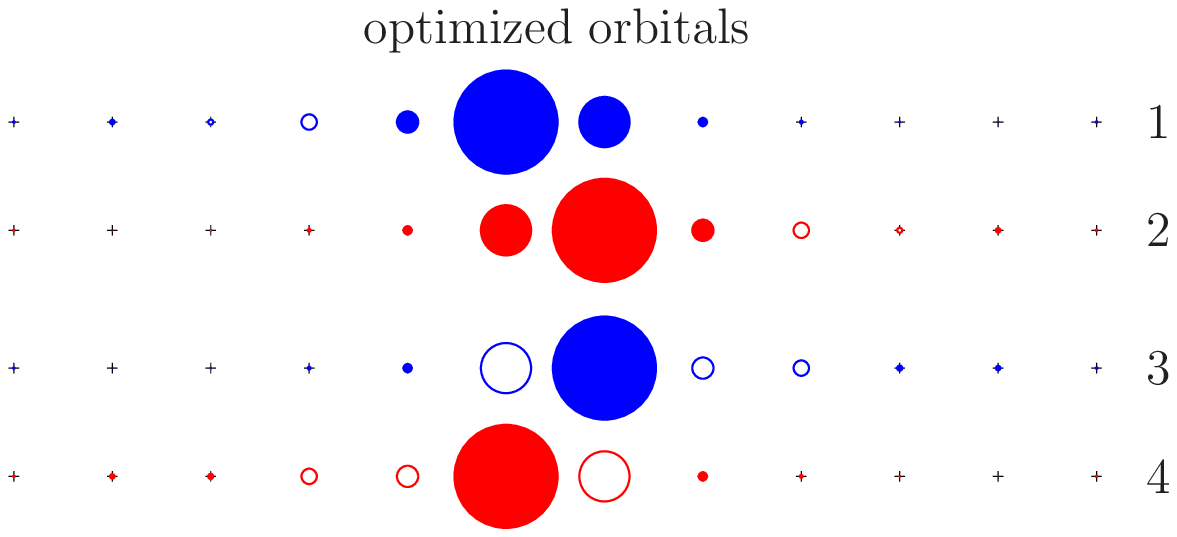}
  \\ \includegraphics[height=1.1cm]{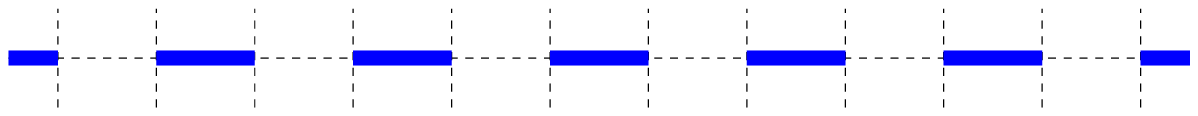}
  \caption{(Color online) In the top schemes we show initial guess
    (left) and optimized (right) spin-orbitals that define a single
    cluster in an U-cMF calculation (12-site half-filled periodic 1D
    lattice, $U/t=4$) using clusters of 2 $\uparrow$- and 2
    $\downarrow$-orbitals with 1 $\uparrow$- and 1
    $\downarrow$-electron. Orbitals are depicted (one per row) in the
    12-site lattice (marked by small $+$ signs) using the following
    conventions: $\uparrow$- ($\downarrow$-) orbitals are plotted in
    blue (red); filled (empty) circles indicate a positive (negative)
    orbital coefficient; the area enclosed by the circle is
    proportional to $|\phi(j)|$. The guess orbitals (left) correspond
    to Boys-localized orbitals of the UHF solution (orbitals 1 and 2
    are occupied; 3 and 4 are empty). The U-cMF optimized orbitals
    (right) displayed are those that diagonalize the $\uparrow$- and
    $\downarrow$- one-particle reduced density matrix, with orbitals 1
    and 2 (3 and 4) having occupation of $0.9$ ($0.1$). Note that the
    orbitals remain fairly local in character within a subset of 2
    lattice sites. Accordingly, the bottom scheme shows a simplified
    representation of the optimized solution expressed in the on-site
    basis. This is characterized by local dimer-like structures, which
    we connect by a solid blue line. \label{fig:illust}}
\end{figure*}

In this section we discuss some practical aspects regarding the
optimization of cMF states. In this way, we hope that the results
presented in subsequent sections will become more transparent to the
reader. We consider a 12-site Hubbard 1D periodic lattice at
half-filling and $U/t=4$. For U-cMF calculations, we typically start
from an unrestricted HF (UHF) solution; we take the resulting orbitals
and perform a Boys localization
\cite{kleier-1974}. Figure~\ref{fig:illust} displays, in the top-left
scheme, (localized) occupied and virtual spin-orbitals mostly tied to
two sites, which are then used to define a cluster of 2-$\uparrow$ and
2-$\downarrow$ orbitals in which a single electron of each spin is
placed. The optimized state in the cluster is expressed as a linear
combination of the 4 possible resulting configurations.

After setting the initial orbitals and the corresponding tiling scheme
({\em i.e.}, defining how orbitals and electrons are grouped into
clusters), the cluster mean-field state is optimized by a
self-consistent diagonalization of the appropriate cluster
Hamiltonians. The orbital gradient (and possibly the Hessian) is then
evaluated which determines how orbitals in different clusters should
be mixed in order to lower the energy. A new orbital basis is defined
by, {\em e.g.}, a steepest-descent step, and the cluster mean-field
step is reoptimized in such basis. The process is repeated until
convergence is reached in both the orbitals and the mean-field
state. The top-right scheme in Fig.~\ref{fig:illust} shows the
converged orbitals that define a single cluster in the calculation. In
particular, the orbitals displayed are the natural orbitals (those
that diagonalize the $\uparrow$- and $\downarrow$- one-particle
reduced density matrix) mostly tied to the original two sites. The
orbitals defining the cluster remain well localized in two lattice
sites (although there is a noticeable spread into neighboring sites
which becomes more pronounced at lower $U/t$). This allows us to, for
simplicity purposes, characterize the optimized solution in terms of a
tiling scheme in the on-site basis (see bottom of the figure),
although we emphasize that this is only approximate. In this case, the
structure is defined by local dimers which are spin polarized (with a
non-zero magnetization in each site) to yield an overall N\'eel-like
configuration.

\subsection{1D: half-filling}
\label{sec:results_1d}

We start by considering the half-filled 1D periodic case. All
calculations in this section, unless explicitly stated, were performed
in a periodic lattice with $L=120$ sites, which we deem large enough
to provide near-thermodynamic limit results for $U/t \geq 1$. Only
uniform tiling schemes were considered; clusters were defined in terms
of a continuous set of $l$ lattice sites, each filled with $l/2$
electrons (for even $l$). For U-cMF calculations with odd $l$, we have
adopted a staggered configuration: if a cluster of size $l$ has
$(l+1)/2$ $\uparrow$-electrons and $(l-1)/2$ $\downarrow$-electrons,
its neighbors have $(l+1)/2$ $\downarrow$-electrons and $(l-1)/2$
$\uparrow$-electrons, respectively. As described in
Sec. \ref{sec:results_illust}, some spreading of the orbitals into
neighboring sites is observed, particularly at low $U/t$. We note that
broken-symmetry U-cMF solutions maintain the overall N\'eel-like
structure observed in UHF, that is, a non-zero magnetization develops
on each lattice site.

\begin{figure*}[!htb]
  \includegraphics[height=6.875cm]{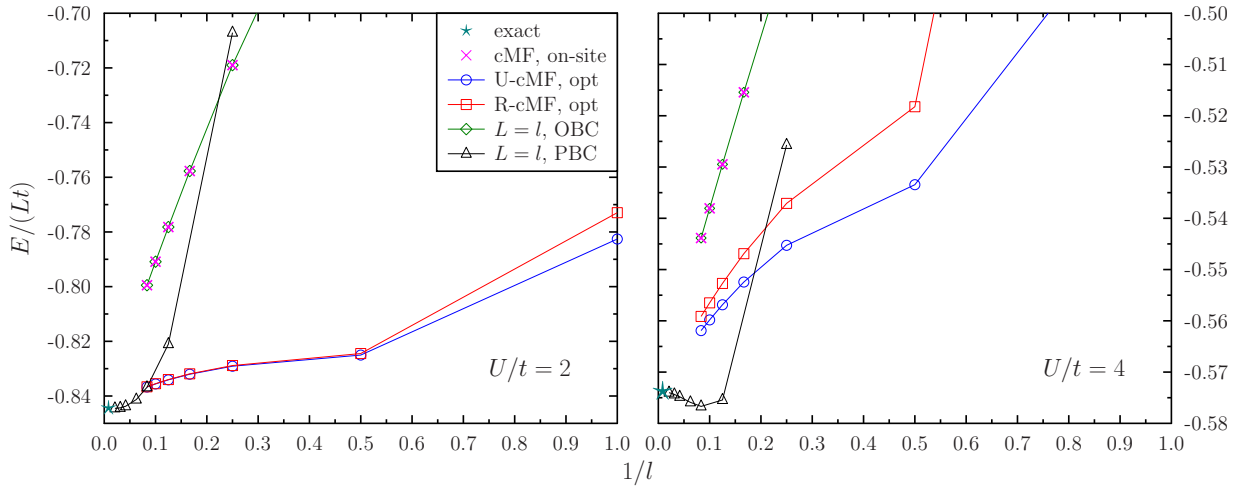}
  \caption{(Color online) Energy per site obtained in cMF calculations
    on a $L=120$ half-filled periodic 1D lattice at $U/t=2$ (left) and
    $U/t=4$ (right) as a function of the inverse of the cluster size
    $l$. $l=1$ R-cMF and U-cMF results correspond to restricted HF and
    UHF, respectively. We consider cMF results in the on-site basis
    and with a fully-optimized single-particle basis (opt). They are
    compared with (exact) calculations on a single cluster ({\em
      i.e.}, $L=l$) using open (OBC) and periodic (PBC) boundary
    conditions. The latter were computed by solving the corresponding
    Lieb-Wu \cite{lieb-1968} equations. The energies of $L=l$
    calculations with OBC exactly match those of cMF calculations in
    the full $L=120$ lattice using the on-site
    basis. \label{fig:1d_extp}}
\end{figure*}

We present in Fig.~\ref{fig:1d_extp} the energy per site obtained in
cMF calculations at $U/t=2$ (left) and $U/t=4$ (right) as a function
of the inverse of the cluster size, using both restricted (R-cMF) and
unrestricted (U-cMF) optimized orbitals, as well as cMF calculations
in the on-site basis. We have also included, for comparison, the
results of (exact) calculations carried out in a single cluster ($L=l$
sites) using both open (OBC) and periodic boundary conditions
(PBC). The former energies exactly match those of cMF calculations in
the $L=120$ lattice performed in the on-site basis, without orbital
optimization. We note that $L=l$ calculations using PBC {\em do not}
provide a variational estimate of the energy per site of the $L=120$
lattice (see Fig.~\ref{fig:1d_extp} at $U/t=4$, where the exact energy
is approached from below).

Comparing the results of cMF calculations that include orbital
optimization with those in the on-site basis, it is evident that
orbital optimization affords a significant improvement in the
variational estimate of the ground state energy. In addition, cMF
calculations using unrestricted orbitals provide a sizable improvement
over the corresponding restricted calculations at $U/t=4$. We note
that, at $U/t=4$, cMF calculations do not converge (in $1/l$) as fast
to the $L=120$ limit as calculations using PBC, though the former have
the advantage of being variational. On the other hand, at $U/t=2$ cMF
provides better estimates than $L=l$ calculations for $l<12$,
suggesting that a finite-size extrapolation with cMF results should be
preferred. We emphasize the significance of this given that, for
arbitrary systems, an exact diagonalization can currently only be
performed up to lattices of size $18$ or so. A linear extrapolation in
$1/l$ of U-cMF energies (using the $l=8$, 10, and 12 results) yield
the following estimates in the $l \to L=120$ limit for the ground
state energy $E/(Lt)$: $-0.5709(2)$ and $-0.8414(2)$ for $U/t=4$ and
2, respectively. These can be compared with the exact energies of
$-0.5738$ and $-0.8444$.

\begin{figure*}[!htb]
  \includegraphics[height=6.875cm]{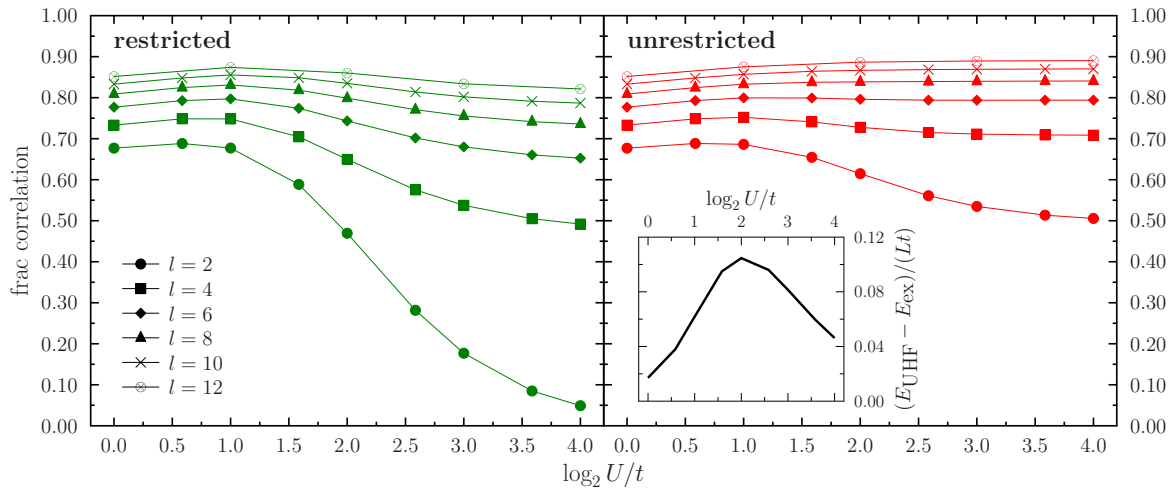}
  \caption{(Color online) Fraction of correlation (with respect to
    UHF) recovered in restricted (left) and unrestricted (right) cMF
    calculations in a $L=120$ periodic 1D lattice, as a function of
    $U/t$. (A log-2 scale is used in $U/t$ for clarity purposes;
    results are shown from $U/t=1$ to $16$.) Cluster sizes from 2 to
    12 were used. The inset in the right panel shows the total
    correlation energy per site, as a function of
    $U/t$. \label{fig:1d_energy}}
\end{figure*}

We will often refer to the fraction of correlation energy in assessing
the quality of the ground state energy. The correlation energy is here
defined as
\begin{equation}
  E_{\textrm{corr}} = E_{\textrm{exact}} - E_{\textrm{UHF}},
\end{equation}
{\em i.e.}, the difference between the exact and the UHF
energies. (Note that this differs from the traditional quantum
chemistry definition based on restricted HF \cite{lowdin-1955}.)
Figure~\ref{fig:1d_energy} shows the fraction of correlation recovered
in R-cMF and U-cMF calculations using clusters of increasing size
($l=2$ to $l=12$) as a function of $U/t$. The inset shows
$-E_{\textrm{corr}}/(Lt)$ as a function of $U/t$. The latter peaks at
$\approx 0.1$ at $U/t=4$.

The fraction of correlation in restricted calculations using $l=2$
seems to vanish at large $U/t$, indicating that the Heisenberg limit
predicted by this method is roughly the same as the UHF limit. This is
not the case in unrestricted $l=2$ calculations, which still recover
around $50\,\%$ of the correlation in the large $U/t$ limit. As $l$
becomes larger, the difference between restricted and unrestricted
calculations gets smaller, as expected. U-cMF calculations with $l=12$
are able to recover $85$ to $90\,\%$ of the correlation energy across
the entire $U/t$ domain plotted, which spans both the weak and the
strongly-correlated regimes. Accordingly, the maximum error in the
energy per site in $l=12$ U-cMF calculations is about $0.01\,t$ at
$U/t=4$, which is remarkable given the simplicity of the approach.

\begin{figure}[!htb]
  \includegraphics[height=6.875cm]{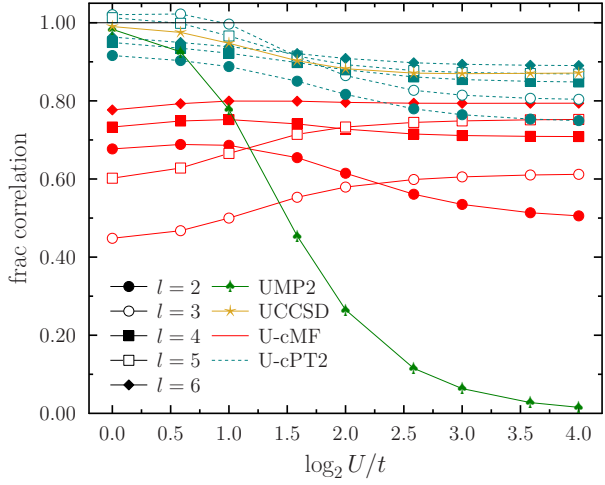}
  \caption{(Color online) Fraction of correlation (with respect to
    UHF) recovered in unrestricted cMF and cPT2 calculations in a
    $L=120$ periodic 1D lattice as a function of $U/t$. UMP2 and UCCSD
    results are provided for comparison
    purposes. \label{fig:1d_energy_pt2}}
\end{figure}

Figure~\ref{fig:1d_energy_pt2} displays the fraction of correlation
recovered in U-cMF and U-cPT2 calculations; results are compared with
UMP2 and UCCSD. U-cPT2 energies significantly improve over U-cMF
results for small cluster sizes. For instance, with $l=2$ U-cPT2
recovers $90\,\%$ ($> 75\,\%$) of the correlation energy at small
(large) $U/t$. Notice also that U-cPT2 results do not overly
deteriorate for large $U/t$ as UMP2 does. UCCSD and U-cPT2 ($l=6$)
recover $\gtrapprox 90\,\%$ of the correlation energy across the
entire $U/t$. It is interesting to point out that at $U/t=1$ UMP2 and
UCCSD results are better than U-cPT2 results with even $l$; U-cPT2
results with odd $l$, on the other hand, slightly overshoot the exact
result. Unfortunately, the steep computational scaling of U-cPT2
rendered calculations with $l>6$ as too expensive with our current
implementation.

\begin{figure}[!htb]
  \includegraphics[height=6.875cm]{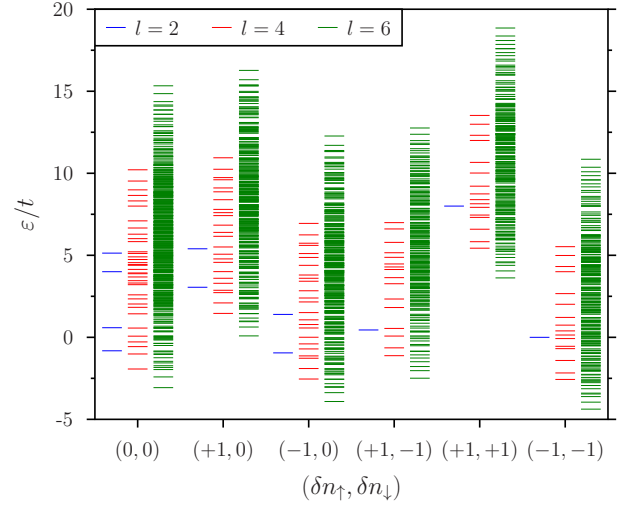}
  \caption{(Color online) Spectrum of the cluster Hamiltonian
    $\hat{H}^0_c$ in U-cMF calculations on a half-filled 1D lattice at
    $U/t=4$, using cluster sizes of 2, 4, and 6. All clusters in the
    $L=120$ lattice display an identical spectrum. The eigenvalues are
    classified according to the $n_\uparrow$ and $n_\downarrow$
    quantum numbers within the cluster. Here, $(0,0)$ corresponds to a
    half-filled cluster, with $l/2$ $\uparrow$- and
    $\downarrow$-electrons. Only those Hilbert sectors relevant to the
    evaluation of the second-order energy are shown. (Certain Hilbert
    sectors are missing as their spectrum is identical to one that is
    actually displayed; {\em e.g.}, the $(+1,0)$ and $(0,+1)$ spectra
    are equivalent, and so are the $(+1,-1)$ and
    $(-1,+1)$.) \label{fig:1d_spectrum}}
\end{figure}

Figure~\ref{fig:1d_spectrum} displays the spectrum of a single cluster
Hamiltonian in U-cMF calculations, using cluster sizes of 2, 4, and
6. As PBC are used along with a uniform tiling scheme, all clusters
end up displaying identical spectra, although we emphasize that this
was not imposed. The eigenvalues are shown in
Fig.~\ref{fig:1d_spectrum} according to the $n_\uparrow$ and
$n_\downarrow$ quantum numbers within the cluster, reference to the
half-filled case. Here, the ground state in the $(0,0)$ sector of each
cluster is used to construct the cMF state $|\Phi_0 \rangle$. Although
not shown in the figure, the spectrum should resemble, as $l$ becomes
larger, that of a lattice of $l$ sites with OBC to the extent that
orbitals remain fully localized. If $\epsilon_0^{(x,y)}$ denotes the
ground state in the $(x,y)$ sector, the perturbation series is stable
({\em i.e.}, all denominators are positive) as long as the following
conditions are met (we indicate in parenthesis the relevant
interactions):
\begin{enumerate}
  \item the $(0,0)$ sector is gapped (two-cluster
    dispersion),
  \item $2 \epsilon_0^{(0,0)} < \epsilon_0^{(+1,0)} +
    \epsilon_0^{(-1,0)}$ (charge-transfer),
  \item $2 \epsilon_0^{(0,0)} < \epsilon_0^{(+1,-1)} +
    \epsilon_0^{(-1,+1)}$ (spin-flip),
  \item $2 \epsilon_0^{(0,0)} < \epsilon_0^{(+1,+1)} +
    \epsilon_0^{(-1,-1)}$ (2 clusters, two-electron charge-transfer).
\end{enumerate}
All the conditions are met in the cases plotted in
Fig.~\ref{fig:1d_spectrum}, although as $l$ becomes larger, $2
\epsilon_0^{(0,0)} \approx \epsilon_0^{(+1,-1)} +
\epsilon_0^{(-1,+1)}$.

\begin{figure}[!htb]
  \includegraphics[height=6.875cm]{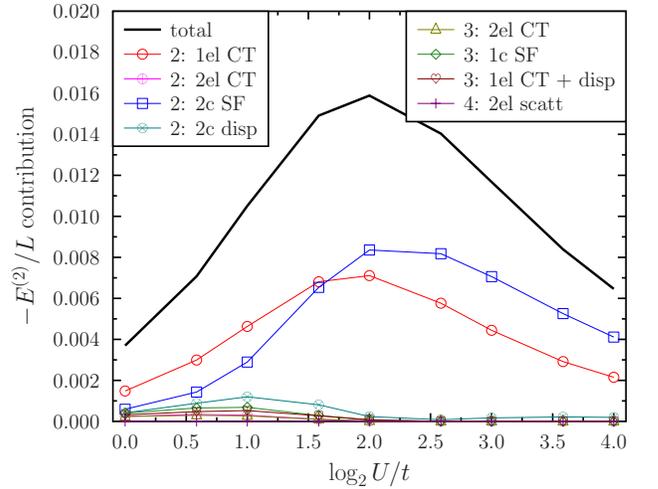}
  \caption{(Color online) Contributions to the second-order energy in
    U-cPT2 calculations ($L=120$ 1D periodic lattice) using a cluster
    size of 4 as a function of $U/t$. The notation used in the key
    takes the form ``\# clusters: interaction type''. Here, CT stands
    for charge transfer and SF refers to spin flip. The interaction
    channels are those from Tab. \ref{tab:pert_channel}; two-electron
    charge transfer processes are of opposite spin due to the nature
    of the Hubbard interaction. \label{fig:pt2_contrib}}
\end{figure}

We show in Fig. \ref{fig:pt2_contrib} the contributions of different
channels to the second-order energy in U-cPT2 calculations as a
function of $U/t$. Results from Fig.~\ref{fig:pt2_contrib} indicate
that two-cluster spin-flip (two neighboring clusters undergoing a spin
flip) and two-cluster 1-electron charge transfer (two clusters
interchanging a single electron) processes are the most important
contributors to the second-order energy. Beyond that, the remaining
two- and three-cluster interactions are small but non-negligible for
$U/t < 4$. Four-cluster interactions are very small across all $U/t$.

\begin{figure*}[!htb]
  \includegraphics[height=6.875cm]{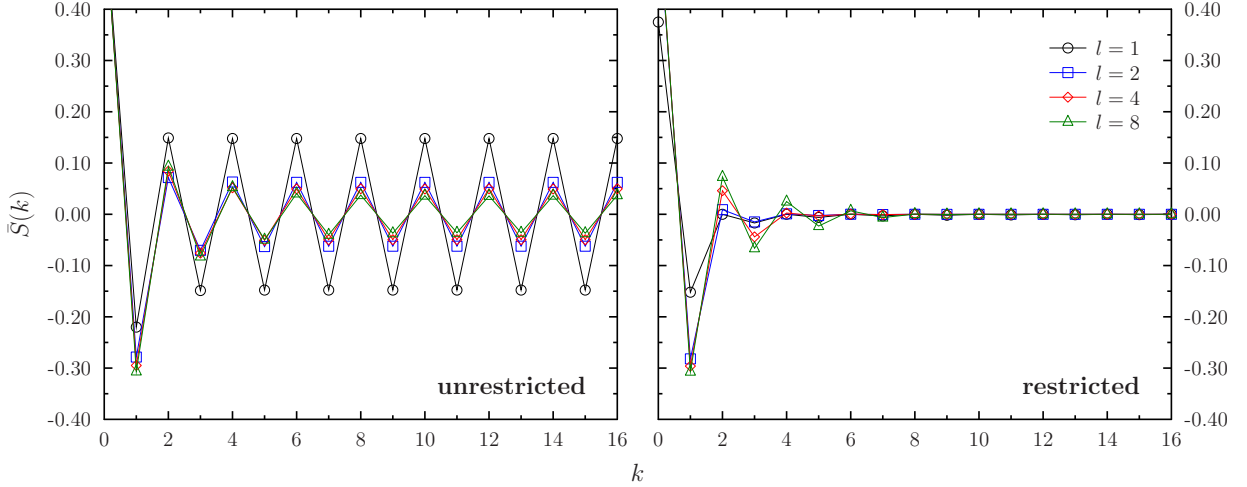}
  \caption{(Color online) Real-space spin-spin correlations computed
    from unrestricted (left) and restricted (right) optimized cMF
    states for a $L=120$ periodic 1D lattice at $U/t=4$. The $l=1$
    result in the left panel corresponds to UHF, while that in the
    right panel corresponds to restricted HF ({\em i.e.}, a product of
    plane-wave states). \label{fig:1d_spin}}
\end{figure*}

We now turn our attention to the spin-spin correlations in the ground
state. As the cMF ansatz breaks the translational invariance of the
Hubbard model, we have computed averaged spin-spin correlations,
defined by
\begin{equation}
  \bar{S}(\mathbf{j}) = \frac{1}{L} \sum_{\mathbf{j'}} \langle
  \mathbf{S}_{\mathbf{j'}} \cdot \mathbf{S}_{\mathbf{j}-\mathbf{j'}}
  \rangle,
\end{equation}
where $\mathbf{j}$ labels a lattice site. We plot in
Fig.~\ref{fig:1d_spin} the (real-space) spin-spin correlations
obtained from R-cMF and U-cMF calculations at $U/t=4$.

It becomes evident that unrestricted calculations (left panel) yield a
structure with long range order. R-cMF, on the other hand, has
non-vanishing spin-spin correlations only within the cluster;
inter-cluster correlations vanish due to the spin singlet character of
each cluster state. The long range spin-spin correlations in U-CMF are
systematically decreased as the size of the cluster is increased. In
the short range, both R-cMF and U-cMF yield significant corrections to
RHF and UHF, respectively. With $l=8$, R-cMF and U-cMF display similar
correlations to the first few neighbors (small $k$), with hints of the
$1/k$ decay \cite{imada-1992,essler} present in the exact solution at
long range.

\begin{figure}[!htb]
  \includegraphics[height=6.875cm]{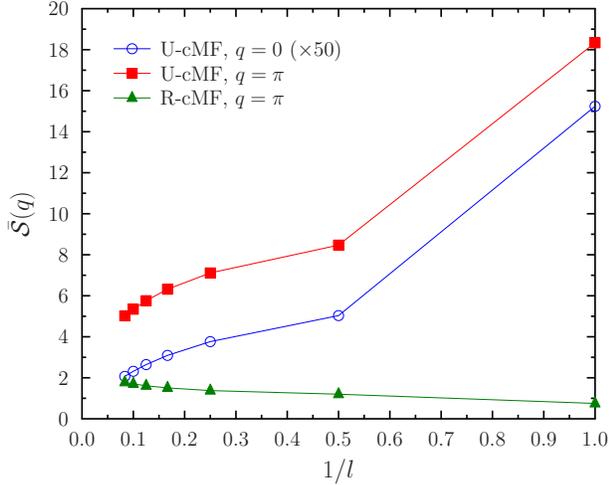}
  \caption{(Color online) Fourier-transformed averaged spin-spin
    correlations ($L=120$ 1D periodic lattice at $U/t=4$) in U-cMF and
    R-cMF calculations as a function of the inverse of the cluster
    size. Two $q$ values are plotted, namely 0 and $\pi$ (note that
    the $q=0$ curve is enhanced by a factor of 50 for clarity
    purposes). U-cMF and R-cMF calculations should tend to the same
    $q=\pi$ finite value as $l \to L$. The $l \to L$ limit at $q=0$ of
    U-cMF results should be 0 as the exact ground state corresponds to
    a true singlet state. \label{fig:1d_spin_ksp}}
\end{figure}

A cleaner picture of the spin-spin correlations can be obtained by
looking at them in reciprocal space. Figure~\ref{fig:1d_spin_ksp}
displays the (discrete) Fourier-transformed spin-spin correlations, at
wave-vectors $q=0$ and $q=\pi$, obtained from R-cMF and U-cMF
calculations. (R-cMF results at $q=0$ are not shown as they
identically vanish.) The two momenta are the most relevant ones: the
$q=0$ result provides $\langle \hat{S}^2 \rangle$, which should
identically vanish for a spin singlet, while $q=\pi$ provides the
anti-ferromagnetic structure factor. We see that both values are
significantly decreased in U-cMF with respect to what UHF
predicts. Note that both R-cMF and U-cMF should converge to the same
value (the exact one) as $l$ tends to the $L=120$ limit.

\subsection{2D: half-filling}
\label{sec:results_2d}

We now consider the periodic two-dimensional, half-filled square
lattice. All the calculations in this section use a $12\times 12$
periodic lattice, which should provide near-thermodynamic limit
estimates for $U/t \geq 2$. In 2D, we are able to find a plethora of
local minima (with respect to orbital rotations and coefficients in
the cMF expansion) corresponding to different (approximate) tiling
patterns. In principle, one could argue that, given a fixed number of
clusters with their associated quantum numbers ($n_\uparrow$,
$n_\downarrow$), an optimal solution (a global minimum) exists which
minimizes the energy. We did not attempt to locate it but we did
consider several uniform-like patterns that converge to different
local minima.

\begin{figure*}[!htb]
  \includegraphics[height=15.6cm]{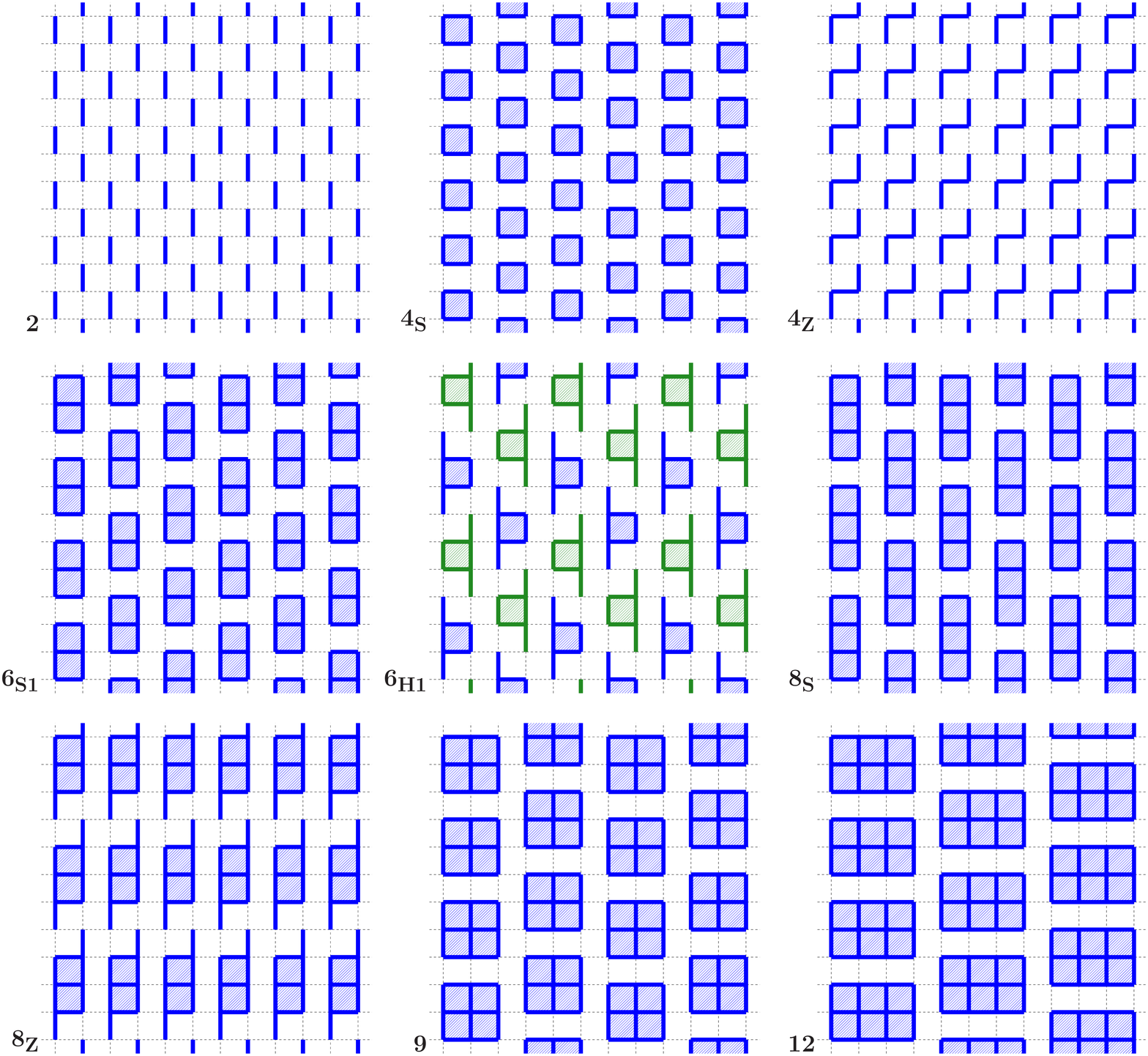}
  \caption{(Color online) Tiling patterns adopted in U-cMF
    calculations on a square $12\times 12$ periodic 2D square lattice
    at half-filling. Sites within the same cluster are connected by
    solid lines; fully enclosed regions are shaded. A key is provided
    to the left of each structure. \label{fig:2d_tile}}
\end{figure*}

We show in Fig.~\ref{fig:2d_tile} the tiling patterns adopted in U-cMF
calculations on the $12\times 12$ lattice. A label used to identify
each pattern is also provided in the figure. In most of the tilings
displayed, a staggered configuration was chosen over an otherwise
uniform tiling as it leads to lower variational energies. A staggered
dimer configuration is used in clusters of size 2. For clusters of
size 4, we discuss results with a square-based (staggered,
$\mathbf{4_S}$) tiling and a z-shaped ($\mathbf{4_Z}$) tiling. We have
considered a staggered configuration in terms of slabs
($\mathbf{6_{S1}}$) and hats ($\mathbf{6_{H1}}$) in connection with
clusters of size 6. Clusters of size 8 with a staggered slab ($4\times
2$, $\mathbf{8_S}$) and a z-shaped ($\mathbf{8_Z}$) configuration are
used, which can be thought of as simple dimers of the considered
size-4 clusters. We finally also examined staggered plaquette
configurations with clusters of size 9 and 12.

We should point that, in constrast to Ref.~\onlinecite{isaev-2009},
all the considered tiling patterns lead to a qualitatively correct
description of the ground state character, {\em i.e.}, all structures
lead to a ground state density with non-zero magnetization in a
N\'eel-type configuration. This is because the ground state of each
cluster was independently optimized, without requiring that all the
clusters share the same ground state.

\begin{figure*}[!htb]
  \includegraphics[height=6.875cm]{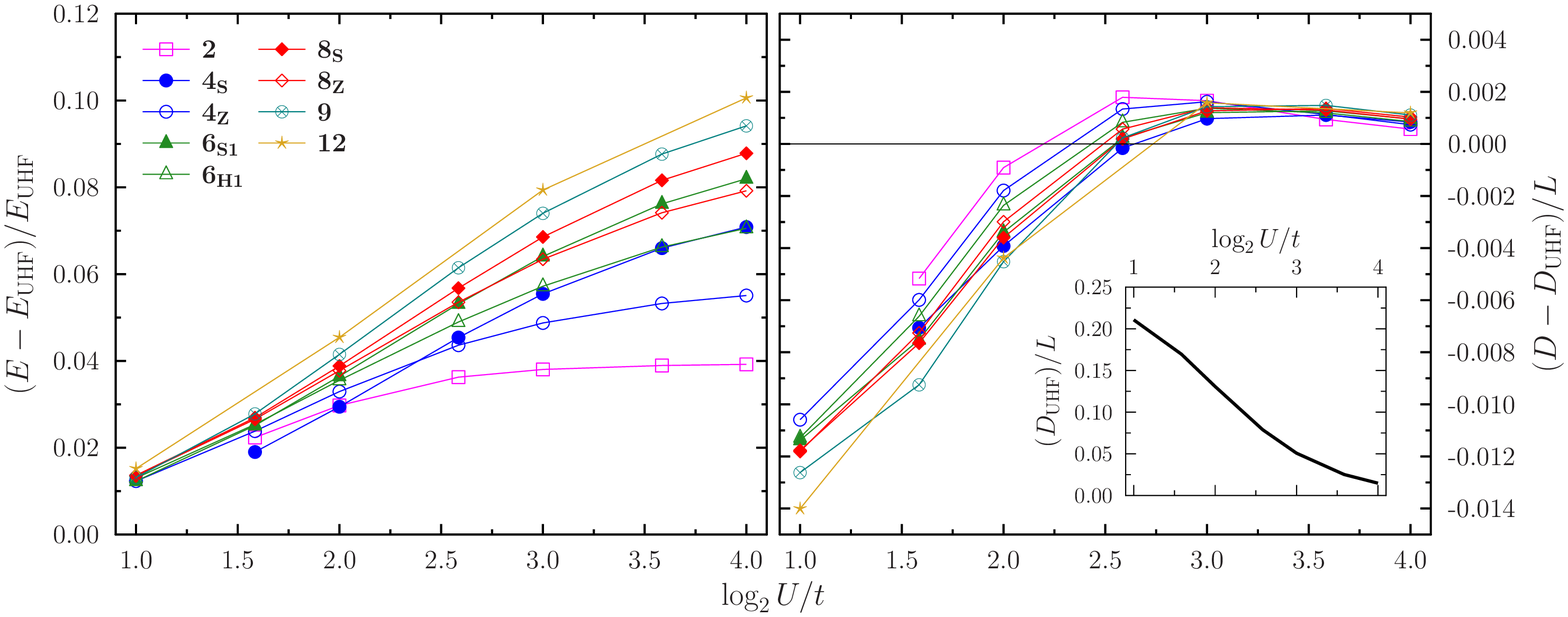}
  \caption{(Color online) (Left) Correlation energy (with respect to
    UHF), divided by the UHF energy, recovered in U-cMF calculations
    in a $12\times 12$ periodic 2D square lattice, as a function of
    $U/t$. The tiling patterns considered are those displayed in
    Fig.~\ref{fig:2d_tile}. (Right) Difference with respect to the
    double occupancy per site predicted by UHF in a variety of U-cMF
    calculations in a $12\times 12$ periodic square lattice. The
    double occupancy of UHF itself is shown in the
    inset. \label{fig:2d_energy}}
\end{figure*}

The left panel of Fig.~\ref{fig:2d_energy} shows the correlation
energy (divided by the UHF energy) predicted by U-cMF as a function of
$U/t$. (These can be contrasted with auxiliary-field quantum Monte
Carlo (AFQMC) results shown in Fig.~\ref{fig:2d_energy-2}.)  Here,
size-12 U-cMF predicts that the correlation energy is $\approx
1.5\,\%$ ($\approx 10\,\%$) of the UHF energy at $U/t=2$
($U/t=16$). Large differences in the correlation energy predicted are
observed as the clusters get larger, particularly at large $U/t$,
which is unsurprising. A larger cluster, irrespective of its shape,
tends to yield a larger correlation energy than a smaller one, but a
few exceptions are observed. When clusters of the same size are
compared in different tiling schemes ({\em e.g.}, $\mathbf{4_S}$ and
$\mathbf{4_Z}$), we observe that the more compact the cluster is, the
better the variational estimate for the ground state energy becomes at
large $U/t$. Thus, the square tiling pattern in size-4 clusters yields
a larger correlation energy than the z-shaped one.

The right panel of Fig.~\ref{fig:2d_energy} shows the difference
between the double occupancy ($D=\sum_i \langle n_{i\uparrow}
n_{i\downarrow} \rangle$) per site predicted in U-cMF calculations
with respect to that of UHF (which is shown in the inset). UHF
overestimates (underestimates) the double occupancy at small (large)
$U/t$. A relatively systematic improvement is observed in the double
occupancy as the cluster becomes bigger.

\begin{figure}[!htb]
  \includegraphics[height=6.875cm]{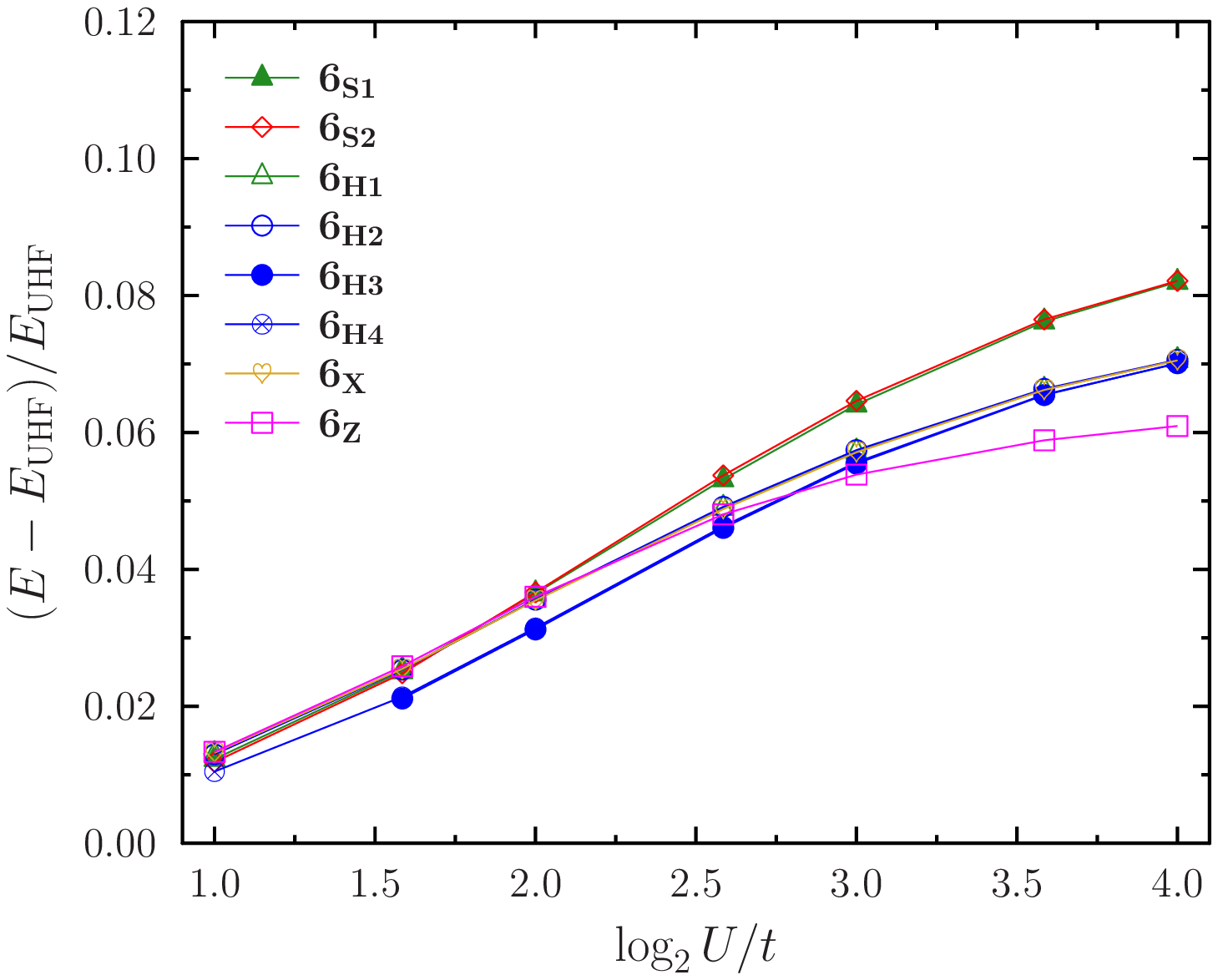}
  \caption{(Color online) Same as Fig. \ref{fig:2d_energy}. Different
    tiling patterns with clusters of size 6 ({\em cf.}
    Figs.~\ref{fig:2d_tile} and \ref{fig:2d_tile2}) are
    displayed. \label{fig:2d_energy_6}}
\end{figure}

\begin{figure*}[!htb]
  \includegraphics[height=10.4cm]{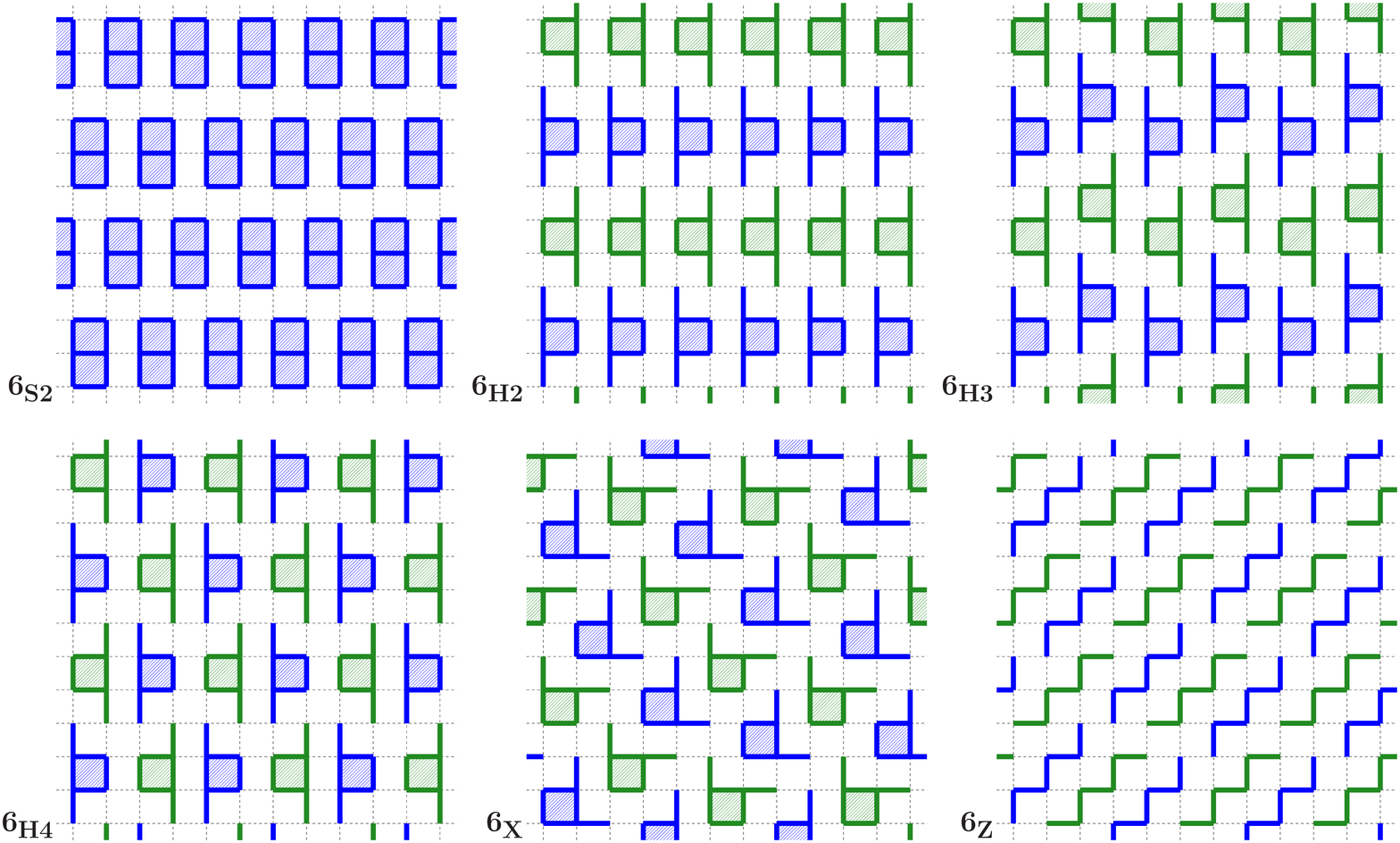}
  \caption{(Color online) Some additional tiling patterns (with
    clusters of size 6) adopted in U-cMF calculations on a square
    $12\times 12$ periodic 2D square lattice at
    half-filling. \label{fig:2d_tile2}}
\end{figure*}

\begin{figure}[!htb]
  \includegraphics[height=6.875cm]{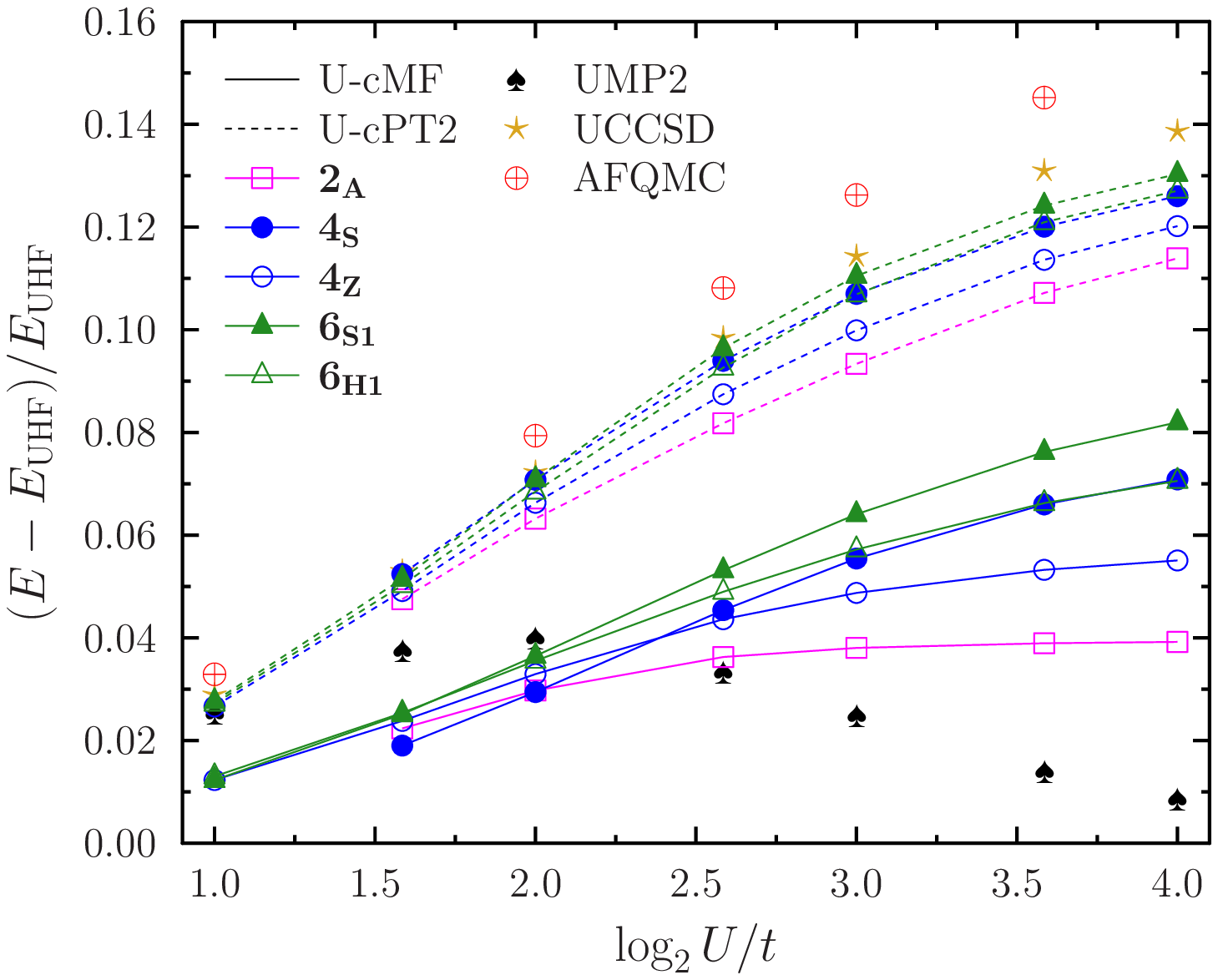}
  \caption{(Color online) Same as Fig. \ref{fig:2d_energy}. U-cMF and
    U-cPT2 calculations are compared with UMP2, UCCSD and AFQMC. AFQMC
    results, from Ref.~\onlinecite{leblanc-2015}, correspond to
    thermodynamic limit estimates. \label{fig:2d_energy-2}}
\end{figure}

To show that other tiling patterns do not lead to fundamentally
different results, we show in Fig. \ref{fig:2d_energy_6} the
correlation energy predicted in U-cMF calculations with different
tiling schemes using clusters of size 6. The additional tiling schemes
(aside from those in Fig.~\ref{fig:2d_tile}) are shown in
Fig.~\ref{fig:2d_tile2}. Several of them lead to approximately the
same ground state energies. As previously discussed, the more compact
the clusters are, the better the variational estimate of the ground
state energy at large $U/t$. The same may not be true at small
$U/t$. For instance, the lowest energies obtained at $U/t=2$
corresponded to structure $\mathbf{6_{Z}}$.

We show in Fig.~\ref{fig:2d_energy-2} the ground state correlation
energies, divided over the UHF energy, obtained from U-cMF and U-cPT2
calculations as a function of $U/t$. Results are compared with UMP2,
UCCSD, and AFQMC \cite{leblanc-2015}, which can be deemed as
numerically exact at half-filling. UMP2 displays the same behavior
observed in 1D, with the correlation energy vanishing for large
$U/t$. It is evident that the U-cMF results are not competitive with
UCCSD or AFQMC, even with the larger clusters from
Fig.~\ref{fig:2d_energy}. For instance, U-cMF using structure
$\mathbf{6_{S1}}$ recovers $< 50\,\%$ of the correlation energy across
all $U/t$. On the other hand, U-cPT2 provides a sizable improvement
over U-cMF results, with structure $\mathbf{6_{S1}}$ capturing
$\approx 85\,\%$ of the correlation energy at $U/t=12$. This is not
far from UCCSD, which recovers $\approx 90\,\%$ of the correlation
energy predicted by AFQMC at $U/t=12$. (Finite size effects account
for most of the difference between UCCSD and AFQMC at $U/t=2$.)

\begin{figure}[!htb]
  \includegraphics[height=7.15cm]{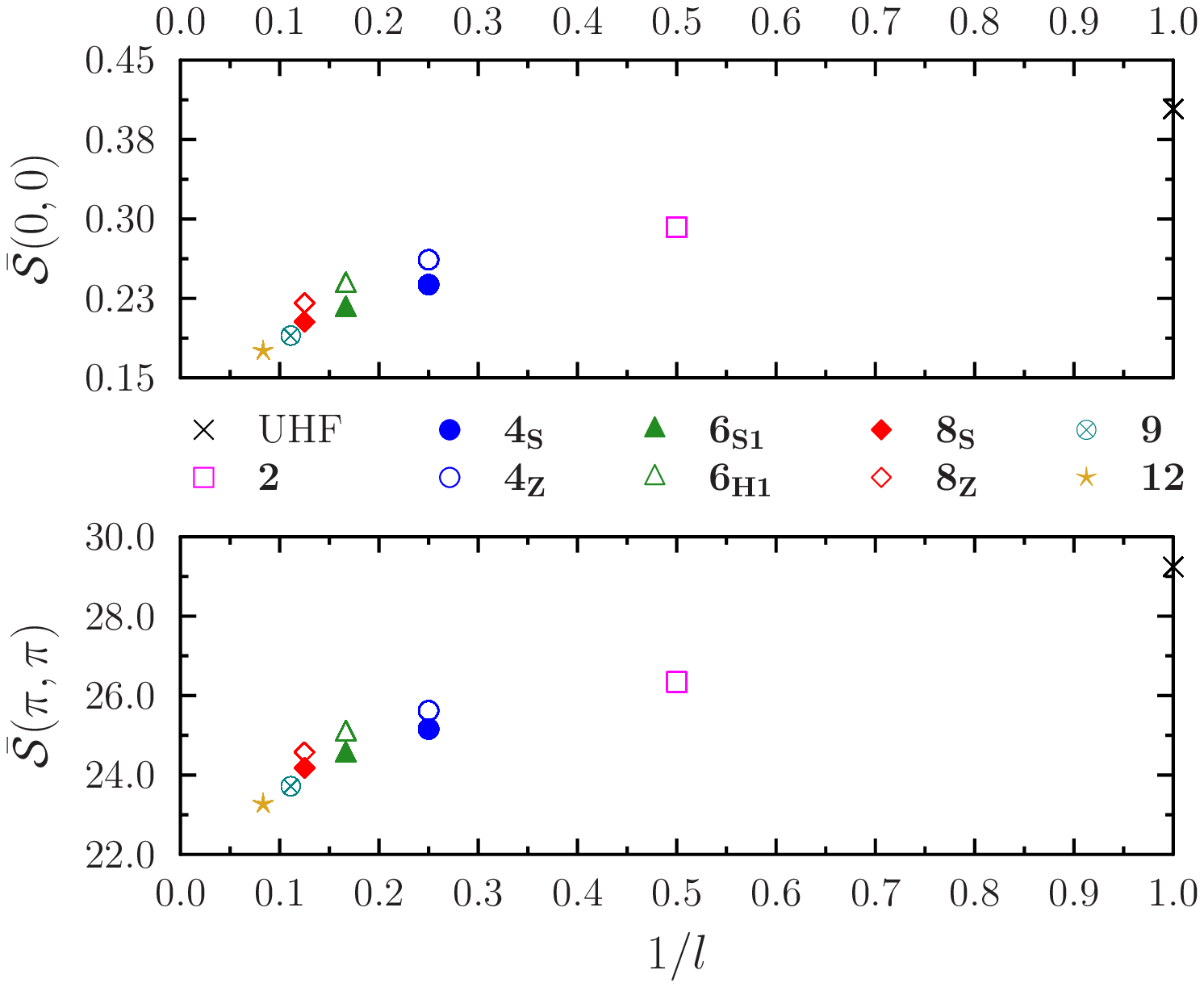}
  \caption{(Color online) Fourier-transformed averaged spin-spin
    correlations ($12\times 12$ periodic 2D square lattice at $U/t=8$)
    in U-cMF calculations as a function of the inverse of the cluster
    size. UHF ($l=1$) results are also displayed. Two $\mathbf{q}$
    values are plotted, namely $(0,0)$ and
    $(\pi,\pi)$. \label{fig:2d_spin_ksp}}
\end{figure}

Figure~\ref{fig:2d_spin_ksp} displays the (discrete)
Fourier-transformed averaged spin-spin correlations, at wave-vectors
$\mathbf{q}=(0,0)$ and $\mathbf{q}=(\pi,\pi)$, obtained from U-cMF
calculations at $U/t=8$ in the 2D lattice. As it was done in the 1D
case, the spin-spin correlations are averaged as the wavefunction
ansatz breaks the translational invariance of the lattice. It becomes
evident that as the cluster becomes bigger (regardless of the specific
tiling pattern), the spin-spin correlations get reduced with respect
to UHF. In particular, large clusters display less than half the
spin-contamination per-site ({\em i.e.}, the deviation of
$\bar{\mathcal{S}}(0,0)$ from 0) of UHF. The anti-ferromagnetic
structure factor is also reduced, though this should converge to a
finite value in the limit $l \to L$.

\subsection{2D: lightly-doped regime}
\label{sec:results_2d_dop}

In the 2D lightly-doped regime, we considered periodic square lattices
with $\langle n \rangle = 0.8$ and $\langle n \rangle = 0.875$,
following Ref.~\onlinecite{leblanc-2015}. We have used a $10 \times
10$ lattice for the former case and a $16 \times 8$ lattice for the
latter case in order to have a lattice commensurate with the striped
order expected to develop, as described below.

Xu {\em et al}. \cite{xu-2011} discussed the UHF phase diagram for the
2D Hubbard model in the half-filled and lightly-doped regime. For
$\langle n \rangle \gtrapprox 0.9$, the system transitions from a
paramagnetic to a linear spin density wave regime at $U/t \approx 1$
({\em cf.} Fig.~17 in Ref.~\onlinecite{xu-2011}). A phase transition
to a regime with diagonal spin density wave character occurs at $U/t
\approx 4$. Finally, the system becomes ferromagnetic at large
$U/t$. This UHF description has guided our U-cMF calculations.

\subsubsection{$\langle n \rangle = 0.8$}

\begin{figure*}[!htb]
  \includegraphics[height=5.5cm]{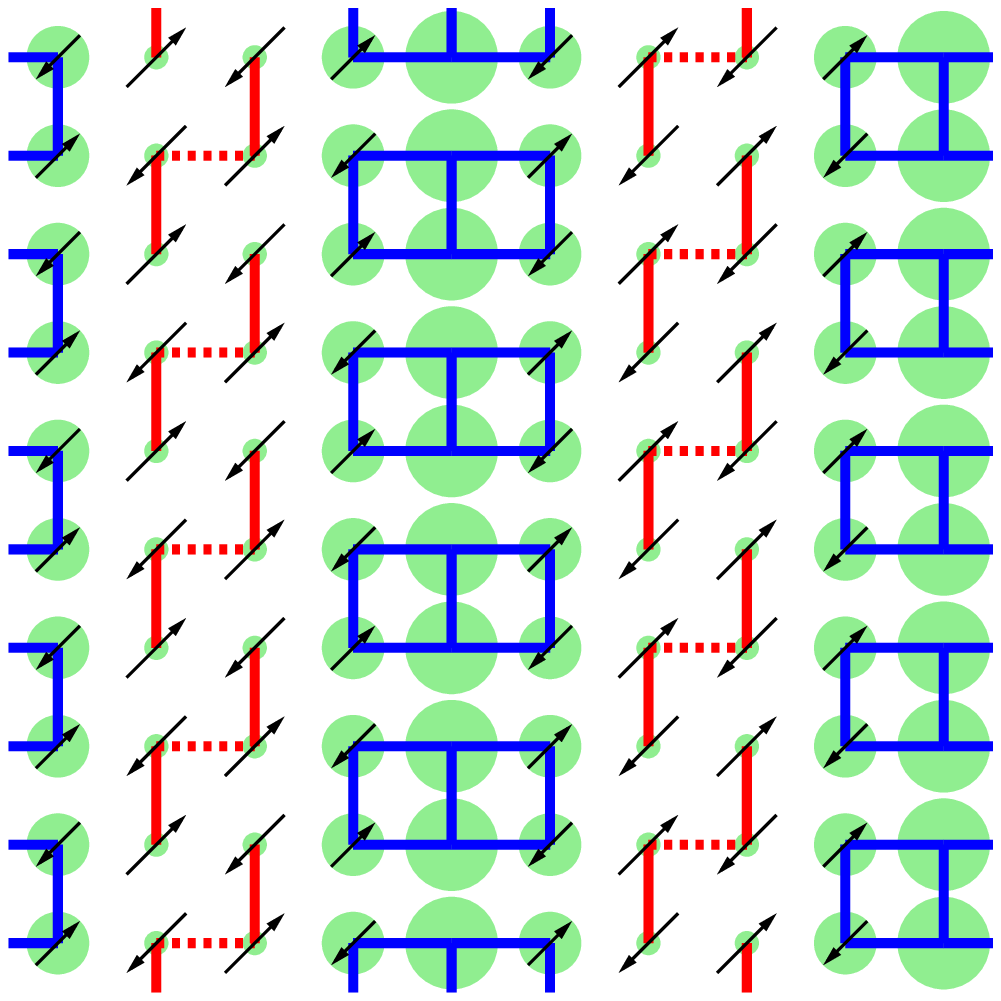}
  \hspace{0.5cm}
  \includegraphics[height=5.5cm]{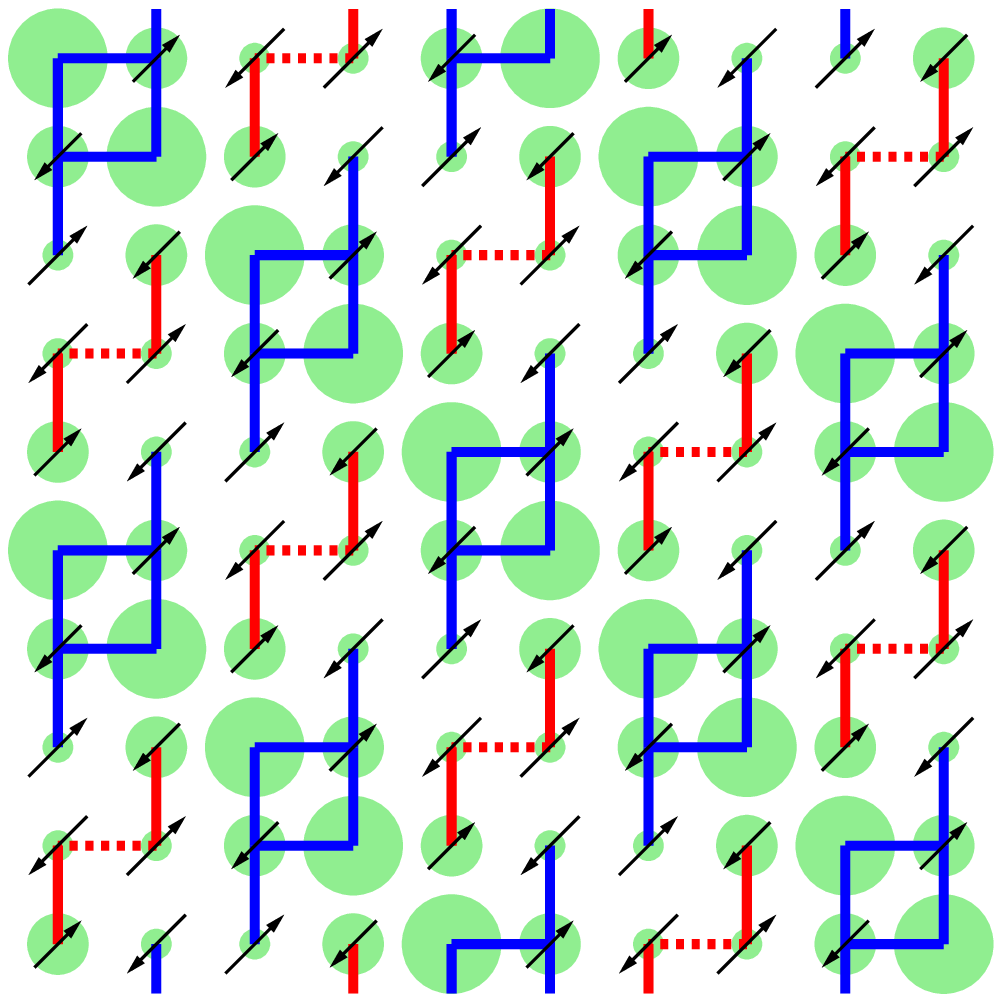}
  \caption{(Color online) Hole and spin density profiles of UHF
    solutions ($10\times 10$ periodic lattice, $\langle n \rangle =
    0.8$) obtained at $U/t=8$ with linear (left) and diagonal (right)
    spin density wave character. The magnitude of the hole density is
    proportional to the area of the green circles. The magnitude of
    the spin density is proportional to the size of the
    arrows. Superimposed on them we display the tiling patterns
    adopted in U-cMF calculations. Clusters with 6 orbitals and 4
    electrons (solid blue) have been used in the high hole density
    sectors, while the N\'eel regions are described in terms of
    staggered dimers (solid red); this leads to structures ${\mathbf
      1}l$ and ${\mathbf 1}d$, respectively, for linear and diagonal
    order. We have additionally considered a tiling pattern where the
    staggered dimers are combined into size 4-clusters, following the
    dotted lines, leading to structures ${\mathbf 2}l$ and ${\mathbf
      2}d$. \label{fig:10x10_uhf}}
\end{figure*}

We show in Fig.~\ref{fig:10x10_uhf} the spin and hole density profiles
of UHF solutions with linear and diagonal spin density wave
character. The (approximate) tiling pattern adopted in U-cMF
calculations is superimposed in the figure. Clusters of size 6 with 4
electrons (2 $\uparrow$ and 2 $\downarrow$) have been used to describe
the sectors with high hole density. On the other hand, the remaining
regions with N\'eel character have been described in terms of a
staggered dimer configuration in structures ${\mathbf 1}l$ and
${\mathbf 1}d$, corresponding to the linear and diagonal spin wave
character, respectively. In ${\mathbf 2}l$ and ${\mathbf 2}d$, the
dimers have been combined into half-filled clusters of size 4 as
depicted in Fig.~\ref{fig:10x10_uhf}.

\begin{figure*}[!htb]
  \includegraphics[height=5.5cm]{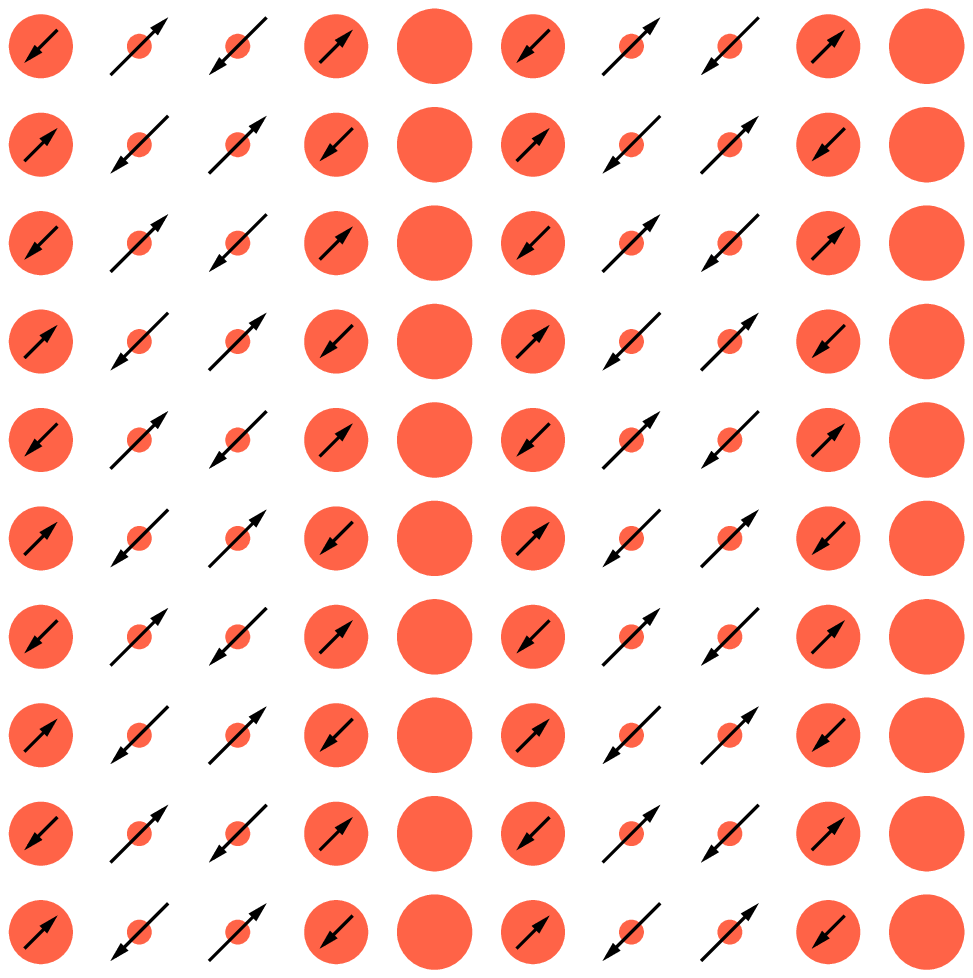}
  \hspace{0.5cm}
  \includegraphics[height=5.5cm]{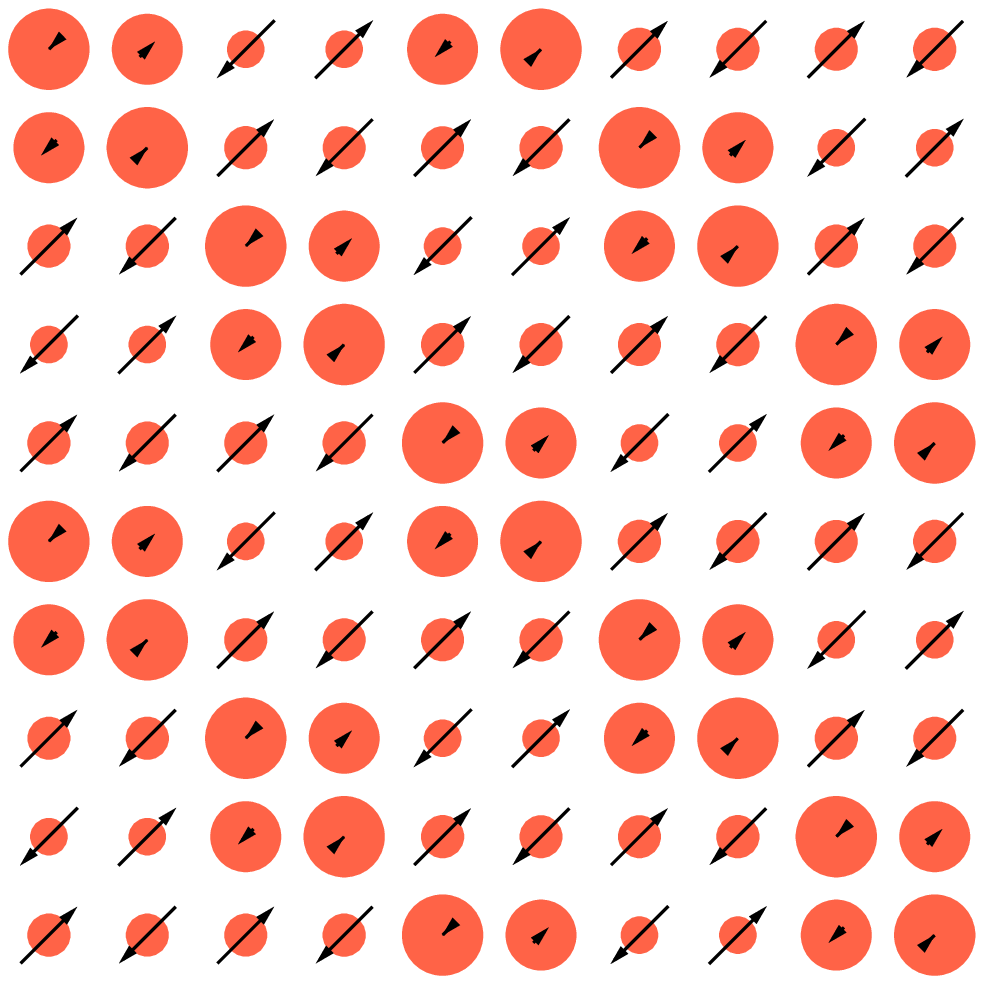}
  \caption{(Color online) Hole and spin density profiles from U-cMF
    calculations ($10\times 10$ lattice, $\langle n \rangle = 0.8$) at
    $U/t=8$ with linear (left, ${\mathbf 1}l$) and diagonal (right,
    ${\mathbf 1}d$) ordering. The magnitude of the hole density is
    proportional to the area of the red circles. The magnitude of the
    spin density is proportional to the size of the
    arrows. \label{fig:10x10str}}
\end{figure*}

The resulting spin- and charge-density profiles obtained from U-cMF
calculations with structures ${\mathbf 1}l$ and ${\mathbf 1}d$ are
displayed in Fig.~\ref{fig:10x10str}. The profiles resemble closely
those obtained from UHF itself and shown in
Fig.~\ref{fig:10x10_uhf}. The main difference is the partial shift of
the hole density away from the main stripes into the neighboring
sites. This suggests that the hole density is too localized in the UHF
solution.

We show in Tab.~\ref{tab:10x10} the resulting ground state energies
from U-cMF and U-cPT2 calculations. Results are compared with UHF,
UMP2, UCCSD, and density matrix embedding (DMET) calculations
\cite{zheng-2015}, which can be deemed as highly accurate.  U-cMF
provides a significant improvement over UHF energies, indicating the
inaccurate treatment of short-range correlations in the simple HF
description. The error in the UHF energies becomes very significant
($> 0.2\,t$) at large $U/t$, sizably larger than the error in the
half-filled regime. Interestingly, the linear spin density wave
character is favored in U-cMF even at relatively large $U/t$. We
cannot rule out, nevertheless, that this is an artifact of the
particular tiling pattern chosen. The use of the size-4 clusters in
place of the N\'eel dimers in U-cMF provides an improvement of about
$10^{-3}\,t$ in the ground state energy for all $U/t$ values quoted.

\begin{table*}[!htb]
  \caption{Ground state energies predicted with a variety of methods
    for a Hubbard 2D periodic $10 \times 10$ lattice with $\langle n
    \rangle = 0.8$. \label{tab:10x10}}
  \begin{ruledtabular}
  \begin{tabular}{l d d d d d}
    method & \multicolumn{1}{r}{$U=2t$}
           & \multicolumn{1}{r}{$U=4t$}
           & \multicolumn{1}{r}{$U=6t$}
           & \multicolumn{1}{r}{$U=8t$}
           & \multicolumn{1}{r}{$U=12t$} \\ \hline
    DMET\ftm{1}
      & -1.3062(4) & -1.108(2) & -0.977(4) & -0.88(3) \\
    UHF (diag)
      & -1.2165 & -0.9646 & -0.7933 & -0.6815 & -0.5501 \\
    UHF (linear) 
      & -1.2678 & -0.9774 & -0.7843 & -0.6597 & \ftm{2} \\
    UMP2\ftm{3}
      & -1.3114 & -1.0760 & -0.8832 & -0.7767 & \ftm{4} \\
    UCCSD\ftm{3}
      & -1.3094 & -1.0925 & -0.9208 & -0.8246 & \ftm{4} \\
    UCCSD(T)\ftm{3}
      & -1.3108 & -1.1045 & -0.9357 & -0.8444 & \ftm{4} \\
    U-cMF(${\mathbf 1} d$)
      & \ftm{5} & \ftm{5} & -0.8417 & -0.7396 & \ftm{5} \\
    U-cMF(${\mathbf 2} d$)
      & \ftm{5} & \ftm{5} & -0.8429 & -0.7406 & \ftm{5} \\
    U-cMF(${\mathbf 1} l$)
      & \ftm{5} & -1.0217 & -0.8520 & -0.7460 & -0.6271 \\
    U-cMF(${\mathbf 2} l$)
      & \ftm{5} & -1.0227 & -0.8536 & -0.7478 & -0.6288 \\
    U-cPT2(${\mathbf 1} l$)
      &         & -1.0865 & -0.9380 & -0.8450 & -0.7394 \\
    U-cPT2(${\mathbf 2} l$)
      &         & -1.0889 & -0.9435 & -0.8526 & -0.7513
    \footnotetext[1]{Results extrapolated to the TDL from
      Refs.~\onlinecite{zheng-2015,leblanc-2015}.}
    \footnotetext[2]{UHF fails to converge with this order.}
    \footnotetext[3]{Calculations use the lowest energy UHF structure
      shown.}
    \footnotetext[4]{Lower energy UHF solutions appear at large $U/t$.}
    \footnotetext[5]{U-cMF optimizations failed to converge.}
  \end{tabular}
  \end{ruledtabular}
\end{table*}

Second-order PT provides a significant improvement over mean-field
energies (both in HF and cMF). The UMP2 results are, however, far from
UCCSD at large $U/t$. The difference between UCCSD(T) and UCCSD is
also large in the strongly-correlated regime, indicating the necessity
of going beyond double excitations in the coupled-cluster ansatz. This
is also evident by comparing UCCSD and DMET results. U-cPT2(${\mathbf
  2} l$) is competitive with UCCSD at $U/t=4$ but outperforms even
UCCSD(T) in the large $U/t$ regime.

\subsubsection{$\langle n \rangle = 0.875$}

\begin{figure}[!htb]
  \includegraphics[height=4.5cm]{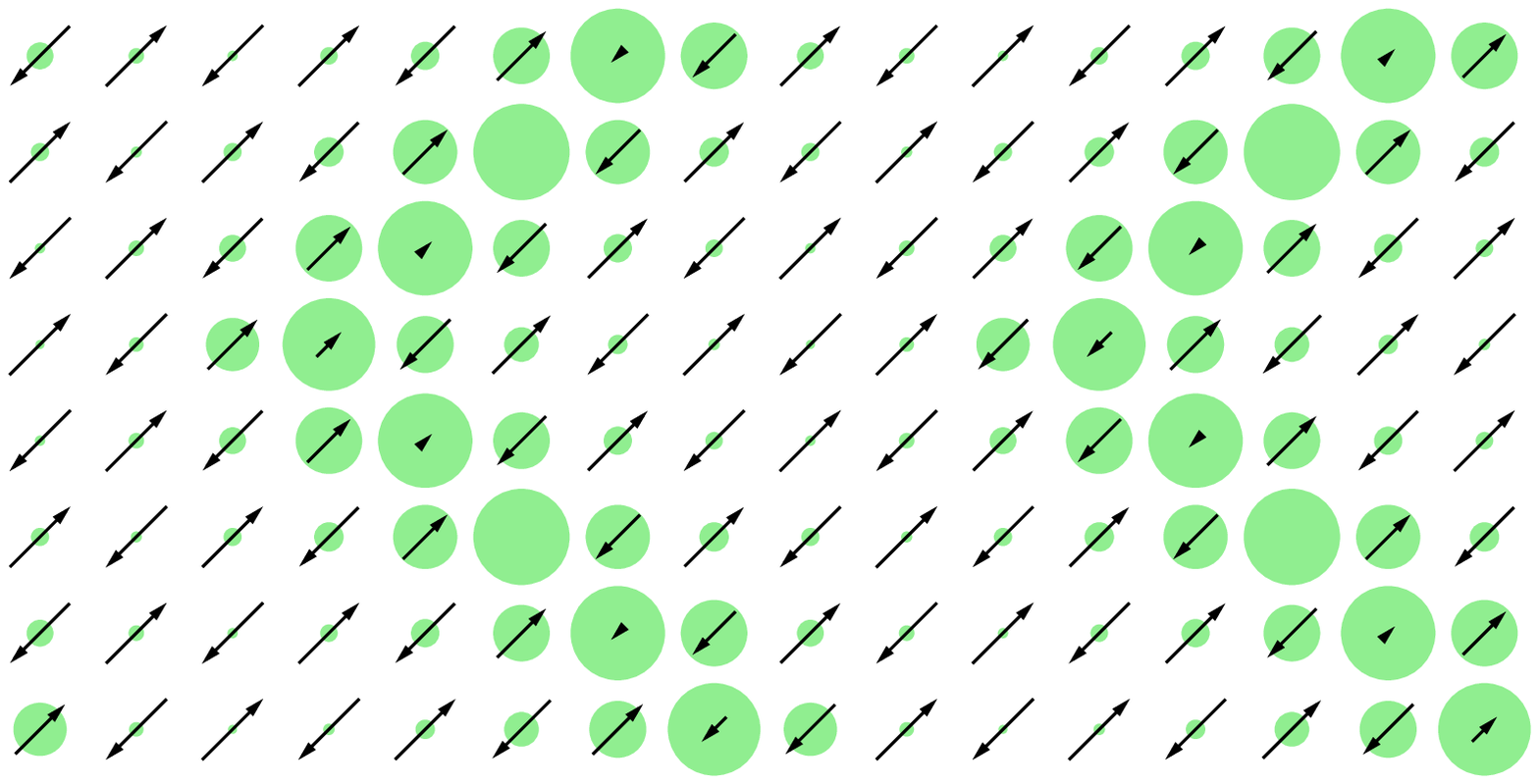}
  \includegraphics[height=4.5cm]{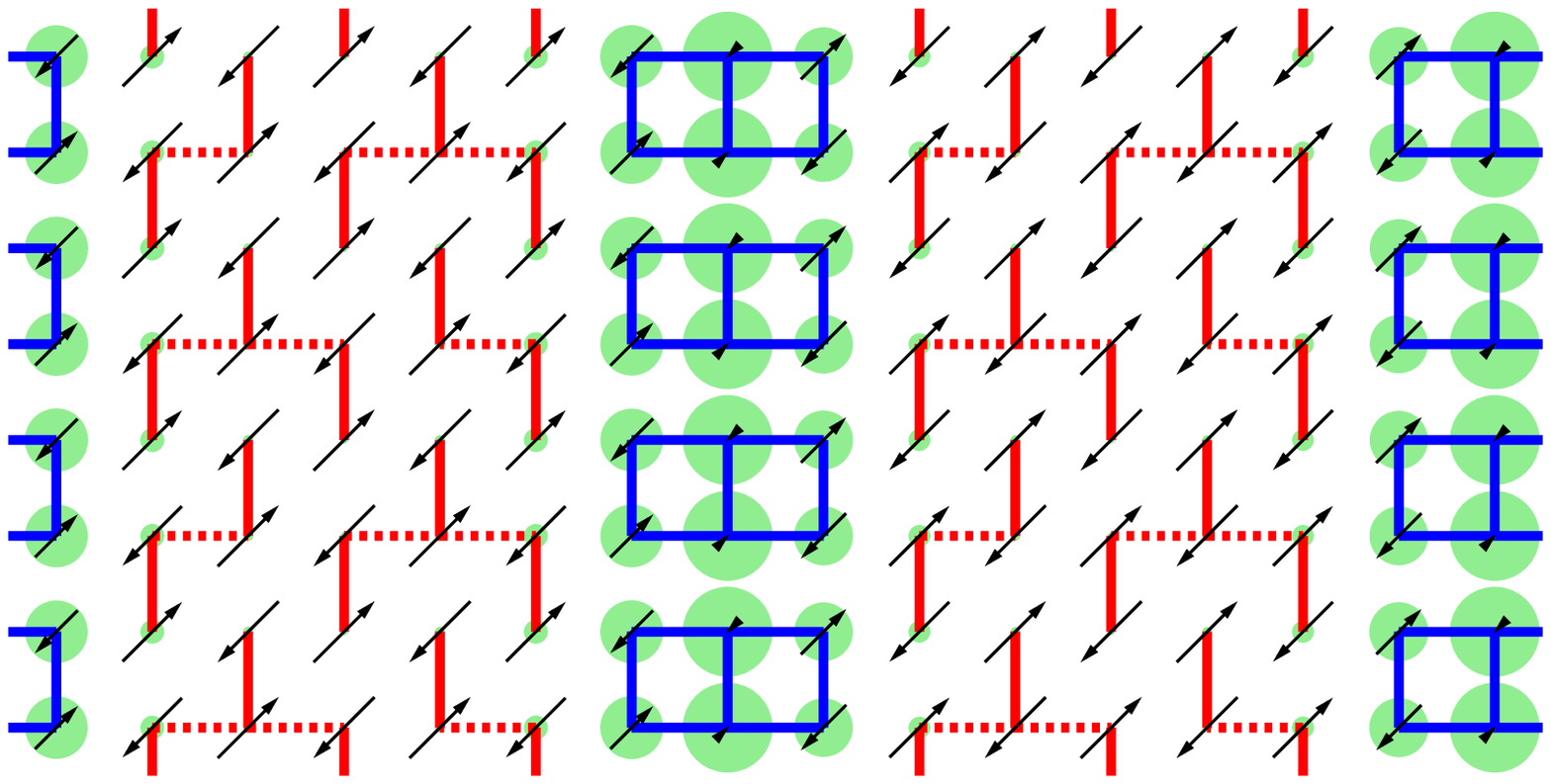}
  \caption{(Color online) Hole and spin density profiles of UHF
    solutions ($16\times 8$ periodic lattice, $\langle n \rangle =
    0.875$) obtained at $U/t=8$ with diagonal (top) and linear
    (bottom) spin density wave character. The magnitude of the hole
    density is proportional to the area of the green circles. The
    magnitude of the spin density is proportional to the size of the
    arrows. Superimposed on the bottom scheme we display the tiling
    pattern adopted in U-cMF calculations. The N\'eel regions are
    described in terms of staggered dimers (solid red), leading to
    structure $\mathbf{1}$, while regions of high hole density are
    described with slab-shaped clusters of size 6 (solid blue). We
    have also considered a pattern (structure $\mathbf{2}$) where the
    dimers are combined into half-filled clusters of size 6 and 4,
    following the dotted lines. \label{fig:16x8_uhf}}
\end{figure}

We now turn our attention to even lighter doping, namely $\langle n
\rangle = 0.875$. We show in Fig.~\ref{fig:16x8_uhf} the spin and hole
density profiles of UHF solutions with linear and diagonal spin
density wave character. Note that kinks are needed in a $16 \times 8$
lattice in the diagonal density wave profile; only a much larger $16
\times 16$ lattice is commensurate with a fully diagonal profile. The
(approximate) tiling pattern adopted in U-cMF calculations (with a
linear density wave character) is superimposed in the figure. As we
did previously in the $10 \times 10$ lattice, the regions with N\'eel
character have been described in terms of a staggered dimer
configuration (structure $\mathbf{1}$), while the regions of high hole
density are tiled into clusters of size 6 (with 2 $\uparrow$- and 2
$\downarrow$-electrons each). The N\'eel dimers have been combined
into half-filled clusters of size 6 and 4 in structure $\mathbf{2}$,
following the pattern indicated in Fig.~\ref{fig:16x8_uhf}.

Table~\ref{tab:16x8} shows the ground state energies obtained from
U-cMF and U-cPT2 calculations. Results are compared with UHF, UMP2,
UCCSD, and DMET. Just as in the $\langle n \rangle = 0.8$ case, U-cMF
improves significantly over UHF. It remains true that second-order PT
provides a nice refinement on top of the mean-field result. The
triples correction in UCCSD(T) becomes significant at large $U/t$,
signaling the deficiencies in UCCSD. U-cPT2 is a bit shy of UCCSD
quality at $U/t=4$, but becomes competitive with UCCSD(T) at $U/t=8$.

\begin{table*}[!htb]
  \caption{Ground state energies predicted with a variety of methods
    for a Hubbard 2D periodic $16 \times 8$ lattice with $\langle n
    \rangle = 0.875$. \label{tab:16x8}}
  \begin{ruledtabular}
  \begin{tabular}{l d d d d d}
    method & \multicolumn{1}{r}{$U=2t$}
           & \multicolumn{1}{r}{$U=4t$}
           & \multicolumn{1}{r}{$U=6t$}
           & \multicolumn{1}{r}{$U=8t$}
           & \multicolumn{1}{r}{$U=12t$} \\ \hline
    DMET\ftm{1}
      & -1.2721(6) & -1.031(3) & -0.86(1) \\
    UHF (diag)
      & \ftm{2} & \ftm{2} & -0.7184 & -0.6008 & -0.4682 \\
    UHF (linear)
      & -1.2270 & -0.9109 & -0.7128 & \ftm{3} & \ftm{3} \\
    UMP2\ftm{4}
      & -1.2732 & -0.9858 & -0.7833 & -0.6594 & \ftm{5} \\
    UCCSD\ftm{4}
      & -1.2719 & -1.0093 & -0.8305 & -0.7147 & \ftm{5} \\
    UCCSD(T)\ftm{4}
      & -1.2738 & -1.0195 & -0.8446 & -0.7299 & \ftm{5} \\
    U-cMF(${\mathbf 1}$)
      & \ftm{6} & -0.9476 & -0.7633 & -0.6477 & -0.5176 \\
    U-cMF(${\mathbf 2}$)
      & \ftm{6} & -0.9500 & -0.7667 & -0.6514 & -0.5210 \\
    U-cPT2(${\mathbf 1}$)
      &         & -1.0004 & -0.8289 & -0.7204 & -0.5965 \\
    U-cPT2(${\mathbf 2}$)
      &         & -1.0040 & -0.8347 & -0.7276 & -0.6068
    \footnotetext[1]{Results extrapolated to the TDL from
      Refs.~\onlinecite{zheng-2015,leblanc-2015}.}
    \footnotetext[2]{UHF (diag) becomes UHF (linear) at low $U/t$.}
    \footnotetext[3]{UHF fails to converge with this order.}
    \footnotetext[4]{Calculations use the lowest energy UHF structure
      shown.}
    \footnotetext[5]{Lower energy UHF solutions appear at large $U/t$.}
    \footnotetext[6]{U-cMF optimizations failed to converge.}
  \end{tabular}
  \end{ruledtabular}
\end{table*}

\begin{figure}[!htb]
  \includegraphics[height=8.91cm]{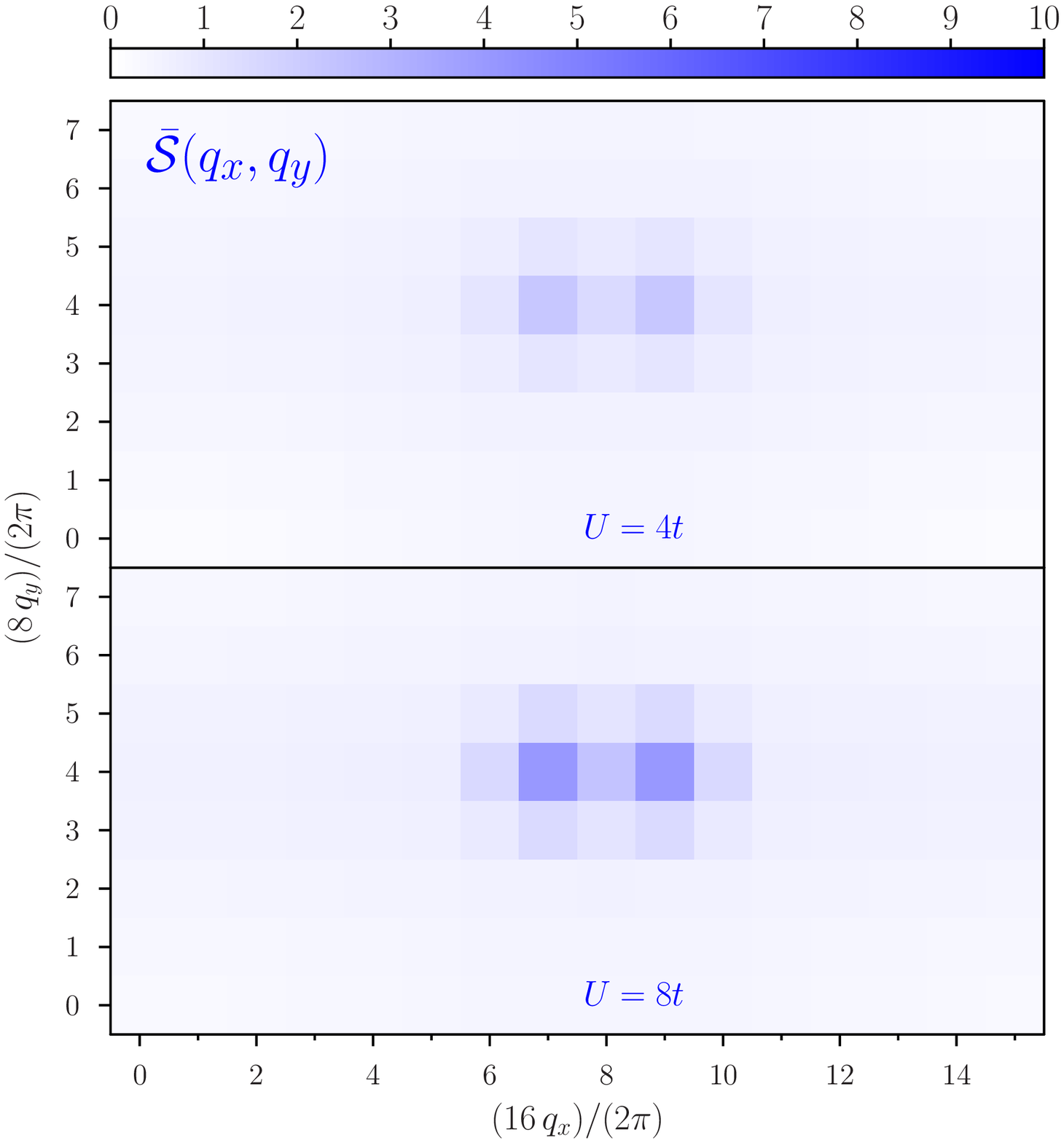}
  \caption{(Color online) Fourier-transform $\bar{\mathcal{S}}(q_x,
    q_y)$ of the averaged spin-spin correlations obtained from U-cMF
    (structure ${\mathbf 1}$) calculations in a $16 \times 8$
    ($\langle n \rangle = 0.875$) lattice at $U/t=4$ (top) and $U/t=8$
    (bottom). \label{fig:16x8_spin}}
\end{figure}

\begin{figure}[!htb]
  \includegraphics[height=8.91cm]{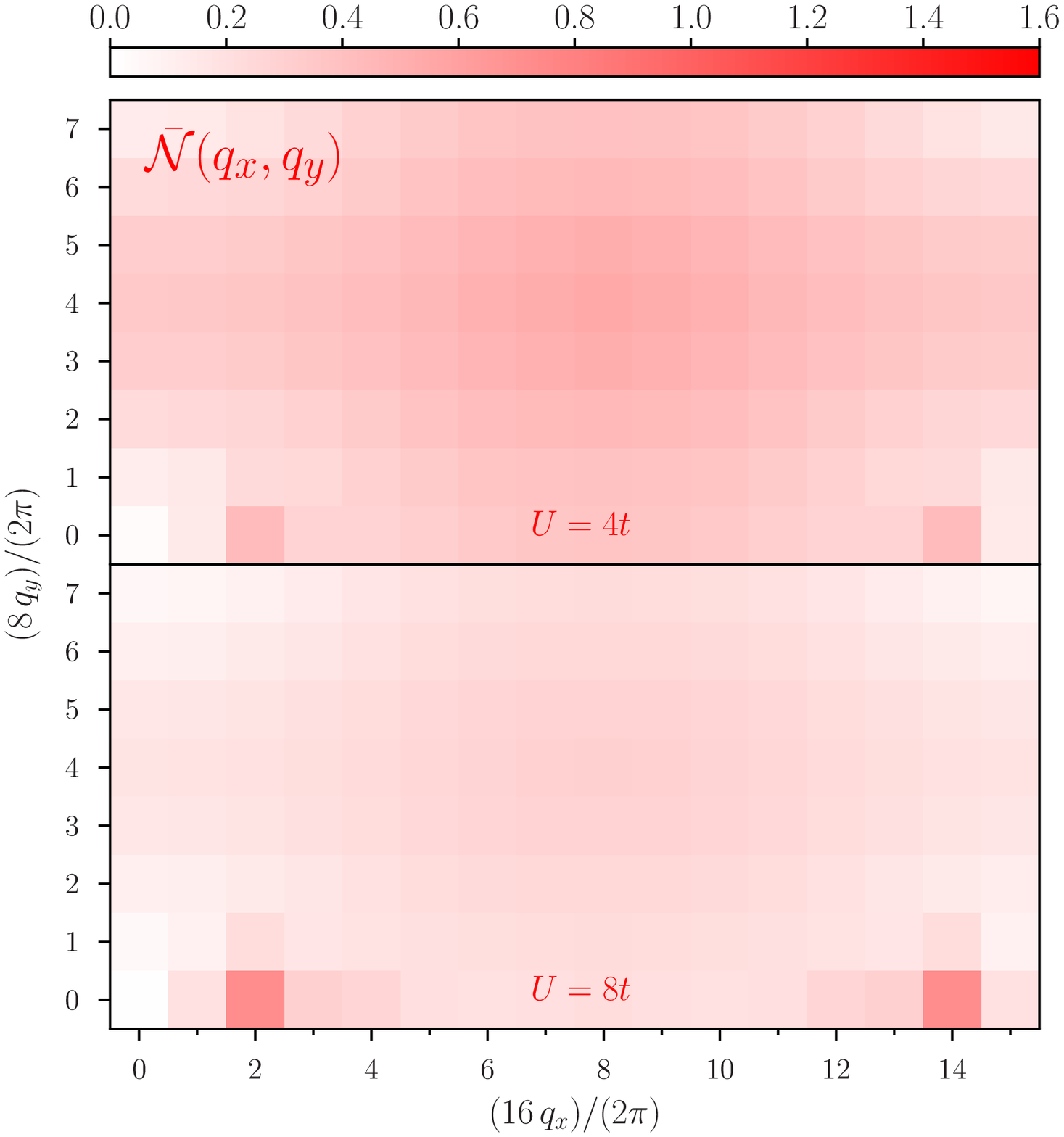}
  \caption{(Color online) Fourier-transform $\bar{\mathcal{N}}(q_x,
    q_y)$ of the averaged density-density correlations obtained from
    U-cMF (structure ${\mathbf 1}$) calculations in a $16 \times 8$
    ($\langle n \rangle = 0.875$) lattice at $U/t=4$ (top) and $U/t=8$
    (bottom). The strong peak at $(0,0)$ ($= N^2/L$) has been removed
    for clarity purposes. \label{fig:16x8_charge}}
\end{figure}

Figures~\ref{fig:16x8_spin} and \ref{fig:16x8_charge} depict the
Fourier-transformed spin-spin and density-density correlations
obtained from U-cMF calculations (using structure $\mathbf{1}$). Just
as in the case of 1D, we have performed a global average over sites in
order to remove the expected fluctuations due to the (spatial)
symmetry broken character of the ansatz. The spin-spin correlations
show a maximum at $\mathbf{q}=(7\pi/8,\pi)$ which becomes more intense
as $U/t$ is increased from 4 to 8. The density-density correlations
display their maximum at $\mathbf{q}=(\pi/4,0)$. These observed
profiles are consistent with the linear spin density wave character of
the UHF charge and spin densities. Symmetry-projected calculations in
Refs.~\onlinecite{rodriguez-2014,juillet-2013} also show the same
features. We refer the reader to Ref.~\onlinecite{chang-2010} for a
discussion of the emergence of spin and charge order in the doped
Hubbard model.

\section{Discussion}
\label{sec:discussion}

In Sec. \ref{sec:formalism}, we have described the cluster mean-field
approach to treat strongly-correlated fermionic systems. A cMF state
(including the orbital optimization degrees of freedom) is used as a
variational ansatz for the ground state wavefunction. This, by
construction, is guaranteed to provide better variational estimates
than HF when the size of the cluster (assuming uniform tiling) is
larger than 1. Because of the simple nature of the ansatz, a RS-PT
scheme can be adopted to account for the missing inter-cluster
correlations. The results presented in Secs.~\ref{sec:results_1d},
\ref{sec:results_2d}, and \ref{sec:results_2d_dop} provide evidence
that a cluster-based approach can provide a (semi)-quantitative
description of the ground state of the half-filled 1D and 2D Hubbard
models, as well as for the lightly-doped regime in 2D square
lattices.

In the half-filled 1D model, results with comparable accuracy to UCCSD
can be obtained by U-cMF using a sufficiently large cluster or by
using U-cPT2 with smaller cluster sizes. Not only the energy is
improved in U-cMF with respect to UHF, but also other ground state
properties such as spin-spin correlations. Due to the local nature of
the interactions in the Hubbard Hamiltonian, contributions to the
second-order energy arise mostly from two-cluster (spin flip and
one-electron charge transfer) interactions.

In the half-filled 2D square model, U-cMF was not as accurate as it
was in the 1D case, even when using clusters of size 12. This
difference can be understood in terms of the missing inter-cluster
correlations and the area-law of entanglement entropy
\cite{eisert-2010}. Whereas in 1D the size of the boundary (which
determines the missing inter-cluster correlations) of a given cluster
remains fixed, in 2D it scales as the perimeter of the cluster
itself. This also explains why more tightly packed clusters provide
better energetic variational estimates for large $U/t$, as the
clusters become more localized. A significant improvement to the
ground state energy is obtained with U-cPT2, where results are again
comparable (although slightly poorer) than UCCSD.

In the lightly-doped regime, we have used {\em ad hoc} tiling schemes
that mostly respect the underlying spin density wave profile obtained
with UHF. Following this strategy, results will be relevant to the
extent that UHF itself provides a qualitatively correct description of
the character of the ground state. Our calculations suggest that it
does. U-cMF results provide a sizable improvement over the UHF
description and U-cPT2 results are competitive with UCCSD at small
$U/t$ and with UCCSD(T) at large $U/t$. The predicted U-cPT2 energies
are still above the DMET estimates, indicating that part of the
long-range correlations expected to develop in the lightly-doped
regime are still unaccounted for. These may be described with higher
order RS-PT or other more powerful many-body approaches.

We note that it is generally true that describing inter-cluster
correlations (here via second-order RS-PT) improves the ground state
energy to a larger extent than enlarging the size of the cluster in
mean-field calculations. (A clear exception to this is the fact that
U-cMF with a cluster of size 2 provides better results at large $U/t$
than UMP2.) The good quality of U-cPT2 results suggest that the
renormalized Hamiltonian (expressed in terms of cluster states) is
more amenable to a perturbative treatment than in the case of the HF
particle-hole transformation. Thus traditional many-body approaches
(such as second-order RS-PT) can be built on top of the cluster
mean-field description to provide a high quality answer. This is
further explored in Appendix \ref{sec:appendix_pert}, were we show the
results of high-order RS-PT calculations in a small lattice. In the
remainder of this section we discuss possible strategies that can
improve the results presented in this manuscript.

Perhaps the simplest strategy is to increase the flexibility in the
mean-field variational ansatz, which can be done in a variety of
ways. The full Hilbert space ({\em i.e.}, not restricted to a given
$m_s$ sector) or, in fact, the full even- or odd-number parity Fock
space within each cluster can be used.\footnote{It is not strictly
  necessary to restrict the number parity of the Fock space in
  mean-field calculations, that is, if a simple product state will be
  considered. Nevertheless, a mixed-number parity description in each
  cluster complicates the evaluation of matrix elements in correlated
  approaches.} In doing this, it is not necessary to use a more
general form for the single-particle transformation that defines the
orbital optimization. The latter can be done in addition to (or in
place of). Thus in systems where local number fluctuations are
essential, a Bogoliubov-de Gennes single-particle transformation can
be used.

The local character of the clusters can be exploited in cPT2
calculations by, {\em e.g.}, truncating the computed interactions
according to some distance criterion. This could alleviate
significantly the computational cost in cPT2 calculations (bringing
them to linear scaling in the number of clusters) while also
facilitating carrying out the cPT$n$ expansion to a higher order. In
order to deal with the large number of interacting clusters and
states, a stochastic sampling of contributing processes can be
performed.

At this point, we would like to comment on the nature of the states
used to carry out the perturbation expansion. In this work, we have
used an energy criterion to truncate the number of cluster states when
this was imperative. Other criteria may be used, such as a density
matrix based criterion (akin to the one used in DMRG
\cite{white-1992,white-1993}). Here, one would diagonalize the
Hamiltonian of the cluster interacting with part of its
environment. The resulting ground state wavefunction is projected into
the cluster states; those states with highest occupation constitute
the optimal subset of states to use. Our main concern regarding this
strategy is that the resulting cluster + (relevant) environment may
become too large to solve (exactly) for its ground state. For
instance, the environment around a four-site square cluster in a 2D
lattice should include at least eight additional sites/orbitals.

Of course, other strategies to account for inter-cluster correlations
may be used. One possible alternative in the case where there are a
few nearly degenerate states in each cluster, is to diagonalize the
full Hamiltonian in the direct product basis spanned by the relevant
cluster states. This is part of the essence of the contractor
renormalization group (CORE) algorithm
\cite{morningstar-1996,capponi-2004,siu-2007} and has also been used
in the active space decomposition (ASD) \cite{parker-2013,parker-2014}
method in quantum chemistry.

We think a coupled-cluster based approach such as the one proposed by
Li in BCCC \cite{li-2004} is among the most promising avenues. In
particular, a coupled-cluster ansatz should provide an improved
description of the missing inter-cluster correlations in the mean
field than low-order RS-PT. Given that the cluster-based Hamiltonian
contains up to four-tile interactions, it appears that the minimal
coupled-cluster model should include up to quadruple
excitations. Nevertheless, we have observed that two-tile interactions
dominate the contribution to the second-order energy. It may not be
unreasonable to restrict the excitation to singles and
doubles. Moreover, locality can also be exploited within a
coupled-cluster framework.

Lastly, we would like to point out that even though we have used the
cMF approach to study strongly interacting systems, it may be used in
other contexts. In particular, systems which can be effectively
represented in terms of weakly interacting fragments of otherwise
strongly-correlated fermions (see Appendix \ref{sec:appendix_dimer})
can be very efficiently described by low-order perturbation theory
based on a cMF state.

\section{Conclusions}
\label{sec:conclusions}

We have introduced a cluster mean-field variational approach and
discussed its applicability to describe the ground state of
strongly-correlated fermion systems. In this work, the full
optimization of the cluster mean-field state has been carried out,
including orbital optimization, with the restriction that the cluster
state has well-defined $n$ and $m_s$ quantum numbers. The restrictions
are imposed in order to preserve $N$ and $M_s$ in the full system. The
cluster product state constitutes an eigenstate of a mean-field
(zero-th order) Hamiltonian, which allows us to formulate a RS
perturbative approach to improve upon the mean-field description.

We have presented mean-field and second-order perturbative results of
the ground state energies (and other observables) of the periodic 1D
and square 2D Hubbard models. In the half-filled 1D case, our U-cMF
results become as accurate as UCCSD across all $U/t$ for sufficiently
large clusters. U-cPT2 results on smaller clusters also provide a
consistent description across all interaction strengths. In 2D at
half-filling, U-cMF is poorer than in the 1D case yet U-cPT2 provides
ground state energies of near UCCSD quality. In the lightly-doped
regime of the 2D model, U-cPT2 results remain competitive with UCCSD
although they are still not competitive with DMET estimates. In
general, we observe that U-cPT2 energies with small clusters are often
better than U-cMF results with significantly larger ones.

Overall, the results of this work suggest that a cluster mean-field
approach can provide an excellent starting point and a path to a
highly accurate, efficient description of strongly-correlated
fermionic systems, and the Hubbard model in particular. Several
strategies to improve the mean-field description as well as correlated
approaches built on top of it have been suggested.

\section*{Acknowledgments}

This work was supported by the Department of Energy, Office of Basic
Energy Sciences, Grant No. DE-FG02-09ER16053. G.E.S. is a Welch
Foundation Chair (C-0036). We thank Jorge Dukelsky for valuable
discussions and we are thankful to the participants in ``The Simons
Collaboration on the Many-Electron Problem'' for providing us with
early access to the results of Ref.~\onlinecite{leblanc-2015}.

\appendix

\section{High-order perturbation theory}
\label{sec:appendix_pert}

\begin{figure*}[!htb]
  \includegraphics[height=6.875cm]{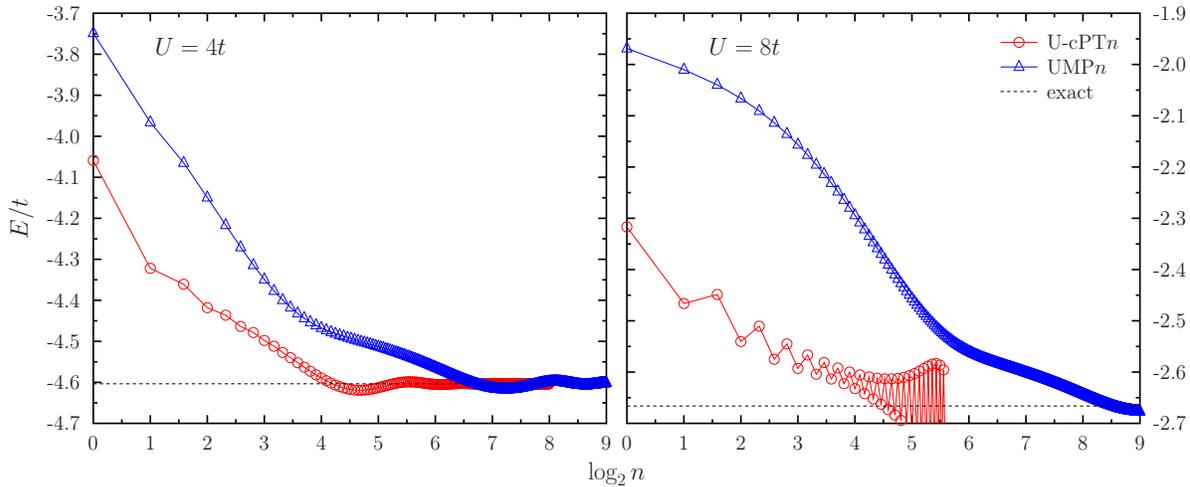}
  \caption{(Color online) Convergence of the UMP$n$ and U-cPT$n$
    perturbation series for a half-filled $L=8$ periodic 1D lattice at
    $U/t=4$ (left) and $U/t=8$ (right). A cluster size of 2 has been
    used in U-cPT$n$ calculations. The exact ground state energy is
    denoted by a dashed line. Due to the divergent nature of the
    U-cPT$n$ series at $U/t=8$, we have truncated at order
    $n=48$. \label{fig:pert}}
\end{figure*}

For sufficiently small systems, the standard RS perturbation series
can be evaluated to high order by direct solution of the RS-PT
equations
\begin{align}
  (\hat{H}_0 - E^{(0)}) |\Psi^{(m)} \rangle = & \, \hat{V}
  |\Psi^{(m-1)} \rangle - \sum_{l=0}^{m-1} E^{(m-l)} |\Psi^{(l)}
  \rangle, \\
  E^{(m)} = & \, \langle \Psi^{(0)} | \hat{V} | \Psi^{(m-1)} \rangle,
\end{align}
where we have assumed intermediate normalization ($\langle \Psi^{(0)}
| \Psi^{(m)} \rangle = 0 \quad \forall \quad m > 0$). We have
evaluated the UMP$n$ ({\em i.e.}, RS-PT using canonical UHF orbitals
and orbital energies) and U-cPT$n$ (as formulated in
Sec. \ref{sec:formalism}, using a cluster of size 2) perturbation
series for a half-filled $L=8$ 1D periodic lattice at $U/t=4$ and
$U/t=8$. The energy as a function of $n$ is displayed in
Fig.~\ref{fig:pert}.

The UMP$n$ series approaches the exact energies very slowly,
particularly at large $U/t$. On the other hand, U-cPT$n$ is much
faster approaching the exact energy, although the series has a
divergent nature at $U/t=8$. This is likely due to the near
degeneracies expected to appear at large $U/t$ in the spectrum of each
cluster. It is possible that a convergent nature can be restored by
tweaking the definition of the zero-th order Hamiltonian. In spite of
that, these results support the premise that once correlations within
the cluster have been described accurately, the ground state of the
resulting renormalized Hamiltonian can be expressed by a many-body
expansion that is more rapidly convergent than the common UMP$n$
series.

\section{Dimerized Hubbard model}
\label{sec:appendix_dimer}

\begin{figure}[!htb]
  \includegraphics[height=6.875cm]{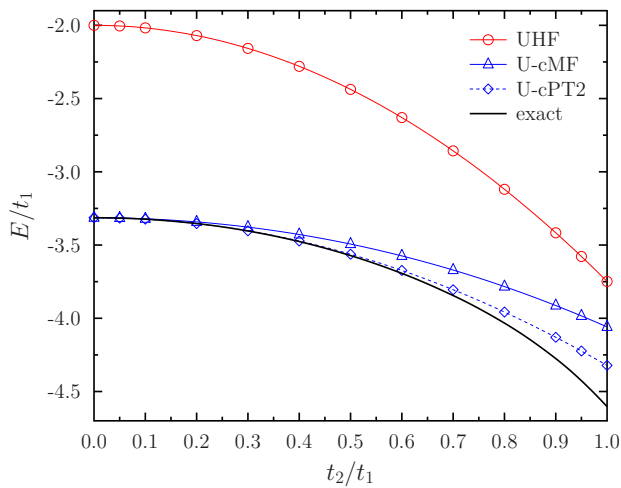}
  \caption{(Color online) UHF, U-cMF, and U-cPT2 ground state energies
    for a half-filled $L=8$ periodic 1D dimerized Hubbard lattice
    ({\em cf.}  Eq.~\ref{eq:dimer}) as a function of the ratio
    $t_2/t_1$, with $U/t_1=4$. U-cMF and U-cPT2 calculations employ
    clusters of size 2. \label{fig:dimer}}
\end{figure}

Consider a Hubbard lattice tiled into clusters. As the clusters become
non-interacting, the cMF approach becomes exact. In this section, we
assess the quality of the cMF and cPT2 ground state energies as a
function of the interaction between clusters. Being more specific, we
consider the dimerized periodic Hubbard 1D model (see, {\em e.g.},
Ref.~\onlinecite{mila-1997}), given by the Hamiltonian
\begin{align}
  \hat{H} = &\, -t_1 \sum_{\textrm{odd } j,\sigma}
  (c^\dagger_{j+1,\sigma} \, c_{j,\sigma} + \textrm{h.c.}) \nonumber
  \\
  &\, -t_2 \sum_{\textrm{even } j,\sigma} (c^\dagger_{j+1,\sigma} \,
  c_{j,\sigma} + \textrm{h.c.})
  + U \sum_j n_{j,\uparrow} \, n_{j,\downarrow}.
  \label{eq:dimer}
\end{align}
Figure~\ref{fig:dimer} shows the UHF, U-cMF, and U-cPT2 ground state
energies obtained in a half-filled $L=8$ periodic 1D lattice, as a
function of the ratio $t_2/t_1$ with $U/t_1 = 4$. Clusters of size 2
have been used in cMF and cPT2 calculations.

At $t_2/t_1 = 0$, the exact ground state energy reduces to four-times
the energy of an $L=2$ half-filled lattice with OBC. Naturally, U-cMF
reproduces this result, while UHF converges to an energy that is equal
to four-times the energy of the corresponding UHF result for the $L=2$
lattice. U-cMF (U-cPT2) remains highly accurate up to $t_2/t_1 \approx
0.3$ ($t_2/t_1 \approx 0.6$). On the other hand, U-cMF and U-cPT2 are
not nearly as accurate in the vicinity of $t_2/t_1 = 1$, yet they
still provide sizable improvements over the UHF description. The
mean-field approach recovers $\approx 35\,\%$ of the correlation
energy, while U-cPT2 is able to capture around $65\,\%$ of it. These
results suggest that a simple cluster-based approach can accurately
describe weak interactions among otherwise strongly-correlated
fragments.

\bibliography{paper}

\begin{thebibliography}{59}%
\makeatletter
\providecommand \@ifxundefined [1]{%
 \@ifx{#1\undefined}
}%
\providecommand \@ifnum [1]{%
 \ifnum #1\expandafter \@firstoftwo
 \else \expandafter \@secondoftwo
 \fi
}%
\providecommand \@ifx [1]{%
 \ifx #1\expandafter \@firstoftwo
 \else \expandafter \@secondoftwo
 \fi
}%
\providecommand \natexlab [1]{#1}%
\providecommand \enquote  [1]{``#1''}%
\providecommand \bibnamefont  [1]{#1}%
\providecommand \bibfnamefont [1]{#1}%
\providecommand \citenamefont [1]{#1}%
\providecommand \href@noop [0]{\@secondoftwo}%
\providecommand \href [0]{\begingroup \@sanitize@url \@href}%
\providecommand \@href[1]{\@@startlink{#1}\@@href}%
\providecommand \@@href[1]{\endgroup#1\@@endlink}%
\providecommand \@sanitize@url [0]{\catcode `\\12\catcode `\$12\catcode
  `\&12\catcode `\#12\catcode `\^12\catcode `\_12\catcode `\%12\relax}%
\providecommand \@@startlink[1]{}%
\providecommand \@@endlink[0]{}%
\providecommand \url  [0]{\begingroup\@sanitize@url \@url }%
\providecommand \@url [1]{\endgroup\@href {#1}{\urlprefix }}%
\providecommand \urlprefix  [0]{URL }%
\providecommand \Eprint [0]{\href }%
\providecommand \doibase [0]{http://dx.doi.org/}%
\providecommand \selectlanguage [0]{\@gobble}%
\providecommand \bibinfo  [0]{\@secondoftwo}%
\providecommand \bibfield  [0]{\@secondoftwo}%
\providecommand \translation [1]{[#1]}%
\providecommand \BibitemOpen [0]{}%
\providecommand \bibitemStop [0]{}%
\providecommand \bibitemNoStop [0]{.\EOS\space}%
\providecommand \EOS [0]{\spacefactor3000\relax}%
\providecommand \BibitemShut  [1]{\csname bibitem#1\endcsname}%
\let\auto@bib@innerbib\@empty
\bibitem [{\citenamefont {Anderson}(1987)}]{anderson-1987}%
  \BibitemOpen
  \bibfield  {author} {\bibinfo {author} {\bibfnamefont {P.~W.}\ \bibnamefont
  {Anderson}},\ }\href@noop {} {\bibfield  {journal} {\bibinfo  {journal}
  {Science}\ }\textbf {\bibinfo {volume} {235}},\ \bibinfo {pages} {1196}
  (\bibinfo {year} {1987})}\BibitemShut {NoStop}%
\bibitem [{\citenamefont {McWeeny}(1959)}]{mcweeny-1959}%
  \BibitemOpen
  \bibfield  {author} {\bibinfo {author} {\bibfnamefont {R.}~\bibnamefont
  {McWeeny}},\ }\href@noop {} {\bibfield  {journal} {\bibinfo  {journal} {Proc.
  R. Soc. Lond. A}\ }\textbf {\bibinfo {volume} {253}},\ \bibinfo {pages} {242}
  (\bibinfo {year} {1959})}\BibitemShut {NoStop}%
\bibitem [{\citenamefont {McWeeny}(1960)}]{mcweeny-1960}%
  \BibitemOpen
  \bibfield  {author} {\bibinfo {author} {\bibfnamefont {R.}~\bibnamefont
  {McWeeny}},\ }\href@noop {} {\bibfield  {journal} {\bibinfo  {journal} {Rev.
  Mod. Phys.}\ }\textbf {\bibinfo {volume} {32}},\ \bibinfo {pages} {335}
  (\bibinfo {year} {1960})}\BibitemShut {NoStop}%
\bibitem [{\citenamefont {Isaev}\ \emph {et~al.}(2009)\citenamefont {Isaev},
  \citenamefont {Ortiz},\ and\ \citenamefont {Dukelsky}}]{isaev-2009}%
  \BibitemOpen
  \bibfield  {author} {\bibinfo {author} {\bibfnamefont {L.}~\bibnamefont
  {Isaev}}, \bibinfo {author} {\bibfnamefont {G.}~\bibnamefont {Ortiz}}, \ and\
  \bibinfo {author} {\bibfnamefont {J.}~\bibnamefont {Dukelsky}},\ }\href@noop
  {} {\bibfield  {journal} {\bibinfo  {journal} {Phys. Rev. B}\ }\textbf
  {\bibinfo {volume} {79}},\ \bibinfo {pages} {024409} (\bibinfo {year}
  {2009})}\BibitemShut {NoStop}%
\bibitem [{\citenamefont {Zhao}\ \emph {et~al.}(2014)\citenamefont {Zhao},
  \citenamefont {{Jim\'enez-Hoyos}}, \citenamefont {Scuseria}, \citenamefont
  {Huerga}, \citenamefont {Dukelsky}, \citenamefont {Rombouts},\ and\
  \citenamefont {Ortiz}}]{zhao-2014}%
  \BibitemOpen
  \bibfield  {author} {\bibinfo {author} {\bibfnamefont {J.}~\bibnamefont
  {Zhao}}, \bibinfo {author} {\bibfnamefont {C.~A.}\ \bibnamefont
  {{Jim\'enez-Hoyos}}}, \bibinfo {author} {\bibfnamefont {G.~E.}\ \bibnamefont
  {Scuseria}}, \bibinfo {author} {\bibfnamefont {D.}~\bibnamefont {Huerga}},
  \bibinfo {author} {\bibfnamefont {J.}~\bibnamefont {Dukelsky}}, \bibinfo
  {author} {\bibfnamefont {S.~M.~A.}\ \bibnamefont {Rombouts}}, \ and\ \bibinfo
  {author} {\bibfnamefont {G.}~\bibnamefont {Ortiz}},\ }\href@noop {}
  {\bibfield  {journal} {\bibinfo  {journal} {J. Phys.: Condens. Matter}\
  }\textbf {\bibinfo {volume} {26}},\ \bibinfo {pages} {455601} (\bibinfo
  {year} {2014})}\BibitemShut {NoStop}%
\bibitem [{\citenamefont {Schr\"odinger}(1926)}]{schrodinger-1926}%
  \BibitemOpen
  \bibfield  {author} {\bibinfo {author} {\bibfnamefont {E.}~\bibnamefont
  {Schr\"odinger}},\ }\href@noop {} {\bibfield  {journal} {\bibinfo  {journal}
  {Ann. Phys.}\ }\textbf {\bibinfo {volume} {385}},\ \bibinfo {pages} {437}
  (\bibinfo {year} {1926})}\BibitemShut {NoStop}%
\bibitem [{\citenamefont {Li}(2004)}]{li-2004}%
  \BibitemOpen
  \bibfield  {author} {\bibinfo {author} {\bibfnamefont {S.}~\bibnamefont
  {Li}},\ }\href@noop {} {\bibfield  {journal} {\bibinfo  {journal} {J. Chem.
  Phys.}\ }\textbf {\bibinfo {volume} {120}},\ \bibinfo {pages} {5017}
  (\bibinfo {year} {2004})}\BibitemShut {NoStop}%
\bibitem [{\citenamefont {Fang}\ \emph {et~al.}(2008)\citenamefont {Fang},
  \citenamefont {Shen},\ and\ \citenamefont {Li}}]{fang-2008}%
  \BibitemOpen
  \bibfield  {author} {\bibinfo {author} {\bibfnamefont {T.}~\bibnamefont
  {Fang}}, \bibinfo {author} {\bibfnamefont {J.}~\bibnamefont {Shen}}, \ and\
  \bibinfo {author} {\bibfnamefont {S.}~\bibnamefont {Li}},\ }\href@noop {}
  {\bibfield  {journal} {\bibinfo  {journal} {J. Chem. Phys.}\ }\textbf
  {\bibinfo {volume} {128}},\ \bibinfo {pages} {224107} (\bibinfo {year}
  {2008})}\BibitemShut {NoStop}%
\bibitem [{\citenamefont {Shen}\ and\ \citenamefont {Li}(2009)}]{shen-2009}%
  \BibitemOpen
  \bibfield  {author} {\bibinfo {author} {\bibfnamefont {J.}~\bibnamefont
  {Shen}}\ and\ \bibinfo {author} {\bibfnamefont {S.}~\bibnamefont {Li}},\
  }\href@noop {} {\bibfield  {journal} {\bibinfo  {journal} {J. Chem. Phys.}\
  }\textbf {\bibinfo {volume} {131}},\ \bibinfo {pages} {174101} (\bibinfo
  {year} {2009})}\BibitemShut {NoStop}%
\bibitem [{\citenamefont {Xu}\ and\ \citenamefont {Li}(2013)}]{xu-2013}%
  \BibitemOpen
  \bibfield  {author} {\bibinfo {author} {\bibfnamefont {E.}~\bibnamefont
  {Xu}}\ and\ \bibinfo {author} {\bibfnamefont {S.}~\bibnamefont {Li}},\
  }\href@noop {} {\bibfield  {journal} {\bibinfo  {journal} {J. Chem. Phys.}\
  }\textbf {\bibinfo {volume} {139}},\ \bibinfo {pages} {174111} (\bibinfo
  {year} {2013})}\BibitemShut {NoStop}%
\bibitem [{\citenamefont {Cirac}\ and\ \citenamefont
  {Verstraete}(2009)}]{cirac-2009}%
  \BibitemOpen
  \bibfield  {author} {\bibinfo {author} {\bibfnamefont {J.~I.}\ \bibnamefont
  {Cirac}}\ and\ \bibinfo {author} {\bibfnamefont {F.}~\bibnamefont
  {Verstraete}},\ }\href@noop {} {\bibfield  {journal} {\bibinfo  {journal} {J.
  Phys. A.: Math. Theor.}\ }\textbf {\bibinfo {volume} {42}},\ \bibinfo {pages}
  {504004} (\bibinfo {year} {2009})}\BibitemShut {NoStop}%
\bibitem [{\citenamefont {Or\'us}(2014)}]{orus-2014}%
  \BibitemOpen
  \bibfield  {author} {\bibinfo {author} {\bibfnamefont {R.}~\bibnamefont
  {Or\'us}},\ }\href@noop {} {\bibfield  {journal} {\bibinfo  {journal} {Ann.
  Phys.}\ }\textbf {\bibinfo {volume} {349}},\ \bibinfo {pages} {117} (\bibinfo
  {year} {2014})}\BibitemShut {NoStop}%
\bibitem [{\citenamefont {White}(1992)}]{white-1992}%
  \BibitemOpen
  \bibfield  {author} {\bibinfo {author} {\bibfnamefont {S.~R.}\ \bibnamefont
  {White}},\ }\href@noop {} {\bibfield  {journal} {\bibinfo  {journal} {Phys.
  Rev. Lett.}\ }\textbf {\bibinfo {volume} {69}},\ \bibinfo {pages} {2863}
  (\bibinfo {year} {1992})}\BibitemShut {NoStop}%
\bibitem [{\citenamefont {White}(1993)}]{white-1993}%
  \BibitemOpen
  \bibfield  {author} {\bibinfo {author} {\bibfnamefont {S.~R.}\ \bibnamefont
  {White}},\ }\href@noop {} {\bibfield  {journal} {\bibinfo  {journal} {Phys.
  Rev. B}\ }\textbf {\bibinfo {volume} {48}},\ \bibinfo {pages} {10345}
  (\bibinfo {year} {1993})}\BibitemShut {NoStop}%
\bibitem [{\citenamefont {Changlani}\ \emph {et~al.}(2009)\citenamefont
  {Changlani}, \citenamefont {Kinder}, \citenamefont {Umrigar},\ and\
  \citenamefont {Chan}}]{changlani-2009}%
  \BibitemOpen
  \bibfield  {author} {\bibinfo {author} {\bibfnamefont {H.~J.}\ \bibnamefont
  {Changlani}}, \bibinfo {author} {\bibfnamefont {J.~M.}\ \bibnamefont
  {Kinder}}, \bibinfo {author} {\bibfnamefont {C.~J.}\ \bibnamefont {Umrigar}},
  \ and\ \bibinfo {author} {\bibfnamefont {G.~K.-L.}\ \bibnamefont {Chan}},\
  }\href@noop {} {\bibfield  {journal} {\bibinfo  {journal} {Phys. Rev. B}\
  }\textbf {\bibinfo {volume} {80}},\ \bibinfo {pages} {245116} (\bibinfo
  {year} {2009})}\BibitemShut {NoStop}%
\bibitem [{\citenamefont {Mezzacapo}\ \emph {et~al.}(2009)\citenamefont
  {Mezzacapo}, \citenamefont {Schuch}, \citenamefont {Boninsegni},\ and\
  \citenamefont {Cirac}}]{mezzacapo-2009}%
  \BibitemOpen
  \bibfield  {author} {\bibinfo {author} {\bibfnamefont {F.}~\bibnamefont
  {Mezzacapo}}, \bibinfo {author} {\bibfnamefont {N.}~\bibnamefont {Schuch}},
  \bibinfo {author} {\bibfnamefont {M.}~\bibnamefont {Boninsegni}}, \ and\
  \bibinfo {author} {\bibfnamefont {J.~I.}\ \bibnamefont {Cirac}},\ }\href@noop
  {} {\bibfield  {journal} {\bibinfo  {journal} {New J. Phys.}\ }\textbf
  {\bibinfo {volume} {11}},\ \bibinfo {pages} {083026} (\bibinfo {year}
  {2009})}\BibitemShut {NoStop}%
\bibitem [{\citenamefont {Mezzacapo}\ and\ \citenamefont
  {Cirac}(2010)}]{mezzacapo-2010}%
  \BibitemOpen
  \bibfield  {author} {\bibinfo {author} {\bibfnamefont {F.}~\bibnamefont
  {Mezzacapo}}\ and\ \bibinfo {author} {\bibfnamefont {J.~I.}\ \bibnamefont
  {Cirac}},\ }\href@noop {} {\bibfield  {journal} {\bibinfo  {journal} {New J.
  Phys.}\ }\textbf {\bibinfo {volume} {12}},\ \bibinfo {pages} {103039}
  (\bibinfo {year} {2010})}\BibitemShut {NoStop}%
\bibitem [{\citenamefont {Neuscamman}\ \emph {et~al.}(2011)\citenamefont
  {Neuscamman}, \citenamefont {Changlani}, \citenamefont {Kinder},\ and\
  \citenamefont {Chan}}]{neuscamman-2011}%
  \BibitemOpen
  \bibfield  {author} {\bibinfo {author} {\bibfnamefont {E.}~\bibnamefont
  {Neuscamman}}, \bibinfo {author} {\bibfnamefont {H.}~\bibnamefont
  {Changlani}}, \bibinfo {author} {\bibfnamefont {J.}~\bibnamefont {Kinder}}, \
  and\ \bibinfo {author} {\bibfnamefont {G.~K.-L.}\ \bibnamefont {Chan}},\
  }\href@noop {} {\bibfield  {journal} {\bibinfo  {journal} {Phys. Rev. B}\
  }\textbf {\bibinfo {volume} {84}},\ \bibinfo {pages} {205132} (\bibinfo
  {year} {2011})}\BibitemShut {NoStop}%
\bibitem [{\citenamefont {Hubbard}(1963)}]{hubbard-1963}%
  \BibitemOpen
  \bibfield  {author} {\bibinfo {author} {\bibfnamefont {J.}~\bibnamefont
  {Hubbard}},\ }\href@noop {} {\bibfield  {journal} {\bibinfo  {journal} {Proc.
  Roy. Soc. Lond. A}\ }\textbf {\bibinfo {volume} {276}},\ \bibinfo {pages}
  {238} (\bibinfo {year} {1963})}\BibitemShut {NoStop}%
\bibitem [{\citenamefont {Lieb}\ and\ \citenamefont {Wu}(1968)}]{lieb-1968}%
  \BibitemOpen
  \bibfield  {author} {\bibinfo {author} {\bibfnamefont {E.~H.}\ \bibnamefont
  {Lieb}}\ and\ \bibinfo {author} {\bibfnamefont {F.~Y.}\ \bibnamefont {Wu}},\
  }\href@noop {} {\bibfield  {journal} {\bibinfo  {journal} {Phys. Rev. Lett.}\
  }\textbf {\bibinfo {volume} {20}},\ \bibinfo {pages} {1445} (\bibinfo {year}
  {1968})}\BibitemShut {NoStop}%
\bibitem [{\citenamefont {Scalapino}(2007)}]{scalapino}%
  \BibitemOpen
  \bibfield  {author} {\bibinfo {author} {\bibfnamefont {D.}~\bibnamefont
  {Scalapino}},\ }in\ \href@noop {} {\emph {\bibinfo {booktitle} {Handbook of
  High-Temperature Superconductivity}}},\ \bibinfo {editor} {edited by\
  \bibinfo {editor} {\bibfnamefont {J.~R.}\ \bibnamefont {Schrieffer}}\ and\
  \bibinfo {editor} {\bibfnamefont {J.~S.}\ \bibnamefont {Brooks}}}\ (\bibinfo
  {publisher} {Springer New York},\ \bibinfo {year} {2007})\ pp.\ \bibinfo
  {pages} {495--526}\BibitemShut {NoStop}%
\bibitem [{\citenamefont {LeBlanc}\ \emph {et~al.}(2015)\citenamefont
  {LeBlanc}, \citenamefont {Antipov}, \citenamefont {Becca}, \citenamefont
  {Bulik}, \citenamefont {Chan}, \citenamefont {Chung}, \citenamefont {Deng},
  \citenamefont {Ferrero}, \citenamefont {Henderson}, \citenamefont
  {Jim\'enez-Hoyos}, \citenamefont {Kozik}, \citenamefont {Liu}, \citenamefont
  {Millis}, \citenamefont {{Prokof'ev}}, \citenamefont {Qin}, \citenamefont
  {Scuseria}, \citenamefont {Shi}, \citenamefont {Svistunov}, \citenamefont
  {Tocchio}, \citenamefont {Tupitsyn}, \citenamefont {White}, \citenamefont
  {Zhang}, \citenamefont {Zheng}, \citenamefont {Zhu},\ and\ \citenamefont
  {Gull}}]{leblanc-2015}%
  \BibitemOpen
  \bibfield  {author} {\bibinfo {author} {\bibfnamefont {J.~P.~F.}\
  \bibnamefont {LeBlanc}}, \bibinfo {author} {\bibfnamefont {A.~E.}\
  \bibnamefont {Antipov}}, \bibinfo {author} {\bibfnamefont {F.}~\bibnamefont
  {Becca}}, \bibinfo {author} {\bibfnamefont {I.~W.}\ \bibnamefont {Bulik}},
  \bibinfo {author} {\bibfnamefont {G.~K.-L.}\ \bibnamefont {Chan}}, \bibinfo
  {author} {\bibfnamefont {C.-M.}\ \bibnamefont {Chung}}, \bibinfo {author}
  {\bibfnamefont {Y.}~\bibnamefont {Deng}}, \bibinfo {author} {\bibfnamefont
  {M.}~\bibnamefont {Ferrero}}, \bibinfo {author} {\bibfnamefont {T.~M.}\
  \bibnamefont {Henderson}}, \bibinfo {author} {\bibfnamefont {C.~A.}\
  \bibnamefont {Jim\'enez-Hoyos}}, \bibinfo {author} {\bibfnamefont
  {E.}~\bibnamefont {Kozik}}, \bibinfo {author} {\bibfnamefont {X.-W.}\
  \bibnamefont {Liu}}, \bibinfo {author} {\bibfnamefont {A.~J.}\ \bibnamefont
  {Millis}}, \bibinfo {author} {\bibfnamefont {N.~V.}\ \bibnamefont
  {{Prokof'ev}}}, \bibinfo {author} {\bibfnamefont {M.}~\bibnamefont {Qin}},
  \bibinfo {author} {\bibfnamefont {G.~E.}\ \bibnamefont {Scuseria}}, \bibinfo
  {author} {\bibfnamefont {H.}~\bibnamefont {Shi}}, \bibinfo {author}
  {\bibfnamefont {B.}~\bibnamefont {Svistunov}}, \bibinfo {author}
  {\bibfnamefont {L.~F.}\ \bibnamefont {Tocchio}}, \bibinfo {author}
  {\bibfnamefont {I.}~\bibnamefont {Tupitsyn}}, \bibinfo {author}
  {\bibfnamefont {S.~R.}\ \bibnamefont {White}}, \bibinfo {author}
  {\bibfnamefont {S.}~\bibnamefont {Zhang}}, \bibinfo {author} {\bibfnamefont
  {B.-X.}\ \bibnamefont {Zheng}}, \bibinfo {author} {\bibfnamefont
  {Z.}~\bibnamefont {Zhu}}, \ and\ \bibinfo {author} {\bibfnamefont
  {E.}~\bibnamefont {Gull}},\ }\href@noop {} {\  (\bibinfo {year} {2015})},\
  \bibinfo {note} {arXiv:1505.02290 [cond-mat.str-el]}\BibitemShut {NoStop}%
\bibitem [{\citenamefont {M\o{}ller}\ and\ \citenamefont
  {Plesset}(1934)}]{moller-1934}%
  \BibitemOpen
  \bibfield  {author} {\bibinfo {author} {\bibfnamefont {C.}~\bibnamefont
  {M\o{}ller}}\ and\ \bibinfo {author} {\bibfnamefont {M.~S.}\ \bibnamefont
  {Plesset}},\ }\href@noop {} {\bibfield  {journal} {\bibinfo  {journal} {Phys.
  Rev.}\ }\textbf {\bibinfo {volume} {46}},\ \bibinfo {pages} {618} (\bibinfo
  {year} {1934})}\BibitemShut {NoStop}%
\bibitem [{\citenamefont {Shavitt}\ and\ \citenamefont
  {Bartlett}(2009)}]{bartlett}%
  \BibitemOpen
  \bibfield  {author} {\bibinfo {author} {\bibfnamefont {I.}~\bibnamefont
  {Shavitt}}\ and\ \bibinfo {author} {\bibfnamefont {R.~J.}\ \bibnamefont
  {Bartlett}},\ }\href@noop {} {\emph {\bibinfo {title} {Many-Body Methods in
  Chemistry and Physics: MBPT and Coupled-Cluster Theory}}},\ Cambridge
  Molecular Science\ (\bibinfo  {publisher} {Cambridge University Press},\
  \bibinfo {address} {New York},\ \bibinfo {year} {2009})\BibitemShut {NoStop}%
\bibitem [{\citenamefont {Parks}\ and\ \citenamefont
  {Parr}(1958)}]{parks-1958}%
  \BibitemOpen
  \bibfield  {author} {\bibinfo {author} {\bibfnamefont {J.~M.}\ \bibnamefont
  {Parks}}\ and\ \bibinfo {author} {\bibfnamefont {R.~G.}\ \bibnamefont
  {Parr}},\ }\href@noop {} {\bibfield  {journal} {\bibinfo  {journal} {J. Chem.
  Phys.}\ }\textbf {\bibinfo {volume} {28}},\ \bibinfo {pages} {335} (\bibinfo
  {year} {1958})}\BibitemShut {NoStop}%
\bibitem [{\citenamefont {Surj\'an}(1999)}]{surjan}%
  \BibitemOpen
  \bibfield  {author} {\bibinfo {author} {\bibfnamefont {P.~R.}\ \bibnamefont
  {Surj\'an}},\ }in\ \href@noop {} {\emph {\bibinfo {booktitle} {Correlation
  and Localization}}},\ \bibinfo {series} {Topics in Current Chemistry}, Vol.\
  \bibinfo {volume} {203},\ \bibinfo {editor} {edited by\ \bibinfo {editor}
  {\bibfnamefont {P.~R.}\ \bibnamefont {Surj\'an}}}\ (\bibinfo  {publisher}
  {Springer Berlin Heidelberg},\ \bibinfo {year} {1999})\ pp.\ \bibinfo {pages}
  {63--88}\BibitemShut {NoStop}%
\bibitem [{\citenamefont {Rassolov}(2002)}]{rassolov-2002}%
  \BibitemOpen
  \bibfield  {author} {\bibinfo {author} {\bibfnamefont {V.~A.}\ \bibnamefont
  {Rassolov}},\ }\href@noop {} {\bibfield  {journal} {\bibinfo  {journal} {J.
  Chem. Phys.}\ }\textbf {\bibinfo {volume} {117}},\ \bibinfo {pages} {5978}
  (\bibinfo {year} {2002})}\BibitemShut {NoStop}%
\bibitem [{\citenamefont {Hartree}\ \emph {et~al.}(1939)\citenamefont
  {Hartree}, \citenamefont {Hartree},\ and\ \citenamefont
  {Swirles}}]{hartree-1939}%
  \BibitemOpen
  \bibfield  {author} {\bibinfo {author} {\bibfnamefont {D.~R.}\ \bibnamefont
  {Hartree}}, \bibinfo {author} {\bibfnamefont {W.}~\bibnamefont {Hartree}}, \
  and\ \bibinfo {author} {\bibfnamefont {B.}~\bibnamefont {Swirles}},\
  }\href@noop {} {\bibfield  {journal} {\bibinfo  {journal} {Phil. Trans. R.
  Soc. Lond. A}\ }\textbf {\bibinfo {volume} {238}},\ \bibinfo {pages} {229}
  (\bibinfo {year} {1939})}\BibitemShut {NoStop}%
\bibitem [{\citenamefont {Shepard}(2007)}]{shepard}%
  \BibitemOpen
  \bibfield  {author} {\bibinfo {author} {\bibfnamefont {R.}~\bibnamefont
  {Shepard}},\ }in\ \href@noop {} {\emph {\bibinfo {booktitle} {Ab Initio
  Methods in Quantum Chemistry -- II}}},\ \bibinfo {series} {Advances in
  Chemical Physics}, Vol.~\bibinfo {volume} {69},\ \bibinfo {editor} {edited
  by\ \bibinfo {editor} {\bibfnamefont {K.~P.}\ \bibnamefont {Lawley}}}\
  (\bibinfo  {publisher} {John Wiley \& Sons},\ \bibinfo {year} {2007})\ pp.\
  \bibinfo {pages} {63--200}\BibitemShut {NoStop}%
\bibitem [{\citenamefont {Roos}\ \emph {et~al.}(1980)\citenamefont {Roos},
  \citenamefont {Taylor},\ and\ \citenamefont {Siegbahn}}]{roos-1980}%
  \BibitemOpen
  \bibfield  {author} {\bibinfo {author} {\bibfnamefont {B.~O.}\ \bibnamefont
  {Roos}}, \bibinfo {author} {\bibfnamefont {P.~R.}\ \bibnamefont {Taylor}}, \
  and\ \bibinfo {author} {\bibfnamefont {P.~E.}\ \bibnamefont {Siegbahn}},\
  }\href@noop {} {\bibfield  {journal} {\bibinfo  {journal} {Chem. Phys.}\
  }\textbf {\bibinfo {volume} {48}},\ \bibinfo {pages} {157} (\bibinfo {year}
  {1980})}\BibitemShut {NoStop}%
\bibitem [{\citenamefont {Roos}(2007)}]{roos}%
  \BibitemOpen
  \bibfield  {author} {\bibinfo {author} {\bibfnamefont {B.~O.}\ \bibnamefont
  {Roos}},\ }in\ \href@noop {} {\emph {\bibinfo {booktitle} {Ab Initio Methods
  in Quantum Chemistry -- II}}},\ \bibinfo {series} {Advances in Chemical
  Physics}, Vol.~\bibinfo {volume} {69},\ \bibinfo {editor} {edited by\
  \bibinfo {editor} {\bibfnamefont {K.~P.}\ \bibnamefont {Lawley}}}\ (\bibinfo
  {publisher} {John Wiley \& Sons},\ \bibinfo {year} {2007})\ pp.\ \bibinfo
  {pages} {399--445}\BibitemShut {NoStop}%
\bibitem [{Note1()}]{Note1}%
  \BibitemOpen
  \bibinfo {note} {Some exploratory calculations were carried out using the
  full Fock space within each cluster. For half-filled systems in the on-site
  basis, the additional flexibility in the ansatz does not result in a lower
  variational estimate of the ground state energy. This, however, may not be
  true for doped systems, or if a full orbital optimization is carried
  out.}\BibitemShut {Stop}%
\bibitem [{\citenamefont {Lehoucq}\ \emph {et~al.}(1998)\citenamefont
  {Lehoucq}, \citenamefont {Sorensen},\ and\ \citenamefont {Yang}}]{lehoucq}%
  \BibitemOpen
  \bibfield  {author} {\bibinfo {author} {\bibfnamefont {R.}~\bibnamefont
  {Lehoucq}}, \bibinfo {author} {\bibfnamefont {D.}~\bibnamefont {Sorensen}}, \
  and\ \bibinfo {author} {\bibfnamefont {C.}~\bibnamefont {Yang}},\ }\href@noop
  {} {\emph {\bibinfo {title} {ARPACK Users' Guide}}}\ (\bibinfo  {publisher}
  {Society for Industrial and Applied Mathematics},\ \bibinfo {year}
  {1998})\BibitemShut {NoStop}%
\bibitem [{\citenamefont {Davidson}(1975)}]{davidson-1975}%
  \BibitemOpen
  \bibfield  {author} {\bibinfo {author} {\bibfnamefont {E.~R.}\ \bibnamefont
  {Davidson}},\ }\href@noop {} {\bibfield  {journal} {\bibinfo  {journal} {J.
  Comput. Phys.}\ }\textbf {\bibinfo {volume} {17}},\ \bibinfo {pages} {87}
  (\bibinfo {year} {1975})}\BibitemShut {NoStop}%
\bibitem [{\citenamefont {Sleijpen}\ and\ \citenamefont {{Van der
  Vorst}}(1996)}]{sleijpen-1996}%
  \BibitemOpen
  \bibfield  {author} {\bibinfo {author} {\bibfnamefont {G.~L.~G.}\
  \bibnamefont {Sleijpen}}\ and\ \bibinfo {author} {\bibfnamefont {H.~A.}\
  \bibnamefont {{Van der Vorst}}},\ }\href@noop {} {\bibfield  {journal}
  {\bibinfo  {journal} {{SIAM} J. Matrix Anal. Appl.}\ }\textbf {\bibinfo
  {volume} {17}},\ \bibinfo {pages} {401} (\bibinfo {year} {1996})}\BibitemShut
  {NoStop}%
\bibitem [{\citenamefont {Yeager}\ and\ \citenamefont {{J\o
  rgensen}}(1979)}]{yeager-1979}%
  \BibitemOpen
  \bibfield  {author} {\bibinfo {author} {\bibfnamefont {D.~L.}\ \bibnamefont
  {Yeager}}\ and\ \bibinfo {author} {\bibfnamefont {P.}~\bibnamefont {{J\o
  rgensen}}},\ }\href@noop {} {\bibfield  {journal} {\bibinfo  {journal} {J.
  Chem. Phys.}\ }\textbf {\bibinfo {volume} {71}},\ \bibinfo {pages} {755}
  (\bibinfo {year} {1979})}\BibitemShut {NoStop}%
\bibitem [{\citenamefont {Douady}\ \emph {et~al.}(1980)\citenamefont {Douady},
  \citenamefont {Ellinger}, \citenamefont {Subra},\ and\ \citenamefont
  {Levy}}]{douady-1980}%
  \BibitemOpen
  \bibfield  {author} {\bibinfo {author} {\bibfnamefont {J.}~\bibnamefont
  {Douady}}, \bibinfo {author} {\bibfnamefont {Y.}~\bibnamefont {Ellinger}},
  \bibinfo {author} {\bibfnamefont {R.}~\bibnamefont {Subra}}, \ and\ \bibinfo
  {author} {\bibfnamefont {B.}~\bibnamefont {Levy}},\ }\href@noop {} {\bibfield
   {journal} {\bibinfo  {journal} {J. Chem. Phys.}\ }\textbf {\bibinfo {volume}
  {72}},\ \bibinfo {pages} {1452} (\bibinfo {year} {1980})}\BibitemShut
  {NoStop}%
\bibitem [{\citenamefont {Siegbahn}\ \emph {et~al.}(1981)\citenamefont
  {Siegbahn}, \citenamefont {Alml\"of}, \citenamefont {Heiberg},\ and\
  \citenamefont {Roos}}]{siegbahn-1981}%
  \BibitemOpen
  \bibfield  {author} {\bibinfo {author} {\bibfnamefont {P.~E.~M.}\
  \bibnamefont {Siegbahn}}, \bibinfo {author} {\bibfnamefont {J.}~\bibnamefont
  {Alml\"of}}, \bibinfo {author} {\bibfnamefont {A.}~\bibnamefont {Heiberg}}, \
  and\ \bibinfo {author} {\bibfnamefont {B.~O.}\ \bibnamefont {Roos}},\
  }\href@noop {} {\bibfield  {journal} {\bibinfo  {journal} {J. Chem. Phys.}\
  }\textbf {\bibinfo {volume} {74}},\ \bibinfo {pages} {2384} (\bibinfo {year}
  {1981})}\BibitemShut {NoStop}%
\bibitem [{\citenamefont {Werner}\ and\ \citenamefont
  {Meyer}(1980)}]{werner-1980}%
  \BibitemOpen
  \bibfield  {author} {\bibinfo {author} {\bibfnamefont {H.}~\bibnamefont
  {Werner}}\ and\ \bibinfo {author} {\bibfnamefont {W.}~\bibnamefont {Meyer}},\
  }\href@noop {} {\bibfield  {journal} {\bibinfo  {journal} {J. Chem. Phys.}\
  }\textbf {\bibinfo {volume} {73}},\ \bibinfo {pages} {2342} (\bibinfo {year}
  {1980})}\BibitemShut {NoStop}%
\bibitem [{Note2()}]{Note2}%
  \BibitemOpen
  \bibinfo {note} {The alternating optimization strategy adopted may have poor
  convergence if the coefficients in the cluster mean-field state couple
  strongly to the orbital optimization degrees of freedom. This problem was
  indeed encountered for certain systems at low $U/t$.}\BibitemShut {Stop}%
\bibitem [{\citenamefont {K\'allay}\ and\ \citenamefont
  {Surj\'an}(2001)}]{kallay-2001}%
  \BibitemOpen
  \bibfield  {author} {\bibinfo {author} {\bibfnamefont {M.}~\bibnamefont
  {K\'allay}}\ and\ \bibinfo {author} {\bibfnamefont {P.~R.}\ \bibnamefont
  {Surj\'an}},\ }\href@noop {} {\bibfield  {journal} {\bibinfo  {journal} {J.
  Chem. Phys.}\ }\textbf {\bibinfo {volume} {115}},\ \bibinfo {pages} {2945}
  (\bibinfo {year} {2001})}\BibitemShut {NoStop}%
\bibitem [{\citenamefont {K\'allay}\ \emph {et~al.}()\citenamefont {K\'allay},
  \citenamefont {Rolik}, \citenamefont {Csontos}, \citenamefont {Ladj\'anszki},
  \citenamefont {Szegedy}, \citenamefont {Lad\'oczki},\ and\ \citenamefont
  {Samu}}]{mrcc}%
  \BibitemOpen
  \bibfield  {author} {\bibinfo {author} {\bibfnamefont {M.}~\bibnamefont
  {K\'allay}}, \bibinfo {author} {\bibfnamefont {Z.}~\bibnamefont {Rolik}},
  \bibinfo {author} {\bibfnamefont {J.}~\bibnamefont {Csontos}}, \bibinfo
  {author} {\bibfnamefont {I.}~\bibnamefont {Ladj\'anszki}}, \bibinfo {author}
  {\bibfnamefont {L.}~\bibnamefont {Szegedy}}, \bibinfo {author} {\bibfnamefont
  {B.}~\bibnamefont {Lad\'oczki}}, \ and\ \bibinfo {author} {\bibfnamefont
  {G.}~\bibnamefont {Samu}},\ }\href@noop {} {\enquote {\bibinfo {title}
  {{MRCC}, a quantum chemical program suite},}\ }\bibinfo {note}
  {\url{www.mrcc.hu}}\BibitemShut {NoStop}%
\bibitem [{\citenamefont {Zheng}\ and\ \citenamefont
  {Chan}(2015)}]{zheng-2015}%
  \BibitemOpen
  \bibfield  {author} {\bibinfo {author} {\bibfnamefont {B.-X.}\ \bibnamefont
  {Zheng}}\ and\ \bibinfo {author} {\bibfnamefont {G.~K.-L.}\ \bibnamefont
  {Chan}},\ }\href@noop {} {\  (\bibinfo {year} {2015})},\ \bibinfo {note}
  {arXiv:1504.01784 [cond-mat.str-el]}\BibitemShut {NoStop}%
\bibitem [{\citenamefont {Kleier}\ \emph {et~al.}(1974)\citenamefont {Kleier},
  \citenamefont {Halgren}, \citenamefont {Hall},\ and\ \citenamefont
  {Lipscomb}}]{kleier-1974}%
  \BibitemOpen
  \bibfield  {author} {\bibinfo {author} {\bibfnamefont {D.~A.}\ \bibnamefont
  {Kleier}}, \bibinfo {author} {\bibfnamefont {T.~A.}\ \bibnamefont {Halgren}},
  \bibinfo {author} {\bibfnamefont {J.~H.}\ \bibnamefont {Hall}}, \ and\
  \bibinfo {author} {\bibfnamefont {W.~N.}\ \bibnamefont {Lipscomb}},\
  }\href@noop {} {\bibfield  {journal} {\bibinfo  {journal} {J. Chem. Phys.}\
  }\textbf {\bibinfo {volume} {61}},\ \bibinfo {pages} {3905} (\bibinfo {year}
  {1974})}\BibitemShut {NoStop}%
\bibitem [{\citenamefont {L\"owdin}(1955)}]{lowdin-1955}%
  \BibitemOpen
  \bibfield  {author} {\bibinfo {author} {\bibfnamefont {P.-O.}\ \bibnamefont
  {L\"owdin}},\ }\href@noop {} {\bibfield  {journal} {\bibinfo  {journal}
  {Phys. Rev.}\ }\textbf {\bibinfo {volume} {97}},\ \bibinfo {pages} {1509}
  (\bibinfo {year} {1955})}\BibitemShut {NoStop}%
\bibitem [{\citenamefont {Imada}\ \emph {et~al.}(1992)\citenamefont {Imada},
  \citenamefont {Furukawa},\ and\ \citenamefont {Rice}}]{imada-1992}%
  \BibitemOpen
  \bibfield  {author} {\bibinfo {author} {\bibfnamefont {M.}~\bibnamefont
  {Imada}}, \bibinfo {author} {\bibfnamefont {N.}~\bibnamefont {Furukawa}}, \
  and\ \bibinfo {author} {\bibfnamefont {T.~M.}\ \bibnamefont {Rice}},\
  }\href@noop {} {\bibfield  {journal} {\bibinfo  {journal} {J. Phys. Soc.
  Jpn.}\ }\textbf {\bibinfo {volume} {61}},\ \bibinfo {pages} {3861} (\bibinfo
  {year} {1992})}\BibitemShut {NoStop}%
\bibitem [{\citenamefont {Essler}\ \emph {et~al.}(2005)\citenamefont {Essler},
  \citenamefont {Frahm}, \citenamefont {G\"ohmann}, \citenamefont {Kl\"umper},\
  and\ \citenamefont {Korepin}}]{essler}%
  \BibitemOpen
  \bibfield  {author} {\bibinfo {author} {\bibfnamefont {F.~H.~L.}\
  \bibnamefont {Essler}}, \bibinfo {author} {\bibfnamefont {H.}~\bibnamefont
  {Frahm}}, \bibinfo {author} {\bibfnamefont {F.}~\bibnamefont {G\"ohmann}},
  \bibinfo {author} {\bibfnamefont {A.}~\bibnamefont {Kl\"umper}}, \ and\
  \bibinfo {author} {\bibfnamefont {V.~E.}\ \bibnamefont {Korepin}},\
  }\href@noop {} {\emph {\bibinfo {title} {The One-Dimensional Hubbard
  Model}}}\ (\bibinfo  {publisher} {Cambridge University Press},\ \bibinfo
  {address} {Cambridge},\ \bibinfo {year} {2005})\BibitemShut {NoStop}%
\bibitem [{\citenamefont {Xu}\ \emph {et~al.}(2011)\citenamefont {Xu},
  \citenamefont {Chang}, \citenamefont {Walter},\ and\ \citenamefont
  {Zhang}}]{xu-2011}%
  \BibitemOpen
  \bibfield  {author} {\bibinfo {author} {\bibfnamefont {J.}~\bibnamefont
  {Xu}}, \bibinfo {author} {\bibfnamefont {C.-C.}\ \bibnamefont {Chang}},
  \bibinfo {author} {\bibfnamefont {E.~J.}\ \bibnamefont {Walter}}, \ and\
  \bibinfo {author} {\bibfnamefont {S.}~\bibnamefont {Zhang}},\ }\href@noop {}
  {\bibfield  {journal} {\bibinfo  {journal} {J. Phys.: Condens. Matter}\
  }\textbf {\bibinfo {volume} {23}},\ \bibinfo {pages} {505601} (\bibinfo
  {year} {2011})}\BibitemShut {NoStop}%
\bibitem [{\citenamefont {Rodr\'iguez-Guzm\'an}\ \emph
  {et~al.}(2014)\citenamefont {Rodr\'iguez-Guzm\'an}, \citenamefont
  {Jim\'enez-Hoyos},\ and\ \citenamefont {Scuseria}}]{rodriguez-2014}%
  \BibitemOpen
  \bibfield  {author} {\bibinfo {author} {\bibfnamefont {R.}~\bibnamefont
  {Rodr\'iguez-Guzm\'an}}, \bibinfo {author} {\bibfnamefont {C.~A.}\
  \bibnamefont {Jim\'enez-Hoyos}}, \ and\ \bibinfo {author} {\bibfnamefont
  {G.~E.}\ \bibnamefont {Scuseria}},\ }\href@noop {} {\bibfield  {journal}
  {\bibinfo  {journal} {Phys. Rev. B}\ }\textbf {\bibinfo {volume} {90}},\
  \bibinfo {pages} {195110} (\bibinfo {year} {2014})}\BibitemShut {NoStop}%
\bibitem [{\citenamefont {Juillet}\ and\ \citenamefont
  {Fr\'esard}(2013)}]{juillet-2013}%
  \BibitemOpen
  \bibfield  {author} {\bibinfo {author} {\bibfnamefont {O.}~\bibnamefont
  {Juillet}}\ and\ \bibinfo {author} {\bibfnamefont {R.}~\bibnamefont
  {Fr\'esard}},\ }\href@noop {} {\bibfield  {journal} {\bibinfo  {journal}
  {Phys. Rev. B}\ }\textbf {\bibinfo {volume} {87}},\ \bibinfo {pages} {115136}
  (\bibinfo {year} {2013})}\BibitemShut {NoStop}%
\bibitem [{\citenamefont {Chang}\ and\ \citenamefont
  {Zhang}(2010)}]{chang-2010}%
  \BibitemOpen
  \bibfield  {author} {\bibinfo {author} {\bibfnamefont {C.-C.}\ \bibnamefont
  {Chang}}\ and\ \bibinfo {author} {\bibfnamefont {S.}~\bibnamefont {Zhang}},\
  }\href@noop {} {\bibfield  {journal} {\bibinfo  {journal} {Phys. Rev. Lett.}\
  }\textbf {\bibinfo {volume} {104}},\ \bibinfo {pages} {116402} (\bibinfo
  {year} {2010})}\BibitemShut {NoStop}%
\bibitem [{\citenamefont {Eisert}\ \emph {et~al.}(2010)\citenamefont {Eisert},
  \citenamefont {Cramer},\ and\ \citenamefont {Plenio}}]{eisert-2010}%
  \BibitemOpen
  \bibfield  {author} {\bibinfo {author} {\bibfnamefont {J.}~\bibnamefont
  {Eisert}}, \bibinfo {author} {\bibfnamefont {M.}~\bibnamefont {Cramer}}, \
  and\ \bibinfo {author} {\bibfnamefont {M.~B.}\ \bibnamefont {Plenio}},\
  }\href@noop {} {\bibfield  {journal} {\bibinfo  {journal} {Rev. Mod. Phys.}\
  }\textbf {\bibinfo {volume} {82}},\ \bibinfo {pages} {277} (\bibinfo {year}
  {2010})}\BibitemShut {NoStop}%
\bibitem [{Note3()}]{Note3}%
  \BibitemOpen
  \bibinfo {note} {It is not strictly necessary to restrict the number parity
  of the Fock space in mean-field calculations, that is, if a simple product
  state will be considered. Nevertheless, a mixed-number parity description in
  each cluster complicates the evaluation of matrix elements in correlated
  approaches.}\BibitemShut {Stop}%
\bibitem [{\citenamefont {Morningstar}\ and\ \citenamefont
  {Weinstein}(1996)}]{morningstar-1996}%
  \BibitemOpen
  \bibfield  {author} {\bibinfo {author} {\bibfnamefont {C.~J.}\ \bibnamefont
  {Morningstar}}\ and\ \bibinfo {author} {\bibfnamefont {M.}~\bibnamefont
  {Weinstein}},\ }\href@noop {} {\bibfield  {journal} {\bibinfo  {journal}
  {Phys. Rev. D}\ }\textbf {\bibinfo {volume} {54}},\ \bibinfo {pages} {4131}
  (\bibinfo {year} {1996})}\BibitemShut {NoStop}%
\bibitem [{\citenamefont {Capponi}\ \emph {et~al.}(2004)\citenamefont
  {Capponi}, \citenamefont {L\"auchli},\ and\ \citenamefont
  {Mambrini}}]{capponi-2004}%
  \BibitemOpen
  \bibfield  {author} {\bibinfo {author} {\bibfnamefont {S.}~\bibnamefont
  {Capponi}}, \bibinfo {author} {\bibfnamefont {A.}~\bibnamefont {L\"auchli}},
  \ and\ \bibinfo {author} {\bibfnamefont {M.}~\bibnamefont {Mambrini}},\
  }\href@noop {} {\bibfield  {journal} {\bibinfo  {journal} {Phys. Rev. B}\
  }\textbf {\bibinfo {volume} {70}},\ \bibinfo {pages} {104424} (\bibinfo
  {year} {2004})}\BibitemShut {NoStop}%
\bibitem [{\citenamefont {Siu}\ and\ \citenamefont
  {Weinstein}(2007)}]{siu-2007}%
  \BibitemOpen
  \bibfield  {author} {\bibinfo {author} {\bibfnamefont {M.~S.}\ \bibnamefont
  {Siu}}\ and\ \bibinfo {author} {\bibfnamefont {M.}~\bibnamefont
  {Weinstein}},\ }\href@noop {} {\bibfield  {journal} {\bibinfo  {journal}
  {Phys. Rev. B}\ }\textbf {\bibinfo {volume} {75}},\ \bibinfo {pages} {184403}
  (\bibinfo {year} {2007})}\BibitemShut {NoStop}%
\bibitem [{\citenamefont {Parker}\ \emph {et~al.}(2013)\citenamefont {Parker},
  \citenamefont {Seideman}, \citenamefont {Ratner},\ and\ \citenamefont
  {Shiozaki}}]{parker-2013}%
  \BibitemOpen
  \bibfield  {author} {\bibinfo {author} {\bibfnamefont {S.~M.}\ \bibnamefont
  {Parker}}, \bibinfo {author} {\bibfnamefont {T.}~\bibnamefont {Seideman}},
  \bibinfo {author} {\bibfnamefont {M.~A.}\ \bibnamefont {Ratner}}, \ and\
  \bibinfo {author} {\bibfnamefont {T.}~\bibnamefont {Shiozaki}},\ }\href@noop
  {} {\bibfield  {journal} {\bibinfo  {journal} {J. Chem. Phys.}\ }\textbf
  {\bibinfo {volume} {139}},\ \bibinfo {pages} {021108} (\bibinfo {year}
  {2013})}\BibitemShut {NoStop}%
\bibitem [{\citenamefont {Parker}\ and\ \citenamefont
  {Shiozaki}(2014)}]{parker-2014}%
  \BibitemOpen
  \bibfield  {author} {\bibinfo {author} {\bibfnamefont {S.~M.}\ \bibnamefont
  {Parker}}\ and\ \bibinfo {author} {\bibfnamefont {T.}~\bibnamefont
  {Shiozaki}},\ }\href@noop {} {\bibfield  {journal} {\bibinfo  {journal} {J.
  Chem. Theory Comput.}\ }\textbf {\bibinfo {volume} {10}},\ \bibinfo {pages}
  {3738} (\bibinfo {year} {2014})}\BibitemShut {NoStop}%
\bibitem [{\citenamefont {Mila}\ and\ \citenamefont {Penc}(1995)}]{mila-1997}%
  \BibitemOpen
  \bibfield  {author} {\bibinfo {author} {\bibfnamefont {F.}~\bibnamefont
  {Mila}}\ and\ \bibinfo {author} {\bibfnamefont {K.}~\bibnamefont {Penc}},\
  }\href@noop {} {\bibfield  {journal} {\bibinfo  {journal} {Phys. Rev. B}\
  }\textbf {\bibinfo {volume} {51}},\ \bibinfo {pages} {1997} (\bibinfo {year}
  {1995})}\BibitemShut {NoStop}%
\end{thebibliography}%

\end{document}